\documentclass[11pt,twoside,a4paper,cmspaper,final,collab]{cms-tdr}
\def\svnVersion{3b0c26a}\def\svnDate{2026/09/04}\def\cmsCernNoTag{CERN-EP-2025-214}\def\cmsCernDate{\today}\def\cmsMessage{Published in Journal of Physics G: Nuclear and Particle Physics as \href{http://dx.doi.org/10.1088/1361-6471/ae8ff1}{\doi{10.1088/1361-6471/ae8ff1}.}}
\begin{document}\cmsNoteHeader{HIG-20-011}

\providecommand{\cmsLeft}{left\xspace}
\providecommand{\cmsRight}{right\xspace}
\newcommand{\vmu}{\ensuremath{\vec{\mu}}\xspace}
\newcommand{\vtheta}{\ensuremath{\vec{\nu}}\xspace}
\newcommand{\qqqq}{\ensuremath{4\Pq}\xspace}
\newcommand{\PA}{\ensuremath{\mathrm{A}}\xspace}
\newcommand{\Phlight}{\ensuremath{\mathrm{h}}\xspace}
\newcommand{\HH}{\ensuremath{{\PH\PH}}\xspace}
\newcommand{\mHH}{\ensuremath{m_{\PH\PH}}\xspace}
\newcommand{\Hgg}{\ensuremath{\PH\to\PGg\PGg}\xspace}
\newcommand{\Hbb}{\ensuremath{\PH\to \bbbar}\xspace}
\newcommand{\hbb}{\ensuremath{\PH\to\bbbar}\xspace}
\newcommand{\ww}{\ensuremath{\PW\PW}\xspace}
\newcommand{\hwwq}{\ensuremath{\PH\to\ww\to\qqqq}\xspace}
\newcommand{\HWW}{\ensuremath{\PH\to \PW\PW}\xspace}
\newcommand{\VVHH}{\ensuremath{\PV\PV\PH\PH}\xspace}

\newcommand{\ttH}{\ensuremath{\ttbar\PH}\xspace}
\newcommand{\ttHgg}{\ensuremath{\ttH(\PGg\PGg)}\xspace}

\newcommand{\GGF}{\ensuremath{\Pg\Pg\text{F}}\xspace}
\newcommand{\VBF}{\ensuremath{\text{VBF}}\xspace}
\newcommand{\sigmaHH}{\ensuremath{\sigma_{\PH\PH}\xspace }}
\newcommand{\sigmaVBF}{\ensuremath{\sigma_{\VBF~\PH\PH}\xspace}}

\newcommand{\bbgg}{\ensuremath{\bbbar\PGg\PGg}\xspace}
\newcommand{\bb}{\ensuremath{\bbbar}\xspace}
\newcommand{\bbbb}{\ensuremath{\bbbar\bbbar}\xspace}
\newcommand{\bbtt}{\ensuremath{\bbbar\PGt\PGt}\xspace}
\newcommand{\bbZZ}{\ensuremath{\bbbar\PZ\PZ}\xspace}
\newcommand{\bbWW}{\ensuremath{\bbbar\PW\PW}\xspace}
\newcommand{\WWgg}{\ensuremath{\PW\PW\PGg\PGg}\xspace}
\newcommand{\ttgg}{\ensuremath{\PGt\PGt\PGg\PGg}\xspace}

\newcommand{\klambda}{\ensuremath{\kappa_\lambda}\xspace}
\newcommand{\ktop}{\ensuremath{\kappa_{\PQt}}\xspace}
\newcommand{\CV}{\ensuremath{\kappa_\PV}\xspace}
\newcommand{\CVV}{\ensuremath{\kappa_{2\PV}}\xspace}
\newcommand{\cg}{\ensuremath{\text{c}_\Pg}\xspace}
\newcommand{\cgg}{\ensuremath{\text{c}_{2\Pg}}\xspace}
\newcommand{\ctwo}{\ensuremath{\text{c}_2}\xspace}
\newcommand{\MX}{\ensuremath{M_\mathrm{X}}\xspace}
\newcommand{\Mbb}{\ensuremath{m_{\bbbar}}\xspace}
\newcommand{\Mgg}{\ensuremath{m_{\PGg\PGg}}\xspace}
\newcommand{\Mbbgg}{\ensuremath{m_{\PQb\PQb\PGg\PGg}}\xspace}
\newcommand{\mreg}{\ensuremath{m_\text{reg}}\xspace}
\newcommand{\Dbb}{\ensuremath{D_{\bbbar}}\xspace}
\newcommand{\mSD}{\ensuremath{m_\text{SD}}\xspace}
\newcommand{\mregbb}{\ensuremath{\mreg^{\bbbar}}\xspace}
\newcommand{\Dbbbb}{\ensuremath{\Dbb^{\bbbar}}\xspace}
\newcommand{\ggfbdt}{\ensuremath{\mathrm{BDT}_\mathrm{ggF}}\xspace}
\newcommand{\vbfbdt}{\ensuremath{\mathrm{BDT}_\mathrm{VBF}}\xspace}
\newcommand{\partmodel}{\textsc{ParT}\xspace}

\newcommand{\PHiggs}{\ensuremath{\PH}\xspace}
\newcommand{\PLepton}{\ensuremath{\ell}\xspace}
\newcommand{\ZZ}{\ensuremath{\PZ\PZ}\xspace}
\newcommand{\WW}{\ensuremath{\PW\PW}\xspace}
\newcommand{\VHH}{\ensuremath{\PV\PHiggs\PHiggs}\xspace}
\newcommand{\WHH}{\ensuremath{\PW\PHiggs\PHiggs}\xspace}
\newcommand{\pWHH}{\ensuremath{\PWp\PHiggs\PHiggs}\xspace}
\newcommand{\mWHH}{\ensuremath{\PWm\PHiggs\PHiggs}\xspace}
\newcommand{\ZHH}{\ensuremath{\PZ\PHiggs\PHiggs}}
\newcommand{\fb}{\ensuremath{\,\text{fb}}\xspace}

\newcommand{\mhfst}{\ensuremath{m_{\PH_{1}}}\xspace}
\newcommand{\mhsnd}{\ensuremath{m_{\PH_{2}}}\xspace}
\newcommand{\mhh}{\ensuremath{m_{\HH}}\xspace}

\newcommand{\muth}{\ensuremath{\PGt_\PGm\tauh}\xspace}
\newcommand{\eleth}{\ensuremath{\PGt_\Pe\tauh}\xspace}
\newcommand{\thth}{\ensuremath{\tauh\tauh}\xspace}
\newcommand{\HHggtt}{\ensuremath{\PH\PH \to \ttgg}\xspace}
\newcommand{\mgg}{\ensuremath{m_{\PGg\PGg}}\xspace}
\ifthenelse{\boolean{cms@external}}{
    \providecommand{\cmsTable}[1]{\resizebox{\textwidth}{!}{#1}}
}{
    \providecommand{\cmsTable}[1]{#1}
}

\cmsNoteHeader{HIG-20-011}

\title{
    Combination of searches for nonresonant Higgs boson pair production in proton-proton collisions at \texorpdfstring{$\sqrt{s}=13\TeV$}{sqrt(s) = 13 TeV}
}
\titlerunning{
    Combination of searches for nonresonant \HH production in $\Pp\Pp$ collisions at 13\TeV
}

\date{\today}

\abstract{
    This paper presents a combination of searches for the nonresonant production of Higgs boson pairs (\HH) in proton-proton collisions at a centre-of-mass energy of 13\TeV.
    The data set was collected by the CMS experiment at the LHC from 2016 to 2018 and corresponds to a total integrated luminosity of 138\fbinv.
    The observed (expected) upper limit on the inclusive \HH production cross section relative to the standard model (SM) prediction is found to be 3.5 (2.5).
    Assuming all other Higgs boson couplings are equal to their SM values, the Higgs boson trilinear self-coupling modifier $\klambda=\lambda_3/\lambda_{3}^\text{SM}$ is constrained in the range $-1.35 \leq \klambda \leq 6.37$ at 95\% confidence level.
    Similarly, for the coupling modifier \CVV, which governs the interaction between two vector bosons and two Higgs bosons, we have excluded $\CVV=0$ at more than 5 standard deviations for all values of \klambda.
    At 95\% confidence level assuming other couplings are equal to their SM values, \CVV is constrained in the range $0.64 \leq \CVV \leq 1.40$.
    This work also studies \HH production in several new physics scenarios, using the Higgs effective field theory (HEFT) framework.
    The HEFT framework is further exploited to study various ultraviolet complete models with an extended Higgs sector and set constraints on specific parameters.
    An extrapolation of the results to the integrated luminosity expected after the high-luminosity upgrade of the LHC is reported as well.
}

\hypersetup{
    pdfauthor={CMS Collaboration},
    pdftitle={Combination of searches for nonresonant Higgs boson pair production in proton-proton collisions at sqrt(s) = 13 TeV},
    pdfsubject={CMS},
    pdfkeywords={CMS, Higgs, HH, BSM}
}

\maketitle

\section{Introduction}
\label{sec:intro}

In 2012, the ATLAS and CMS Collaborations at the CERN, LHC discovered a new particle with a mass of approximately 125\GeV~\cite{Higgs-Discovery_CMS, Higgs-Discovery_CMS_long, Higgs-Discovery_ATLAS}.
Extensive measurements of its properties, including the mass, spin, parity, couplings to other standard model (SM) particles, and decay width, have revealed that this particle is compatible with the SM Higgs boson (\PHiggs)~\cite{Nature2022, 2022ATLASNature}.
However, the question of the Higgs boson self-interaction remains unresolved.
Measurements of the Higgs boson trilinear ($\lambda_3$) and quartic self-couplings ($\lambda_4$) provide valuable information about the Brout--Englert--Higgs mechanism, the process by which a complex scalar doublet field---the Higgs doublet---generates the masses of the \PW and \PZ bosons through spontaneous electroweak (EW) symmetry breaking.
The findings will confirm the SM or point to new physics scenarios, such as extended Higgs sectors~\cite{Dawson:2017vgm,deBlas:2014mba,PhysRevD.8.1226,Branco:2011iw,Haber:2015pua,Kling:2016opi,Giudice:2007fh,Grober:2010yv,Contino:2013kra,Contino:2010rs}.
Beyond particle physics, the true nature of electroweak phase transition will shed light on cosmological questions related to inflation, gravitational waves, and baryogenesis~\cite{Huang:2016,RamseyMusolf:2024xxx,Mazumdar:2018dfl,Zhang:2023xxx,Bednyakov:2015xxx,Basler:2023xxx,Morrissey:2012db}.
While the quartic coupling is challenging to measure at the LHC and requires future colliders~\cite{Plehn:2005nk,Cepeda:2019klc,Chiesa:2020awd,Brigljevic:2024vuv}, the trilinear coupling can be directly accessed by studying Higgs boson pair production.

\begin{figure*}[htb]
    \centering
    \includegraphics[width=0.32\textwidth]{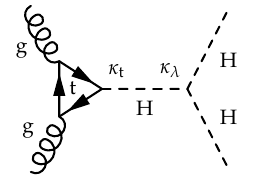}
    \includegraphics[width=0.32\textwidth]{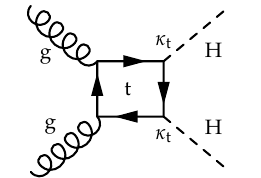}
    \caption{
        Feynman diagrams at leading order of nonresonant Higgs boson pair production via gluon-gluon fusion in the SM.
        The modifiers of the Higgs boson coupling with the top quark and the Higgs boson trilinear self-coupling are shown as \ktop and \klambda, respectively.
    }
    \label{fig:dihiggs-production-diagrams-ggf_SM}
\end{figure*}

At the LHC, pairs of SM Higgs bosons (\HH) are primarily produced via gluon-gluon fusion (\GGF), with a cross section of $31.05^{+0.06}_{-0.23}\fb$~\cite{2016_GGFXS,2019_GGFXS} at next-to-next-to-leading order (NNLO) in quantum chromodynamics (QCD) for a centre-of-mass energy of 13\TeV and a Higgs boson mass of 125\GeV, which is the assumed value for the studies in this paper.
At leading order (LO), two amplitudes contribute to the \GGF \HH production cross section, namely the Higgs boson self-interaction (`triangle diagram') and the top quark loop (`box diagram'), represented by the Feynman diagrams in Fig.~\ref{fig:dihiggs-production-diagrams-ggf_SM}.
The triangle diagram amplitude is proportional to the Higgs boson trilinear self-coupling, or in the theoretical framework described in Section~\ref{sec:MC}, to the coupling modifier $\klambda=\lambda_3/\lambda_{3}^\text{SM}$, where $\lambda_3^\text{SM}$ is the Higgs boson trilinear self-coupling in the SM.
The triangle and box diagram amplitudes interfere destructively, leading to a significant reduction in the \GGF cross section for certain values of \klambda.

\begin{figure*}[htb]
    \centering
    \includegraphics[width=0.32\textwidth]{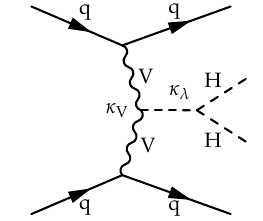}
    \includegraphics[width=0.32\textwidth]{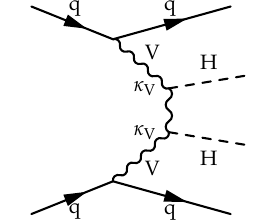}
    \includegraphics[width=0.32\textwidth]{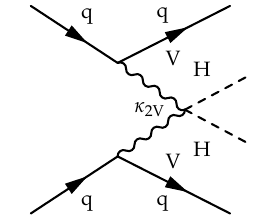}
    \caption{
        Feynman diagrams at leading order of nonresonant Higgs boson pair production via vector boson fusion in the SM.
        The modifiers of the Higgs boson coupling with a vector boson, the Higgs boson trilinear self-coupling, and the coupling between two Higgs bosons and two vector bosons are shown as \CV, \klambda, and \CVV, respectively.
    }
    \label{fig:dihiggs-production-diagrams-vbf}
\end{figure*}

\begin{figure*}[htb]
    \centering
    \includegraphics[width=0.32\textwidth]{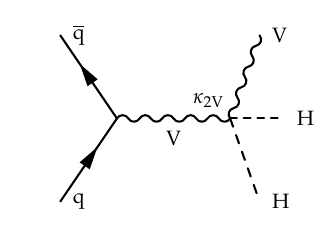}
    \includegraphics[width=0.32\textwidth]{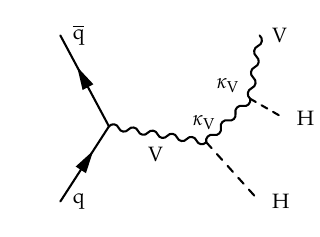}
    \includegraphics[width=0.32\textwidth]{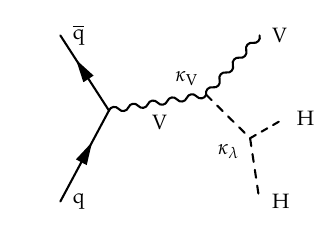}
    \caption{
        Feynman diagrams at leading order of nonresonant Higgs boson pair production associated with a vector boson in the SM.
        The modifiers of the Higgs boson coupling with a vector boson, the Higgs boson trilinear self-coupling, and the coupling between two Higgs bosons and two vector bosons are shown as \CV, \klambda, and \CVV, respectively.
    }
    \label{fig:dihiggs-production-diagrams-vhh}
\end{figure*}

The second largest production mechanism for \HH events is vector boson fusion (\VBF) with a cross section of $1.726\pm0.036\fb$~\cite{2018_VBFXS} in the SM at $\sqrt{s}=13\TeV$ and next-to-NNLO (N$^3$LO) in QCD.
The \VBF production mode, shown in Fig.~\ref{fig:dihiggs-production-diagrams-vbf}, gives access to the coupling modifier of the interaction between two vector bosons and two Higgs bosons (\CVV) and the interaction between a Higgs boson and two vector bosons (\CV).
The characteristic signature of \VBF, two jets in the forward direction, with a large pseudorapidity ($\eta$) separation, allows us to identify the process despite the small cross section.
Higgs boson pairs can also be produced in association with a vector boson $\PV=\PW$ or \PZ (\VHH) with a cross section of approximately half that of the \VBF mode, $0.36\pm0.03\fb$ for \ZHH, $0.33^{+0.03}_{-0.04}\fb$ for \pWHH and $0.17\pm0.01\fb$ for \mWHH~\cite{Baglio:2012np}, as shown in Fig.~\ref{fig:dihiggs-production-diagrams-vhh}.
The \VHH production is affected differently by the destructive interference compared to \GGF and \VBF because of the presence of additional diagrams at next-to-LO (NLO), offering complementary constraints on \klambda especially for $\klambda>0$~\cite{Grazzini_2018}.

\begin{figure*}[htb]
    \centering
    \includegraphics[width=0.32\textwidth]{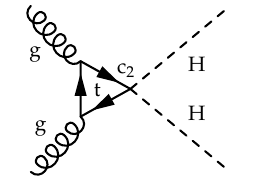}
    \includegraphics[width=0.32\textwidth]{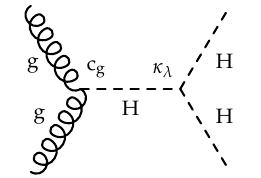}
    \includegraphics[width=0.32\textwidth]{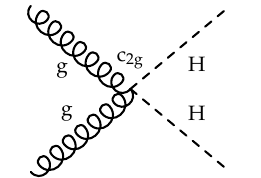}
    \caption{
        Feynman diagrams at leading order of nonresonant Higgs boson pair production via gluon-gluon fusion with anomalous Higgs boson couplings \ctwo, \cg, and \cgg.
        The Higgs boson trilinear self-coupling modifier is shown as \klambda.
    }
    \label{fig:dihiggs-production-diagrams-ggf_BSM}
\end{figure*}

In new physics models, additional diagrams that include couplings not present in the SM may contribute to \HH production.
These can be studied in the context of the Higgs effective field theory (HEFT)~\cite{Grober:2843280}.
The HEFT Lagrangian extends the SM Lagrangian with additional operators of dimension five or greater that are invariant under $\text{SU}(3)_C \times \text{U}(1)_\text{EM}$ and their corresponding couplings.
Figure~\ref{fig:dihiggs-production-diagrams-ggf_BSM} shows the anomalous couplings $\text{c}$ investigated in this paper.
In particular, \ctwo is the coupling between two top quarks and two Higgs bosons, \cg between two gluons and one Higgs boson, and \cgg between two gluons and two Higgs bosons.
The total \HH HEFT Lagrangian as a function of the anomalous coupling modifiers is:

\begin{equation}
    \label{eqn:LHEFT}
    \begin{aligned}
        \Delta \mathcal{L}_{\text{HEFT}} & =
        - m_\PQt \left( \ktop \frac{h}{v} + \ctwo \frac{h^2}{v^2} \right) \bar{t} t
        - \klambda \frac{m_\PH^2}{2v} h^3
        + \frac{\alpS}{8\pi} \left( \cg \frac{h}{v} + \cgg \frac{h^2}{v^2} \right)
        G^a_{\mu\nu} G^{a,\mu\nu}                                                             \\
                                         & + \left(2\CV\frac{h}{v}+\CVV\frac{h^2}{v^2}\right)
        \left( m_\PW^2 W_\mu^+W^{-\mu} + \frac{1}{2} m_\PZ^2Z_\mu Z^\mu \right).
    \end{aligned}
\end{equation}

The ATLAS and CMS Collaborations have performed previous \HH searches at $\sqrt{s} = 8$ and 13\TeV.
The ATLAS Collaboration has combined \HH searches in the \bbgg, \bbtt, and \bbbb final states at 13\TeV and set an upper limit on the inclusive production cross section of 2.9 times the SM expectation at 95\% confidence level (\CL), with 2.4 expected under the background-only hypothesis~\cite{ATLASCombination2024}.
A combined search in the \bbgg, \bbtt, \bbbb, \bbZZ, and multilepton final states by CMS at 13\TeV constrained (expected to constrain) the inclusive \HH production cross section to less than 3.4 (2.5) times the SM expectation at 95\% \CL~\cite{Nature2022}.
This paper supersedes previous results by introducing additional channels, including the \bbWW, \WWgg, and \ttgg final states and \VHH production mode in the \bbbb final state, to perform a variety of parameter scans, as well as searches for anomalous Higgs boson couplings in the HEFT approach.
The HEFT parametrisation is further exploited to constrain parameters in specific ultraviolet (UV) complete models.
The results presented in this paper and provided in the corresponding HEPData record~\cite{hepdata} are based on the data set of proton-proton ($\Pp\Pp$) collisions at 13\TeV collected by the CMS experiment from 2016 to 2018 (Run~2), corresponding to an integrated luminosity of 138\fbinv.

\section{The CMS detector and event reconstruction}
\label{sec:detector_reco}

The CMS apparatus~\cite{Chatrchyan:2008zzk,CMS:2023gfb} is a multipurpose, nearly hermetic detector, designed to trigger on~\cite{CMS:2020cmk,Khachatryan_2017,CMS:2024psu} and identify electrons, muons, photons, and (charged and neutral) hadrons~\cite{Khachatryan:2015hwa,CMS:2018rym,CMS:2014pgm}.
Its central feature is a superconducting solenoid of 6\unit{m} internal diameter, providing a magnetic field of 3.8\unit{T}.
Within the solenoid volume are a silicon pixel and strip tracker, a lead tungstate crystal electromagnetic calorimeter (ECAL), and a brass and scintillator hadron calorimeter (HCAL), each composed of a barrel and two endcap sections.
Forward calorimeters extend the pseudorapidity coverage provided by the barrel and endcap detectors.
Muons are reconstructed using gas-ionisation detectors embedded in the steel flux-return yoke outside the solenoid.
More detailed descriptions of the CMS detector, together with a definition of the coordinate system used and the relevant kinematic variables, can be found in Refs.~\cite{Chatrchyan:2008zzk,CMS:2023gfb}.

Events of interest are selected using a two-tiered trigger system.
The first level, composed of custom hardware processors, uses information from the calorimeters and muon detectors to select events at a rate of around 100\unit{kHz} within a fixed latency of 4\mus~\cite{CMS:2020cmk}.
The second level, known as the high-level trigger, consists of a farm of processors running a version of the full event reconstruction software optimised for fast processing, and reduces the event rate to a few kHz before data storage~\cite{Khachatryan_2017,CMS:2024psu}.

A particle-flow (PF) algorithm~\cite{PRF-14-001} aims to reconstruct and identify each particle in an event (PF candidate), with an optimised combination of information from the various elements of the CMS detector.
The primary vertex (PV) is taken to be the vertex corresponding to the hardest scattering in the event, evaluated using tracking information alone, as described in Section 9.4.1 of Ref.~\cite{CMS-TDR-15-02}.
The energy of photons is obtained from the ECAL measurement.
The energy of electrons is determined from a combination of the track momentum at the PV, the corresponding ECAL cluster energy, and the energy sum of all bremsstrahlung photons attached to the track.
The momentum of muons is obtained from the curvature of the corresponding track.
The energy of charged hadrons is determined from a combination of the track momentum and the corresponding ECAL and HCAL energies, corrected for the response function of the calorimeters to hadronic showers.
Finally, the energy of neutral hadrons is obtained from the corresponding corrected ECAL and HCAL energies.

Small-radius jets are reconstructed from PF candidates, using the anti-\kt clustering algorithm~\cite{Cacciari:2008gp,Cacciari:2011ma} with a distance parameter $R = 0.4$ (AK4 jets).
Charged particles not originating from the PV are excluded from the jet clustering, and an offset correction is applied to correct for remaining contributions.
Higgs boson decays into a pair of \PQb quarks with high transverse momentum (\pt) result in final states with large Lorentz boost, and as a result, the \PQb jets are overlapping, forming one large merged jet (``large-radius jet'') and substructure, \ie, the two overlapping jets are ``subjets'' of the large-radius jet.
The jets are reconstructed using the anti-\kt algorithm with a distance parameter $R = 0.8$ (AK8 jets).

The missing transverse momentum vector \ptvecmiss is computed as the negative vector \pt sum of all the PF candidates in an event, and its magnitude is denoted as \ptmiss~\cite{CMS:2019ctu}.
The \ptvecmiss is modified to account for corrections to the energy scale of the reconstructed jets in the event.

The searches presented in this paper use a variety of object identification techniques and algorithms, tailored to their specific characteristics.

\section{Signal modelling}
\label{sec:MC}

The \GGF \HH cross section as a function of \klambda and \ktop is
\begin{equation}
    \label{eqn:xsGGFHH}
    \sigma(\klambda, \ktop)=  \klambda^2 \ktop^2 \mathit{t} + \ktop^4 \mathit{b} + \klambda\ktop^3 \mathit{i},
\end{equation}
where $\mathit{t} = \abs{T}^2$, $\mathit{b} = \abs{B}^2$, and $\mathit{i} = \abs{T B^* + B^* T}$, are the three amplitudes that correspond to the triangle and box diagrams shown in Fig.~\ref{fig:dihiggs-production-diagrams-ggf_SM}, and their interference, respectively~\cite{Heinrich:2019bkc}.

The same formula holds for every differential cross section $d\sigma/dx$ for \GGF \HH production.
At higher order in the QCD perturbative expansion, $T$ and $B$ correspond to the sum of all diagrams of the same order in \klambda and \ktop, and the polynomial relation in Eq.~(\ref{eqn:xsGGFHH}) remains valid.
Since the \GGF cross section can be expressed as a polynomial in \klambda and \ktop with three independent terms, we can model the \GGF signal over a wide range of \klambda and \ktop by evaluating the cross section at three different combinations of values for (\klambda, \ktop) and performing a linear combination.
In our case, we use the information from three simulated data samples for the linear combination and that allows us to perform scans in the \klambda and \ktop parameters.

An event-level reweighting method is used to study modified SM couplings and couplings not present in the SM.
The reweighting parametrises the differential cross section as a function of the invariant mass of the \HH pair and their angular separation in the azimuthal plane, which fully characterise the system.
This enables modelling different combinations of (\klambda, \ktop, \ctwo, \cg, \cgg) present in $\Delta\mathcal{L}_{\text{HEFT}}$ of Eq.~(\ref{eqn:LHEFT}), including combinations without a corresponding generated sample.
The method used to derive the reweighting factors is described in Refs.~\cite{deFlorian:2021azd,Carvalho:2017vnu}.
The procedure is similar to the recommendation by the LHC Higgs Working Group 4~\cite{Grober:2843280}.
In our derivation, we use finer bins for high values of the invariant mass of the \HH system.
The results are validated using simulated samples at the generator level with custom combinations of the coupling parameters.
The method is used for all \HH decay channels except the \bbbb channel with high Lorentz boost, where the reweighting factors are not precise enough because of the limited number of simulated events in the high \mHH bins.
Therefore, all necessary Monte Carlo (MC) simulated samples for each benchmark are produced.

Two sets of benchmarks are produced using the reweighting method.
The first set defines thirteen benchmark scenarios~\cite{Carvalho_2016} spanning a broad range of the coupling values in the HEFT parameterisation at LO precision, noted as JHEP04(2016)126.
The second set~\cite{Capozi_2020}, noted as JHEP03(2020)091, was derived with NLO precision and considers direct and indirect constraints on the allowed range for the coupling values.
The aim of both sets of benchmarks is to sample the parameter space describing the signal according to kinematic features of the final state.
The values of the coupling modifiers corresponding to each benchmark are shown in Tables~\ref{tab:JHEP04}~and~\ref{tab:JHEP03}.

We further study the \ctwo modifier and perform scans with \ctwo as the parameter of interest.
While \ctwo, \ktop, and \klambda are free parameters in the HEFT Lagrangian it is of interest to check for correlations.

Finally, constraints are set on specific UV complete models.
The parameters of each model can be written in terms of \klambda, \ktop, and \ctwo based on a mapping described in Ref.~\cite{Carvalho:2017vnu}.
Three types of models are studied in this paper: models with one additional scalar~\cite{Dawson:2017vgm,deBlas:2014mba}, models with two Higgs doublets (2HDM)~\cite{PhysRevD.8.1226,Branco:2011iw,Haber:2015pua,Kling:2016opi}, and composite Higgs models~\cite{Giudice:2007fh,Grober:2010yv,Contino:2013kra,Contino:2010rs}.

The scans in the parameters of interest \klambda, \ktop, \CV, and \CVV are performed using a linear combination of a set of \VBF and \VHH samples, similar to what is done for \GGF.
Modifications of Higgs boson couplings to other particles that affect \HH and single Higgs boson production are taken into account in all cases~\cite{HIG-23-006}.

The simulated events for \GGF production are produced using NLO matrix elements with the model described in Ref.~\cite{Heinrich:2019bkc}, implemented in \POWHEG v2~\cite{POWHEG1,POWHEG2,POWHEG3}.
The cross section is corrected to the NNLO value as a function of \klambda according to Refs.~\cite{leshouches2019physics, Grazzini_2018}.
Four \GGF samples are produced with different values of the trilinear coupling modifier ($\klambda = 0$, 2.45, 5.0, and 1.0).
The simulated signal samples for \VBF and \VHH production modes are generated at LO using \MGvATNLO 2.6.5~\cite{MadGraph5_aMCatNLO,Alwall_2007,Frederix_2012}.
Seven \VBF samples are generated with different values of the coupling modifiers $(\CV, \CVV, \klambda) = (1, 1, 1)$, (1, 1, 0), (1, 1, 2), (1, 0, 1), (1, 2, 1), (0.5, 1, 1), and (1.5, 1, 1), and eight \VHH samples are generated with the same coupling modifier values plus $(\CV, \CVV, \klambda) = (1, 1, 20)$. This last benchmark is used in the \VHH case where \klambda is less constrained, therefore we need to model large values of \klambda.
The total cross section is corrected to the corresponding NNLO cross section~\cite{Dreyer:2018rfu,Baglio:2012np}.
The parton shower and hadronisation are simulated with \PYTHIA 8.230~\cite{pythia8} using the {CUETP8M1}~\cite{CMS_Pythia8_tunes_2020} set of tuned parameters for 2016 simulations and the {CP5}~\cite{CMS_Pythia8_CR_tunes_2023} tune set in 2017 and 2018 simulations.
The simulated samples produced with \PYTHIA and the \textsc{CUETP8M1} (\textsc{CP5}) tune use the \textsc{NNPDF3.0} (\textsc{NNPDF3.1}) parton distribution functions (PDFs)~\cite{NNPDF1,NNPDF3,Ball:2017nwa}, while samples produced with \MGvATNLO or \POWHEG v2 use the \textsc{NNPDF3.1} set.
The CMS detector response is modelled using \GEANTfour~\cite{geant4}.

\begin{table*}[htpb]
    \centering
    \topcaption{
        Values of the effective Lagrangian couplings for the Higgs effective field theory benchmarks proposed in Ref.~\cite{Carvalho_2016} and referred to in this paper as JHEP04(2016)126.
    }
    \label{tab:JHEP04}
    \cmsTable{
        \begin{tabular}{lccccccccccccc}
            \hline
            {}       & {1}    & {2}    & {3}    & {4}    & {5}    & {6}    & {7}    & {8}    & {9}    & {10}   & {11}   & {12} & {8a} \\
            \hline
            \klambda & 7.5    & 1.0    & 1.0    & $-3.5$ & 1.0    & 2.4    & 5.0    & 15.0   & 1.0    & 10.0   & 2.4    & 15.0 & 1.0  \\
            \ktop    & 1.0    & 1.0    & 1.0    & 1.5    & 1.0    & 1.0    & 1.0    & 1.0    & 1.0    & 1.5    & 1.0    & 1.0  & 1.0  \\
            \ctwo    & $-1.0$ & 0.5    & $-1.5$ & $-3.0$ & 0.0    & 0.0    & 0.0    & 0.0    & 1.0    & $-1.0$ & 0.0    & 1.0  & 0.5  \\
            \cg      & 0.0    & $-0.8$ & 0.0    & 0.0    & 0.8    & 0.2    & 0.2    & $-1.0$ & $-0.6$ & 0.0    & 1.0    & 0.0  & 0.27 \\
            \cgg     & 0.0    & 0.6    & $-0.8$ & 0.0    & $-1.0$ & $-0.2$ & $-0.2$ & 1.0    & 0.6    & 0.0    & $-1.0$ & 0.0  & 0.0  \\
            \hline
        \end{tabular}}
\end{table*}

\begin{table*}[htpb]
    \centering
    \topcaption{
        Values of the effective Lagrangian couplings for the Higgs effective field theory benchmarks proposed in Ref.~\cite{Capozi_2020} and referred to in this paper as JHEP03(2020)091.
    }
    \label{tab:JHEP03}
    \begin{tabular}{lccccccc}
        \hline
        {}       & {1}     & {2}  & {3}     & {4}     & {5}     & {6}     & {7}     \\
        \hline
        \klambda & 3.94    & 6.84 & 2.21    & 2.79    & 3.95    & 5.68    & $-0.10$ \\
        \ktop    & 0.94    & 0.61 & 1.05    & 0.61    & 1.17    & 0.83    & 0.94    \\
        \ctwo    & $-0.33$ & 0.33 & $-0.33$ & 0.33    & $-0.33$ & 0.33    & 1.0     \\
        \cg      & 0.75    & 0.0  & 0.75    & $-0.75$ & 0.25    & $-0.75$ & 0.25    \\
        \cgg     & $-1$    & 1    & $-1.5$  & $-0.5$  & 1.5     & $-1$    & 1       \\
        \hline
    \end{tabular}
\end{table*}

\section{Analysis strategy}
\label{sec:strategy}

The list of \HH analyses considered in this combination, along with a summary of results, are listed in Table~\ref{tab:HHanalyses}.
A brief description of each analysis is given later in this section.
More details can be found in the respective publications.
For the purposes of this combination, extensive studies were performed in order to avoid double counting of events among the different analysis.
Specific actions taken by the various analyses are described in each section.
Two channels, \bbWW, $\PW\PW\to\qqqq$ and \WWgg, have not been previously published, and therefore more details are provided in the corresponding sections.
The \bbWW analysis is the first search for $\HH\to\bbWW$ in the all-hadronic final state.

\begin{table*}[ht]
    \centering
    \topcaption{
        Summary of results for the \HH analyses included in this combination.
        The second column is the observed (expected) 95\% \CL upper limit on the inclusive signal strength $\mu= \sigmaHH/\sigmaHH^\text{SM}$, with one exception where $\mu_{\VHH}= \sigma_{\VHH}/\sigma_{\VHH}^\text{SM}$ is shown.
        The third (fourth) column is the allowed 68\% \CL interval for the coupling modifier \klambda (\CVV).
        The last column indicates whether the analysis is included in the results using the HEFT parametrisation.
    }
    \label{tab:HHanalyses}
    \cmsTable{
        \begin{tabular}{ l  c  c  c  c}
            \hline
            {Analysis}                                                            & {$\mu$}                  & {$\klambda$}    & {\CVV}          & {HEFT}     \\
            \hline
            \bbbb, resolved jets~\cite{CMS_HHbbbb_resolved_R2Legacy}              & 3.9 (7.8)                & [$-2.3$, 9.4]   & [$-0.1$, 2.2]   & \checkmark \\
            \bbbb, merged jets~\cite{CMS_HHbbbb_boosted_R2Legacy}                 & 9.9 (5.1)                & [$-9.9$, 16.9]  & [0.62, 1.41]    & \checkmark \\
            \VHH, $\PH\PH\to\bbbb$~\cite{Hayrapetyan:2895302}                     & $\mu_{\VHH} < 294$ (124) & [$-37.7$, 37.2] & [$-12.2$, 8.9]  & \NA        \\
            \\
            \bbtt~\cite{CMS_bbtautau_R2Legacy}                                    & 3.3 (5.2)                & [$-1.7$, 8.7]   & [$-0.4$, 2.6]   & \checkmark \\
            \\
            \bbgg~\cite{CMS_HHbbgg_R2Legacy}                                      & 7.7 (5.2)                & [$-3.3$, 8.5]   & [$-1.3$, 3.5]   & \checkmark \\
            \\
            \bbWW, $\PW\PW\to\ell\nu\Pq\Pq/2\ell 2\nu$~\cite{CMS_HHbbWW_R2Legacy} & 14 (18)                  & [$-7.2$, 13.8]  & [$-8.7$, 15.2]  & \checkmark \\
            \bbWW, $\PW\PW\to\qqqq$                                               & 141 (69)                 & \NA             & [$-0.04$, 2.05] & \NA        \\
            \\
            \HH multilepton~\cite{CMS_HHmultilepton_R2Legacy}                     & 21.3 (19.4)              & [$-6.9$, 11.1]  & \NA             & \checkmark \\
            \\
            \ttgg~\cite{CMS-PAS-HIG-22-012}                                       & 33 (26)                  & [$-13$, 18]     & \NA             & \NA        \\
            \\
            \bbZZ~\cite{CMS_HHbbZZ4l_R2Legacy}                                    & 32.4 (39.6)              & [$-8.8$, 13.4]  & \NA             & \NA        \\
            \\
            \WWgg                                                                 & 97 (53)                  & [$-25.8$, 14.4] & \NA             & \checkmark \\
            \hline
        \end{tabular}}
\end{table*}

Where required, the event selection of each analysis was modified for the combination to prevent double counting of events.
The strategy for this overlap removal is based upon work done for the combinations in Ref.~\cite{Nature2022}.

\subsection{\texorpdfstring{$\HH\to\bbbb$}{HH to bbbb}}
\begin{figure*}[htb]
    \centering
    \includegraphics[width=0.3\textwidth]{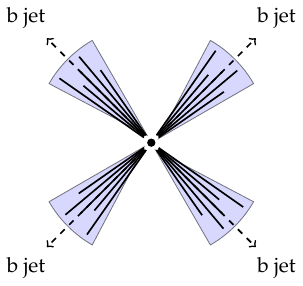}
    \includegraphics[width=0.6\textwidth]{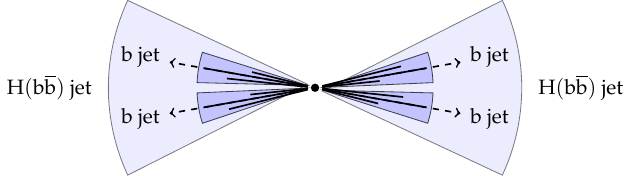}\\
    \includegraphics[width=0.3\textwidth]{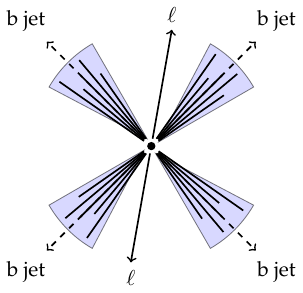}
    \includegraphics[width=0.3\textwidth]{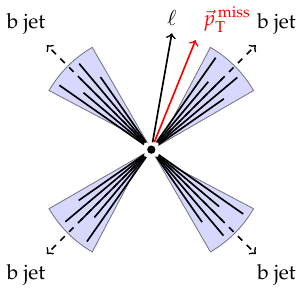}
    \caption{
        Illustrations of the resolved (upper left) and boosted (upper right) topologies of the \bbbb decay channel.
        The topology of \VHH production when all Higgs bosons decay to \PQb jets and the vector boson is either a \PZ boson that decays into two leptons (lower left) or a \PW boson decaying into a lepton and a neutrino, giving missing transverse momentum (lower right).
    }
    \label{4b_sketches}
\end{figure*}

The \bbbb decay channel has the largest \HH branching fraction (33.6\%) among the \HH decays to SM particles.
Three analyses target the \bbbb decay channel.
The first two search for \HH pair produced via \GGF or \VBF.
In one case the topology of the \PQb jets is resolved while in the other the \PQb jets are produced with high Lorentz boost resulting in a merged signature, where pairs of jets are reconstructed as a single large radius jets.
The third analysis is dedicated to the search for \VHH production.
In this case, the final state is characterised by the decay mode of the vector boson.
Figure~\ref{4b_sketches} illustrates the topology of each analysis.

\subsubsection{\texorpdfstring{$\HH\to\bbbb$}{HH to bbbb} resolved}

The resolved search~\cite{CMS_HHbbbb_resolved_R2Legacy} focuses on the kinematic phase space where each Higgs boson is reconstructed from two small-radius \PQb-tagged jets.
It explores both the \GGF and \VBF \HH production modes.
The \VBF topology is characterised by two additional jets with high invariant mass and large pseudorapidity separation $\Delta\eta$.
The online trigger selection requires the presence of at least four jets, satisfying thresholds on jet \pt and \HT, defined as the scalar \pt sum of all the jets.
These thresholds vary depending on the data-taking year between 30 and 90\GeV for \pt and 300 to 330\GeV for \HT.
Consequently, data collected in 2016 are analysed separately from those collected in 2017 and 2018.

Events selected offline contain at least four jets.
Jets originating from a \PQb quark decay are identified using the \textsc{DeepJet}~\cite{Bols:2020bkb} algorithm.
The four jets with the highest \textsc{DeepJet} score are chosen as \PQb jet candidates.
The \pt of the four \PQb jets is corrected using a multivariate regression method based on a deep neural network (DNN)~\cite{b_regression}.
At least three \PQb jets must satisfy the \textsc{DeepJet} medium working point (WP) \textsc{DeepJet}~\cite{Bols:2020bkb}, which corresponds to a \PQb jet identification efficiency of about 75\% and a misidentification rate for light-flavour quark and gluon jets of about 1\%.
The \VBF jet candidates are the two highest \pt non-\PQb jets located in opposite $\eta$ hemispheres.
Events that contain any isolated electrons or muons are rejected.

There are three possible ways to pair the four \PQb jets in order to reconstruct two Higgs boson candidates.
For each pairing, a distance parameter $d = \abs{\mhfst - k \mhsnd}/\sqrt{1 + k^{2}}$ is calculated, where \mhfst is the mass of the Higgs boson candidate with the highest \pt, \mhsnd is the mass of the other Higgs boson, and $k=1.04$.
For correctly paired events in MC simulation, $k$ is the expected ratio of the peak positions of the reconstructed Higgs boson masses.
Its value is greater than one because the energy regression depends on jet \pt, which has a larger effect on the less energetic \PH candidate.
If the difference between the two smallest distance parameters is larger than 30\GeV, then the pairing with the smallest distance parameter is selected.
Otherwise, between the two pairings with the smallest distance, the one maximising the \pt of the two \PH candidates in the four-jet centre-of-mass frame is chosen.
This procedure results in a correct jet pairing of about 82--96 (91--98)\% of the selected events for the different couplings studied in \GGF (\VBF) simulated signal events.

Events that do not have a \VBF jet pair are assigned to the \GGF event category.
A $\text{BDT}_{\GGF/\VBF}$ discriminant is trained to separate \GGF and \VBF \HH signals in events that contain the \VBF jet pair.
The $\text{BDT}_{\GGF/\VBF}$ is trained using SM \GGF signal and \VBF signal with $\CVV=2$.
A threshold on the $\text{BDT}_{\GGF/\VBF}$ score is used to classify those events into the \GGF or \VBF event categories.
The \GGF and \VBF categories are each further divided into subcategories to optimise the search sensitivity to anomalous coupling hypotheses.
In the \GGF category, events are separated into low-mass and high-mass categories using a 450\GeV threshold on the Higgs boson pair invariant mass \mhh.
In the \VBF category, events are divided into ``SM-like" and anomalous-\CVV categories using a threshold on $\text{BDT}_{\GGF/\VBF}$.

The dominant background source is from SM events composed uniquely of jets produced through the strong interaction, referred to as QCD multijet events.
The analysis signal region (SR) contains at least four \PQb jet candidates satisfying the \textsc{DeepJet} medium WP.
The background of this `4\PQb' region is estimated using events from an orthogonal region in which the fourth-highest \PQb-tagged jet fails the medium WP (or `3\PQb' region).
The transfer function to model the 4\PQb background from the 3\PQb data, comprises a BDT reweighting and a transfer factor (TF) for reshaping and normalisation corrections.

To remove overlapping events between this analysis and the analysis of \bbbb events in the boosted topology described in Section~\ref{sec:bbbb_boosted}, an event veto is applied for the combination described in this paper.
Events containing two massive large-radius jets with $\pt>300\GeV$ and at least two subjets are removed from this analysis.
The \pt threshold was optimised based on the expected \HH signal strength.

\subsubsection{\texorpdfstring{$\HH\to\bbbb$}{HH to bbbb} boosted}
\label{sec:bbbb_boosted}

In the \bbbb boosted final state~\cite{CMS_HHbbbb_boosted_R2Legacy}, both Higgs bosons are highly Lorentz boosted, decay to \bbbar pairs, and are reconstructed as large-radius jets.
The search is sensitive to modified coupling scenarios that enhance the production of highly boosted Higgs boson pairs, including through \VBF production.

Several trigger algorithms with requirements on \HT or large-radius jet \pt are used to select events online.
This search uses the soft-drop (SD) algorithm~\cite{Larkoski:2014wba} to reconstruct the Higgs boson candidate masses (\mSD).

The graph neural network (NN) algorithm~\cite{neuraljets,Qu:2019gqs,Moreno:2019bmu,Moreno:2019neq} known as \textsc{ParticleNet}~\cite{Qu:2019gqs,CMS-PAS-BTV-22-001} is used to discriminate between \Hbb and QCD jets.
A regression algorithm based on \textsc{ParticleNet} is also used to improve the jet mass estimation (\mreg)~\cite{CMS-DP-2021-017}.

Events with two or more large-radius jets are selected and grouped into mutually exclusive \GGF and \VBF categories.
Events with two additional small-radius jets in the opposing forward regions of the detector, consistent with the characteristic \VBF signature, are sorted into the \VBF categories.
Three \VBF event categories (high, medium, and low purity) are defined using the \textsc{ParticleNet} discriminant (\Dbb) scores of the Higgs boson candidate jets.

The \GGF categories consist of events that fail the criteria for the \VBF categories.
A BDT is trained to discriminate the \HH signal from QCD multijet or \ttbar backgrounds based on the \mSD, $\eta$, and \Dbb of the \Dbb-leading jet, and additional kinematic and substructure properties of both Higgs boson candidate jets.
Three SRs targeting the \GGF production are constructed depending on the BDT discriminant and the \Dbb-subleading large-radius jet \Dbb score.
The \mreg distribution of the \Dbb-subleading jet in each \GGF category is fit simultaneously to extract the signal.

The dominant SM backgrounds are \ttbar and QCD multijet production.
Control regions (CRs) for the QCD multijet process are defined by inverting or loosening the requirement on the \Dbb score.
The QCD multijet background in the SRs is estimated using the data in the CRs and fitted TFs~\cite{Sirunyan:2017isc,Sirunyan:2018qca,Sirunyan:2020hwz}.
Corrections for the \ttbar background estimate are derived from a data sample enriched in \ttbar events with one \PW boson decaying leptonically~\cite{Sirunyan:2020lcu}.

\subsubsection{\texorpdfstring{$\VHH$, $\HH\to\bbbb$}{VHH, HH to bbbb}}
\label{sec:vhh}

This analysis~\cite{Hayrapetyan:2895302} focuses on the final state with both Higgs bosons decaying into a \PQb quark-antiquark pair.
All decay modes of vector bosons \PZ and \PW are considered.
Experimentally, the events are divided into four categories based on the presence of a light lepton \Pell (=$\Pgm,\,\Pe$) and jets: 2 leptons, 1 lepton, \ptmiss and fully hadronic channel, corresponding to the $\PZ\to \Pell\Pell$, $\PW\to \Pell\PGn$, $\PZ\to \PGn\PGn$, and $\PZ/\PW\to \Pq\Pq$ processes, respectively.
Because of the large overlap with the \GGF and \VBF \bbbb channels, the \VHH fully hadronic channel is removed from this combination.
Two event topologies are explored in this analysis: one involving four small-radius \PQb jets, and the other with two large-radius jets, each from an $\PH\to\bb$ decay.

In the leptonic channels, events are selected using triggers requiring isolated leptons or large \ptmiss.
In the resolved topology, at least three AK4 jets are required to pass the medium WP of the \textsc{DeepJet} \PQb tagging algorithm.
If there are more than four \PQb-tagged jets, the ones with the highest \PQb tagging scores are selected.
An energy-regression method~\cite{b_regression} is applied to the \PQb jets in order to improve the resolution of the dijet mass.
The two Higgs boson candidates are reconstructed by pairing the four jets with the same method used in the resolved \bbbb analysis.
In the boosted topology, two AK8 jets are required to have \textsc{ParticleNet} \Dbb scores greater than 0.8.
Given the higher sensitivity in the boosted topology, it is given priority if an event can be assigned to both topologies.

A categorisation BDT is trained to divide events into regions sensitive to \klambda and \CVV.
In each region, another BDT is trained using all the signal processes in order to optimise the signal versus background discrimination.
The final signal strength is extracted by fitting the BDT score distribution in the signal and CRs.

\subsection{\texorpdfstring{$\HH\to\bbbar{\PGt}{\PGt}$}{HH to bbtautau}}
\label{sec:bbtautau}
In the \bbtt channel~\cite{CMS_bbtautau_R2Legacy}, the events are characterised by the decay mode of the two $\PGt$ leptons. With branching fraction 7.3\%, this channels combines relatively large branching fraction and easy to reconstruct final states.
This analysis studies final states where at least one of the $\PGt$ leptons decays hadronically, \ie \muth, \eleth, and \thth.
The light leptons and $\tauh$ candidates must have opposite electric charges.
The $\tauh$ candidates are reconstructed using the \textsc{DeepTau} algorithm~\cite{CMS:2022prd}, an NN method discriminating $\tauh$ candidates from jets, electrons, and muons.
Events are required to have at least two \PQb jets identified using the \textsc{DeepJet} algorithm.

The events are selected by triggers requiring a single light lepton, or a lepton accompanied by a \tauh candidate or two \tauh candidates.
A trigger dedicated to the VBF topology was introduced in late 2017, requiring two \tauh candidates and two additional jets.

For each event, all \PQb jet candidates are assigned a score by an NN trained to recognise the $\PH\to\bbbar$ topology.
The two jets with the highest score are selected as the \PQb jets from the Higgs boson decays.
The events that have two additional small-radius jets are identified as originating from the \VBF process.

After the event selection, the events are divided into categories.
The first set of categories splits the events according to the number and topology of the jets.
Events with two overlapping small-radius jets are classified in the boosted category.
All remaining events are classified in the resolved categories.
The resolved category is further split according to whether one or both jets pass the medium WP of the \textsc{DeepJet} \PQb tagging algorithm.
Finally, an additional \VBF category selects events with at least two \VBF jet candidates.
The contamination of this category from \GGF events with an extra jet is significant; therefore, a multiclassifier DNN is trained.
The DNN splits the events in the \VBF category into signal-like \VBF events, \GGF contamination, \ttH, \ttbar, and Drell--Yan (DY) background events.

All selected events are subject to a selection on the two-dimensional plane of the invariant mass of the dijet (\Mbb) and ditau ($m_{\PGt\PGt}$) systems, where the $m_{\PGt\PGt}$ is reconstructed using the \textsc{SVFIT} algorithm~\cite{Bianchini_2017}.
This selection is optimised separately for the resolved and the boosted categories, and is not applied to the \VBF categories.

The eight event categories defined earlier are used for the signal extraction using a DNN developed to identify $\HH\to\bbtt$ events.
The DNN is trained to classify the events from each event category as signal or background-like by assigning a single prediction per event.
The maximum likelihood fit for signal extraction is performed using the DNN distribution simultaneously for all categories and all channels.

The main background processes for this search, \ttbar, DY, and QCD multijet, are estimated using control samples in data.
The shapes of the \ttbar and DY processes are taken from the simulation and then scaled using background-enriched regions.
The scaling is extracted with a simultaneous fit over all the CRs and applied to the simulated events in the SR.

The QCD multijet process is entirely determined from data in jet-enriched regions where two selection requirements for the \PGt pair are inverted.
The event yield and shape are obtained in a region where the \tauh is not required to be isolated from other tracks.
In \thth events, only the \tauh with the lowest \pt is allowed to not be isolated.
Then, the yield is scaled by the ratio of event yield in a region where the opposite charge requirement is inverted and a region where both the charge and the isolation are inverted.
The contributions from other backgrounds are subtracted.
The QCD multijet background is estimated after the \ttbar background has been scaled.

\subsection{\texorpdfstring{$\HH\to\bbbar{\PGg}{\PGg}$}{HH to bbgg}}
The \bbgg final state~\cite{CMS_HHbbgg_R2Legacy} despite the small branching fraction (0.26\%), profits from the good mass resolution of the diphoton system.
The parameters of interest of the model are extracted from a two-dimensional fit to the invariant mass distributions of the photon pair (\Mgg) and \PQb jet pair (\Mbb).
The \HH signal can be identified as a peak in the (\Mgg, \Mbb) distribution at the value of the Higgs boson mass.
Jets originating from the \PQb quark decay are identified using the \textsc{DeepJet} algorithm.
The background is dominated by the $\PGg\PGg$+jets and $\PGg$+jets continuum, and is modelled from the data sidebands of the \Mgg and \Mbb distributions.
Other important background sources are the single Higgs boson production processes, in which the Higgs boson decays into a photon pair (\Hgg).
The largest \Hgg background is the \ttHgg process because of the sizable cross section and the \PQb jets in the final state from the \ttbar decay.
The \Hgg processes are modelled using MC simulations.

Exclusive categories targeting specific \HH production modes are defined to maximise the sensitivity to the signal.
The SR targeting the \VBF mechanism has requirements on the two additional jets.
The \VBF event candidates are classified into two exclusive categories depending on the value of the reduced four-body mass \MX, defined as $\MX = \Mbbgg - \Mgg - \Mbb + 250\GeV$, where \Mbbgg is the four-body invariant mass of the pairs of photons and \PQb jets.
A BDT classifier is defined to isolate the \VBF signal from the \GGF signal and the continuum background.
For the \GGF-enriched categories, a BDT classifier is used to reject the continuum background.
The values of the BDT score and \MX are used to classify the events in twelve \GGF-enriched categories.
The number and boundaries of the categories are optimised to maximise the expected significance of the SM \HH signal.

Furthermore, a DNN is trained to separate the \HH signal from the \ttHgg process.
The selection on the DNN output score reduces the \ttH background by between 80 and 85\% and provides an efficiency between 90 and 95\% for the \HH signal, depending on the analysis category.

\subsection{\texorpdfstring{$\HH\to\bbWW$}{HH to bbWW general}}

\begin{figure*}[htb]
    \centering
    \includegraphics[width=0.4\textwidth]{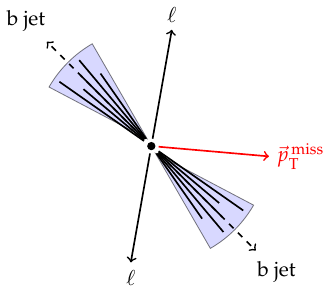}
    \includegraphics[width=0.4\textwidth]{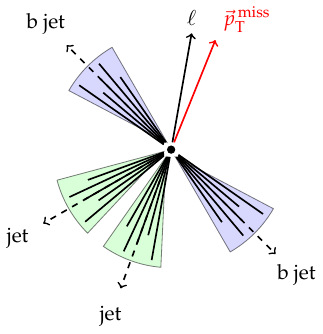}\\
    \includegraphics[width=0.7\textwidth]{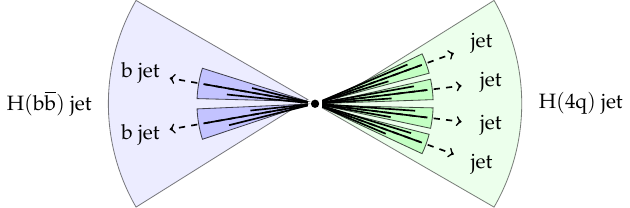}
    \caption{
        Illustrations of the resolved topology of the \bbWW decay channel in the final state with two leptons (upper left) and one lepton (upper right).
        The boosted topology of the fully hadronic \bbWW decay channel (lower).
    }
    \label{bbWW_sketches}
\end{figure*}

The \bbWW decay mode has the second largest branching fraction (24.8\%) among all \HH decay modes.
Two different analyses target the \bbWW decay channel.
The analyses address distinct final states and experimental signatures.
The first is a search for the $\PW\PW\to\ell\nu\Pq\Pq/2\ell 2\nu$ final state.
The Higgs boson candidates decaying to \PQb quarks can lead to either resolved or boosted topologies, \ie forming two narrow-radius \PQb jets or one large-radius \PQb jet.
The second analysis is a search for $\PW\PW\to\qqqq$  studied only in  boosted topologies.
All four jets from the \PW decays are expected to be overlapping and form one large-radius jet.
The accompanying pair of \PQb jets is also merged into one large-radius jet.
Figure~\ref{bbWW_sketches} illustrates the final states of the two analyses.

\subsubsection{\texorpdfstring{$\HH\to\bbWW$, $\PW\PW\to\ell\nu\Pq\Pq/2\ell 2\nu$}{HH to bbVV, VV to lnuqq/2l2nu}}

While \bbWW is the dominant channel in this final state, \bbZZ and \bbtt decays of the \HH system that result in a pair of \PQb quarks and one or two light leptons are also included in the signal.
The analysis~\cite{CMS_HHbbWW_R2Legacy} considers two final states: single-lepton (1\Pell) $\PQb\PQb \ell \nu \PQq\PQq$ and dilepton (2\Pell) $\PQb\PQb 2\ell 2\nu$.
The most significant background contribution is the \ttbar process, while other backgrounds include events with misidentified leptons, and the DY process and $\PW$+jets process for the 2\Pell and 1\Pell channels, respectively.

The data are collected using single- and double-lepton triggers.
The event selection criteria require the presence of one isolated lepton in the 1\Pell channel or two isolated leptons of opposite charge in the 2\Pell channel, originating from the \PW leptonic decays.
The \Hbb system is characterised by two jets.
In the resolved category these are small-radius jets, and for the signal they are both originating from \PQb quark. In the event selection at least one must be classified as a \PQb jet by the \textsc{DeepJet} algorithm.
If the \Hbb system is produced with high momentum, the two \PQb jets are Lorentz boosted and can be reconstructed as one large-radius jet with substructure.
In the 1\Pell case, the events must have at least one more small-radius jet present, corresponding to the hadronic \PW decay.

To avoid event overlap with the \bbtt analysis, this analysis vetoes events containing at least one \tauh candidate, as defined in Ref.~\cite{CMS_bbtautau_R2Legacy}.
The events are classified using multiclass DNNs, trained separately for the 1\Pell and 2\Pell cases.
The DNNs are trained using high-level features, such as invariant masses, hadronic activity, and the output of a Lorentz-boost network that performs automated feature engineering using the four-momenta of selected leptons and jets.
The DNN training data set includes signal samples with SM and anomalous \klambda values to ensure good performance on the full model spectrum.

The multiclass DNNs have different output nodes corresponding to several background processes and two signal processes: \GGF and \VBF \HH production.
The DNNs provide a score for each node that can be interpreted as the probability of each event to belong to that class.
The events are categorised according to the node with the highest score.
The resulting background categories are: DY or $\PW$+jets for the 2\Pell or 1\Pell channel, respectively, and \ttbar for both channels.
The two signal categories are further divided into subcategories according to the $\PQb$ jet multiplicity and topology into resolved-1$\PQb$, resolved-2$\PQb$, and boosted.

The background contribution from misidentified leptons is estimated using the misidentification factor method~\cite{Sirunyan:2018shy}.
In the 2\Pell channel, the DY background is estimated using data.
Events with zero $\PQb$-tagged jets are weighted by TFs extracted from events with the dilepton invariant mass inside the $\PZ$ boson mass window.
All other backgrounds are estimated using simulated events.
The signal extraction is performed using a simultaneous maximum likelihood fit to the distributions of the DNN outputs in the signal and background event categories.

\subsubsection{\texorpdfstring{$\HH\to\bbWW$, $\PW\PW\to\qqqq$}{HH to bbVV, VV to 4q}}

The all-hadronic \bbWW search targets highly Lorentz-boosted pairs of Higgs bosons produced via \GGF and \VBF.
The \bb quark pair is reconstructed as a large-radius jet and all four quarks from the \PW decays form another large-radius jet.
This analysis focuses mainly on the \bbWW decays, however, \HH decays into \bbZZ in the fully hadronic final state are also included in the signal.
The MC samples used for the modeling of the backgrounds are the same as in Ref.~\cite{CMS_HHbbWW_R2Legacy}.

Each Higgs boson is reconstructed as an AK8 jet.
Additional AK4 jets are used to identify the \VBF \HH topology, while events containing isolated muons or electrons are vetoed.
Events are selected online using triggers that require high hadronic activity and the presence of AK8 jets.

Offline, two AK8 jets with $\pt > 300\GeV$ and $\abs{\eta} < 2.4$ are required, representing the \hbb and \HWW decays.
The \textsc{ParticleNet} algorithm~\cite{Qu:2019gqs} is used to isolate the signal.
The \hbb jets are separated from QCD jets using the \textsc{ParticleNet}~\cite{Qu:2019gqs} discriminant \Dbb, while the \hwwq jets are identified using an attention-based NN, based on the ``particle transformer'' (\partmodel) architecture~\cite{Qu:2022mxj,CMS-PAS-JME-25-001}.
To enable the background estimation and their calibration, both discriminants are decorrelated from the mass of their respective AK8 jets by training the NNs with jets sampled from a nearly uniform distribution in mass and \pt~\cite{CMS-PAS-BTV-22-001}.
Further, the \textsc{ParticleNet} mass regression \mreg~\cite{CMS-DP-2021-017} is used to improve the AK8 jet mass resolution.
The \bb- (\ww-) candidate jet is defined as the AK8 jet with the higher (lower) \Dbb score.

The \VBF topology consists of two jets in opposite forward regions of the detector, with large invariant mass and $\Delta\eta$.
To identify events consistent with this topology, AK4 jets with $\pt > 25\GeV$ and $\abs{\eta} < 4.7$ that are well separated from the \bbbar- (\ww-) candidate AK8 jet are selected.
Out of the selected jets, the two highest \pt AK4 jets are considered the \VBF jet candidates, and their invariant mass and $\Delta\eta$ are used as inputs to a BDT trained to discriminate signal from the QCD multijet and \ttbar backgrounds.
Other BDT inputs include the \bbbar- and \ww-candidate jet kinematics and \partmodel tagger outputs.
The inputs are chosen to optimise the BDT performance without inducing correlations with the \bbbar-candidate jet mass.

The BDT is trained to classify both SM \GGF and $\CVV=0$ \VBF signal events as well as QCD multijet and \ttbar background events.
The ``\GGF'' and ``\VBF'' SRs are defined using discriminants, \ggfbdt and \vbfbdt, built from the output probabilities for the corresponding signal and background classes.
Separate BDTs for the two processes were also tested, and the difference in performance was negligible.

The \VBF SR selection consists of thresholds on the \Dbb score of the \bbbar-candidate jet (\Dbbbb) and the \vbfbdt discriminant.
These thresholds correspond to $\CVV=0$ \VBF signal (background) efficiencies of 40\% ($\approx$0.1\%) and 20\% ($\approx$0.003\%), respectively, and are chosen to minimise the expected upper limit on the $\CVV=0$ \VBF signal.
Likewise, the \GGF SR selection vetoes events in \VBF SR, and consists of thresholds on \Dbbbb and \ggfbdt, corresponding to \GGF signal (background) efficiencies of 60\% ($\approx$0.3\%) and 7\% ($\approx$0.01\%), respectively.
The dominant uncertainty in the BDT signal efficiency is related to the \partmodel tagger calibration~\cite{CMS-PAS-JME-23-001}, which applies data/MC scale factors to the primary Lund jet plane~\cite{Dreyer:2018nbf} densities of each quark subjet.

\begin{figure*}[htbp]
    \centering
    \includegraphics[width=0.485\textwidth]{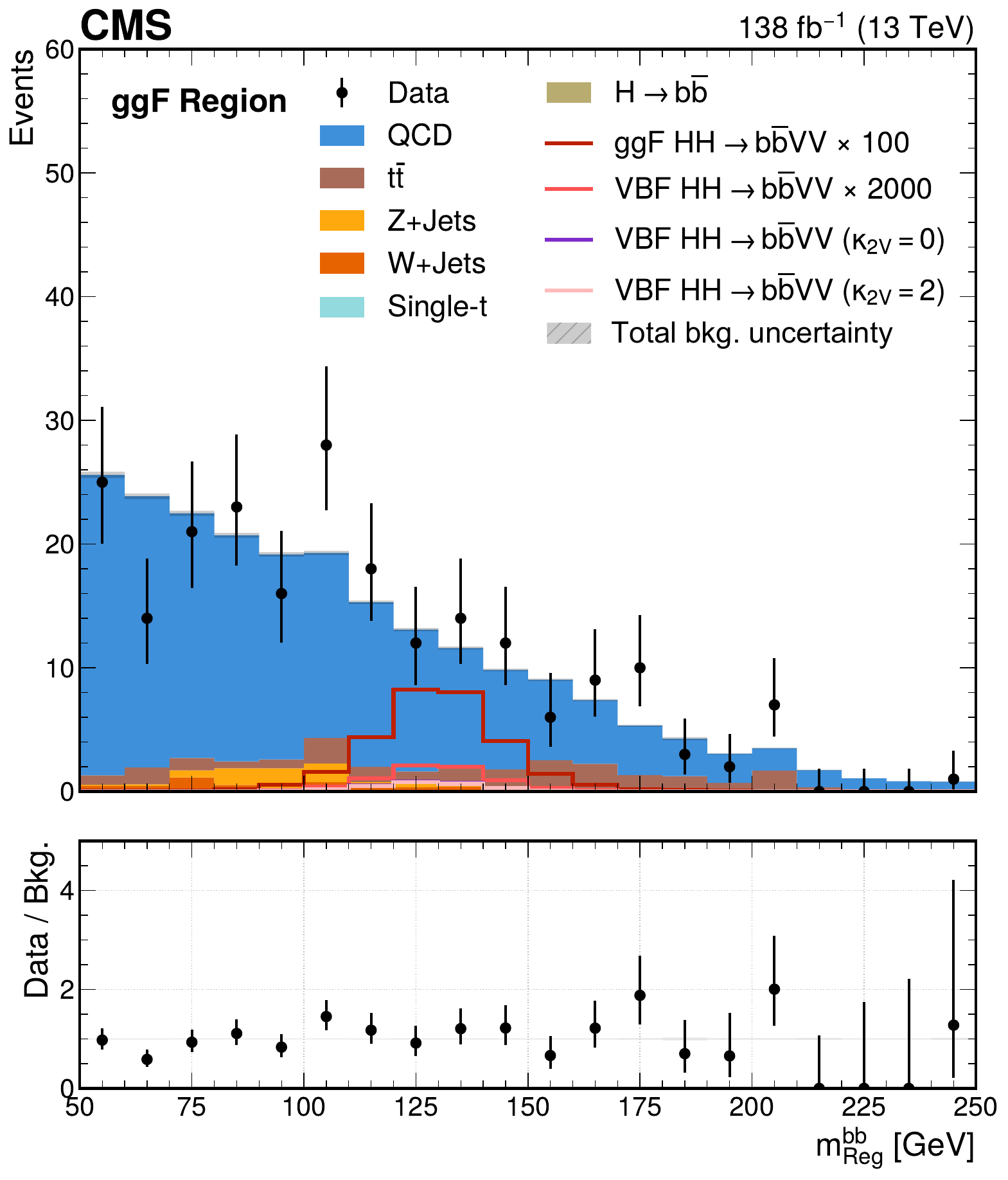}
    \includegraphics[width=0.495\textwidth]{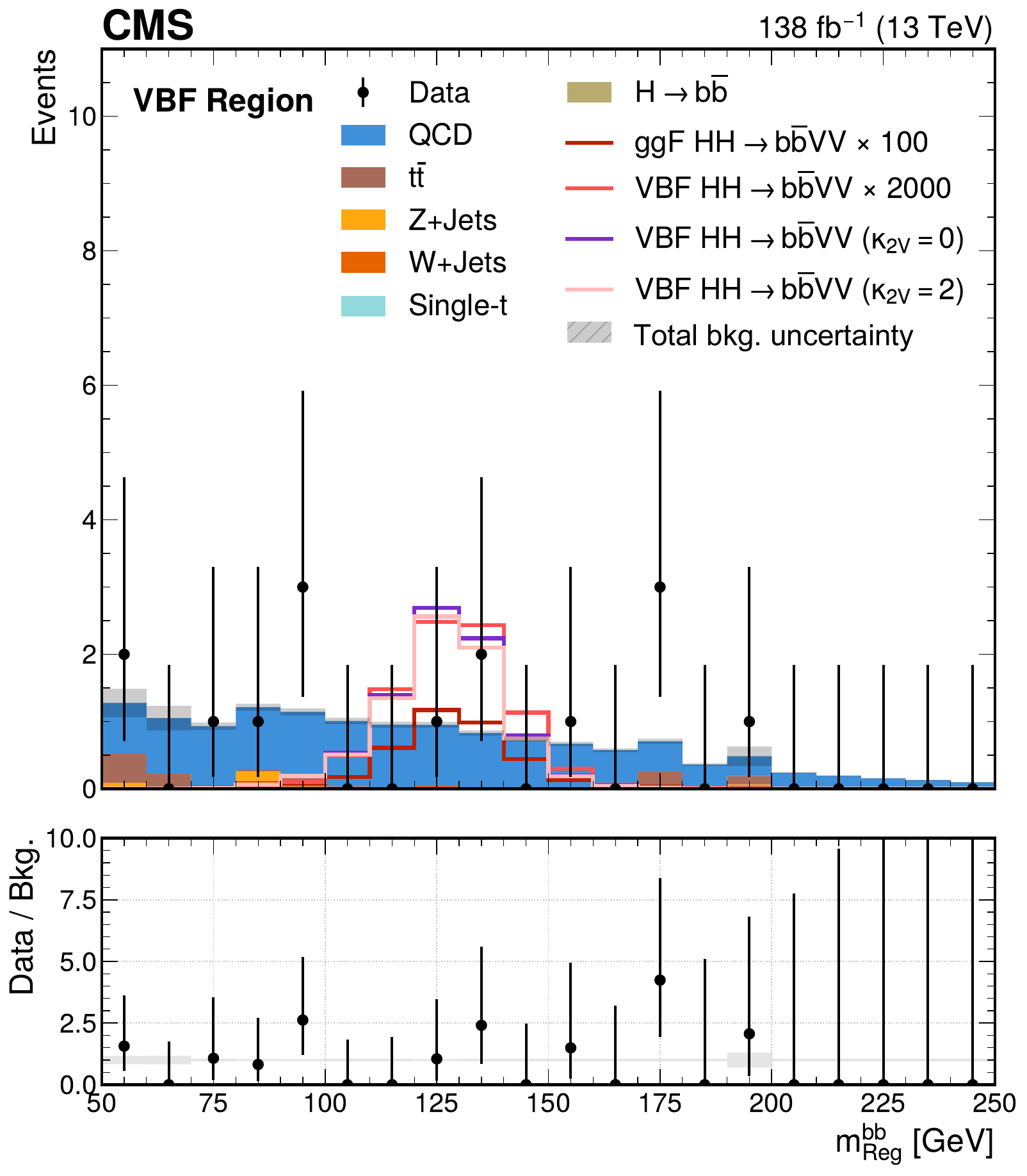}
    \caption{
        Distributions of the \mregbb observable in the \GGF (left) and \VBF (right) signal regions of the all-hadronic \bbWW search, after a maximum likelihood fit of the background and SM \HH signal to the data.
    }
    \label{fig:bbvv_regions}
\end{figure*}

The search is performed in the two SRs using a one-dimensional likelihood model binned in the \mreg of the \bbbar-candidate jet (\mregbb).
The background in these regions is dominated by QCD multijet events, and is estimated through data in a QCD multijet CR defined by inverting the \Dbb selection.
The QCD multijet yield in the SRs is estimated as the product of data in the fail region and polynomial TFs~\cite{Sirunyan:2017isc,Sirunyan:2018qca,Sirunyan:2020hwz}.
The $\ttbar$, $\PV$+jets, and other minor backgrounds are estimated using MC simulation and their systematic and statistical uncertainties are incorporated in the final statistical analysis.
The resulting \mregbb in the SRs, after a maximum likelihood fit to the data of the background model and SM \HH signal, are shown in Fig.~\ref{fig:bbvv_regions}.

\subsection{\texorpdfstring{$\HH\to\text{multilepton}$}{HH to multilepton}}

The \HH multilepton analysis~\cite{CMS_HHmultilepton_R2Legacy} focuses on \HH decays into the $4\PW$, $2\PW 2\Pgt$, and $4\Pgt$ channels.
Events are selected using a set of single, double, and triple lepton triggers, as well as lepton and \tauh triggers and double \tauh triggers.
The leptons and \tauh candidates are reconstructed with a set of special identification (ID) criteria originally developed for the CMS \ttH multilepton analysis~\cite{Sirunyan:2020icl}.
These IDs are customised for the use of the misidentification factor method~\cite{Sirunyan:2018shy} used for the estimation of the background with misidentified lepton or \tauh candidates from control samples in data.
For electrons and muons, these two IDs also make use of a specialised multivariate analysis specifically trained to recognise genuine leptons that originate from vector boson and \PGt lepton decays.
All categories veto events containing at least one \PQb jet identified with the medium WP of the \textsc{DeepJet} algorithm and events containing at least two \PQb jets identified with the loose WP, corresponding to a \PQb jet selection efficiency of about 84\% and a misidentification rate for light-flavour quark and gluon jets of about 11\%.
This reduces contributions from top quark related backgrounds, such as \ttbar and makes this analysis orthogonal to the \HH analyses requiring \PQb jets.

The events are split into seven mutually exclusive event categories with final states containing multiple leptons \PLepton ($\Pgm,\,\Pe$) or hadronically decaying \PGt leptons (\tauh): 2\PLepton with the same electric charge (SS), 3\PLepton, 4\PLepton, 3$\PLepton$+1\tauh, 2$\PLepton$+2\tauh, 1$\PLepton$+3\tauh and 4\tauh.
All categories require the specified number of final-state light leptons and \tauh.
The 2\PLepton SS and 3\PLepton categories additionally require the presence of jets originating from hadronic \PW boson decays.
The primary background in most categories is given by events containing misidentified \PLepton or \tauh candidates.
All categories also contain a sizable contribution of genuine multiboson backgrounds from $\PZ\PZ$ and $\PW\PZ$ decays.
Therefore, two kinematic distributions in $\PW\PZ$ and $\PZ\PZ$ CRs are included in the fit.
The categories are constructed by inverting the \PZ boson veto in the 3\PLepton and 4\PLepton categories, respectively.
The inclusion of these two CRs helps to constrain the two main diboson backgrounds, as well as the corresponding systematic uncertainties.

The analysis makes use of a parameterised BDT tuning the sensitivity to the SM and to the EFT benchmark set JHEP04(2016)126 described in Section~\ref{sec:MC}.
The maximum likelihood fit for the signal extraction is performed simultaneously on the seven event categories.
The results for each benchmark point are extracted from the corresponding BDT output.
For the benchmarks not included in the BDT training, the most similar one based on the shape of the parton-level \mHH spectrum is used during inference.

\subsection{\texorpdfstring{$\HH\to{\PGt}{\PGt}{\PGg}{\PGg}$}{HH to tautaugammagamma}}

The search for \HH production in the \ttgg final state aims to cover both the hadronic and leptonic decay modes of the \PGt lepton.
Despite the small \HHggtt branching fraction in the SM ($2.85\times10^{-4}$), the diphoton pair offers a clean experimental signature to trigger on with a good mass resolution, while the additional tau leptons in the event help to further isolate signal from background.
The target of the search is the nonresonant \HH production via \GGF.
The \VBF production is not considered in this analysis.

Events are selected online using a diphoton trigger with asymmetric photon \pt thresholds of 30 and 18\GeV.
All events are required to have at least one diphoton candidate, and in the case where more than one candidate exists, the candidate with the highest scalar \pt sum of the photons is chosen.

Events are required to have at least one \PGt lepton candidate.
Both hadronic and leptonic decay modes of the \PGt lepton are considered.
A $\PGt\PGt$ candidate can be identified from any of the following pairs of reconstructed objects: $\tauh\tauh$, $\tauh\PGm$, $\tauh\Pe$, $\PGm\Pe$, $\PGm\PGm$, $\Pe\Pe$, $\tauh$+isolated track.
The \textsc{DeepTau} discriminant is used to select \tauh candidates.
Events consistent with a $\PZ\to\ell\ell$ or $\PZ\to\ell\ell \PGg$ decay are rejected.

A BDT classifier is used to further isolate events with signal-like characteristics from the background.
Hadronic jets in the event are only used in the BDT to reject backgrounds like \ttbar.
To reduce the probability of creating an artificially peaking structure, the \mgg is not included as an input feature to the BDT, and the \pt of the photon candidates are divided by \mgg to reduce correlations between the photon \pt and \mgg.
Sequential boundaries are placed on the BDT output to define event categories of different signal purity, where the boundary positions are chosen to maximise the signal sensitivity.

The \mgg distribution is used in the maximum likelihood fit for each event category.
The smoothly falling background (continuum) is modelled directly from the data.
The signal and the single Higgs boson production, which is a resonant background, are modelled using simulation.

\subsection{\texorpdfstring{$\HH\to\bbZZ$}{HH to bbZZ}}
This analysis~\cite{CMS_HHbbZZ4l_R2Legacy} focuses on \GGF \HH production where one Higgs boson decays to a \bbbar pair, while the other decays to a $\PZ\PZ$ pair resulting in a $4\ell$ final state.
The $4\ell$ decay channel provides a clean signature despite the small branching fraction, which is partially compensated by the high branching fraction of the \bbbar decay channel.
The final state comprises two jets and two pairs of oppositely charged isolated electrons or muons.
The $4\ell$ SR is defined by requiring that the four-lepton invariant mass $m(4\ell)$ is within the Higgs boson mass window, while events with $m(4\ell)$ outside this range comprise the $4\ell$ CR.

The dominant $\PZ$+$X$ background is characterised by nonprompt and misidentified leptons, mainly from decays of heavy-flavour hadrons, misidentified jets, and electrons from photon conversions.
The $\PZ$+$X$ background is estimated from a data CR, and the probabilities of a light lepton to be misidentified.
The main contributions to the systematic uncertainty in the reducible background estimation are the limited sizes of the CR and the sample used to estimate the misidentification rates.
Uncertainties related to the difference in composition between the SR and the misidentification rate estimation sample are also considered.

A BDT is trained using MC simulated events in the $4\ell$ SR to further discriminate the \GGF \HH signal process from all background processes.
The BDT output distribution is fit to constrain the parameters of interest.

\subsection{\texorpdfstring{$\HH\to\PW\PW{\PGg}{\PGg}$}{HH to WWgg}}

This analysis studies three $\PW\PW$ decay modes: semileptonic (1\Pell), fully hadronic, and fully leptonic (2\Pell) decays.
Events are selected using double-photon triggers with thresholds on the leading (subleading) photon \pt of $\pt^{\PGg} > 30$ (18)\GeV for the data collected during 2016 and $\pt^{\PGg} > 30$ (22)\GeV for 2017 and 2018.
The MC samples used in this analysis are the same as the ones used in the \bbgg analysis~\cite{CMS_HHbbgg_R2Legacy}.
The continuum background is modelled from data using the \mgg distribution.
The single Higgs boson background is modeled using simulated samples.

\begin{figure*}[htbp]
    \centering
    \includegraphics[width=0.412\textwidth]{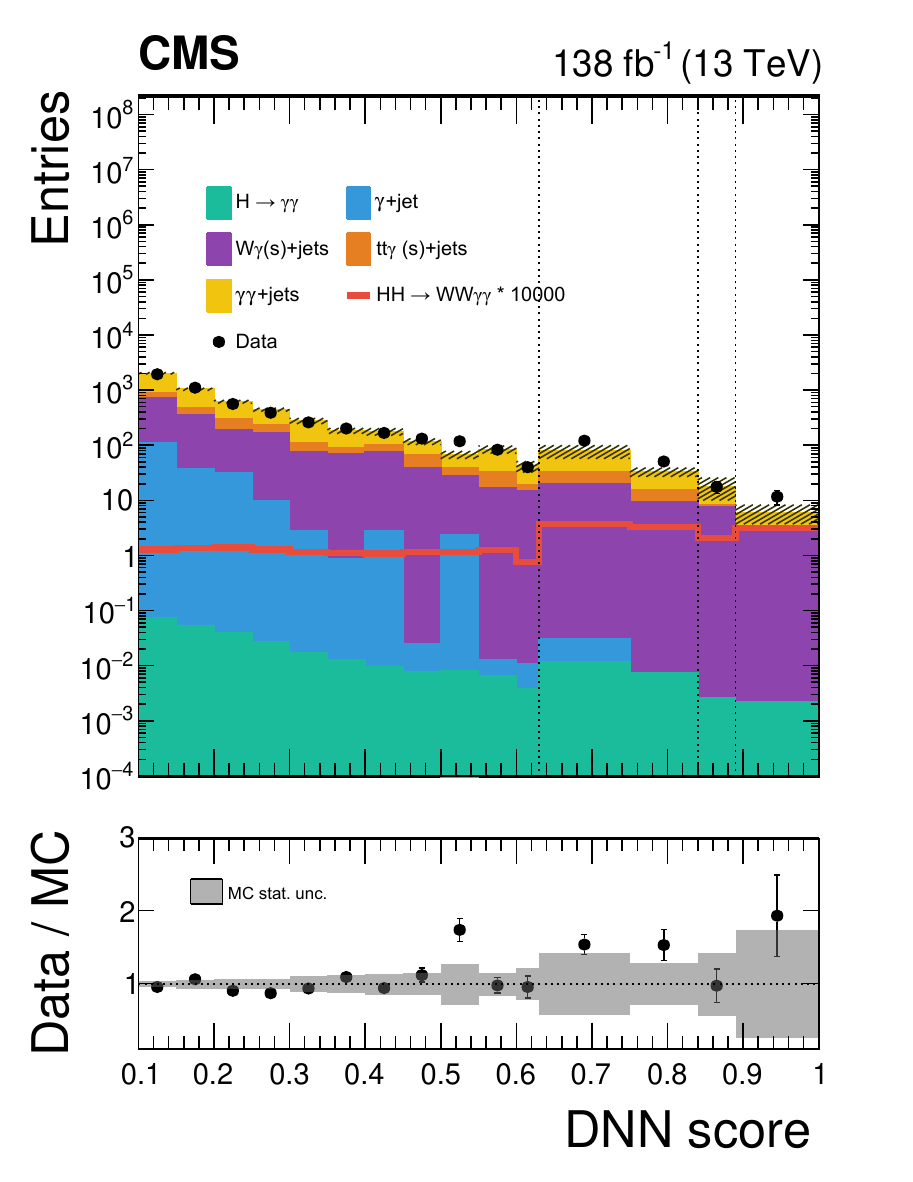}
    \includegraphics[width=0.408\textwidth]{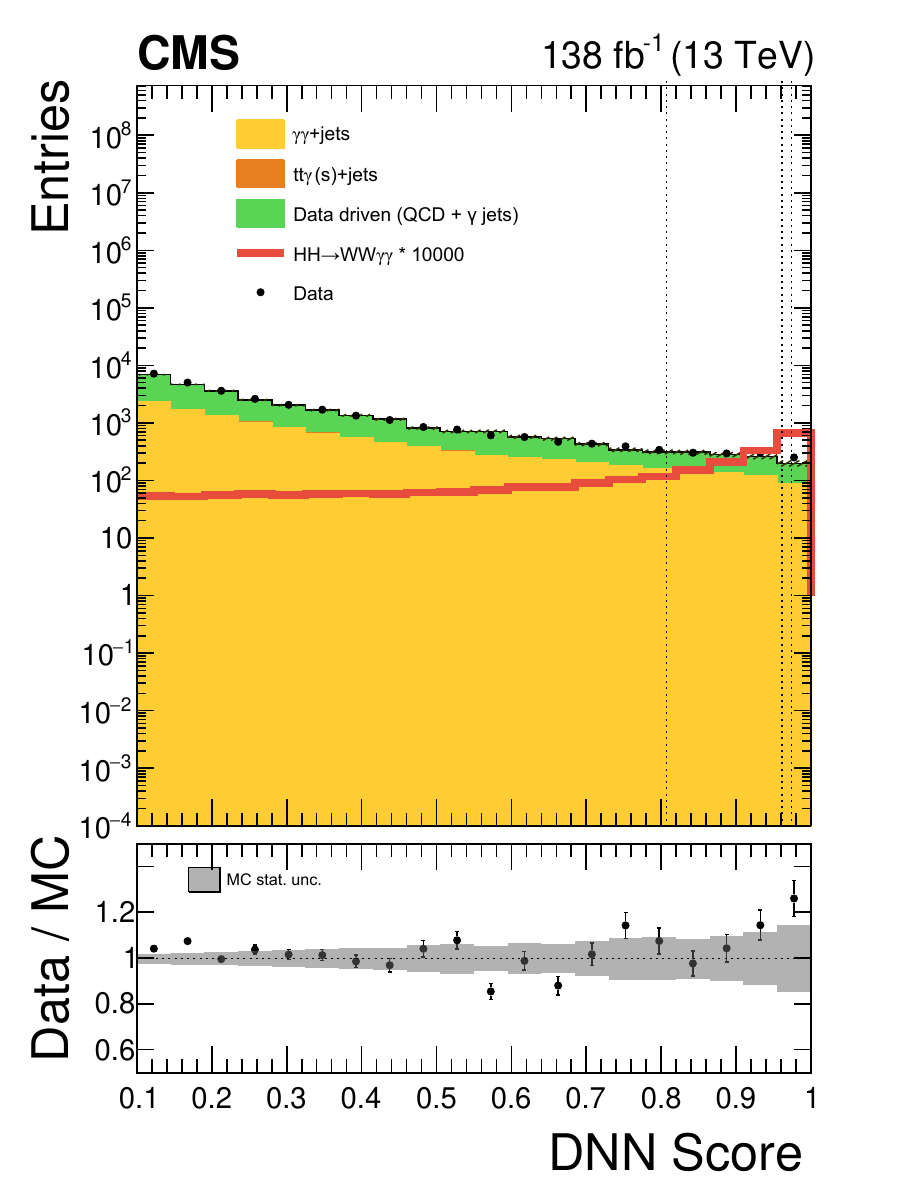}\\
    \includegraphics[width=0.4\textwidth]{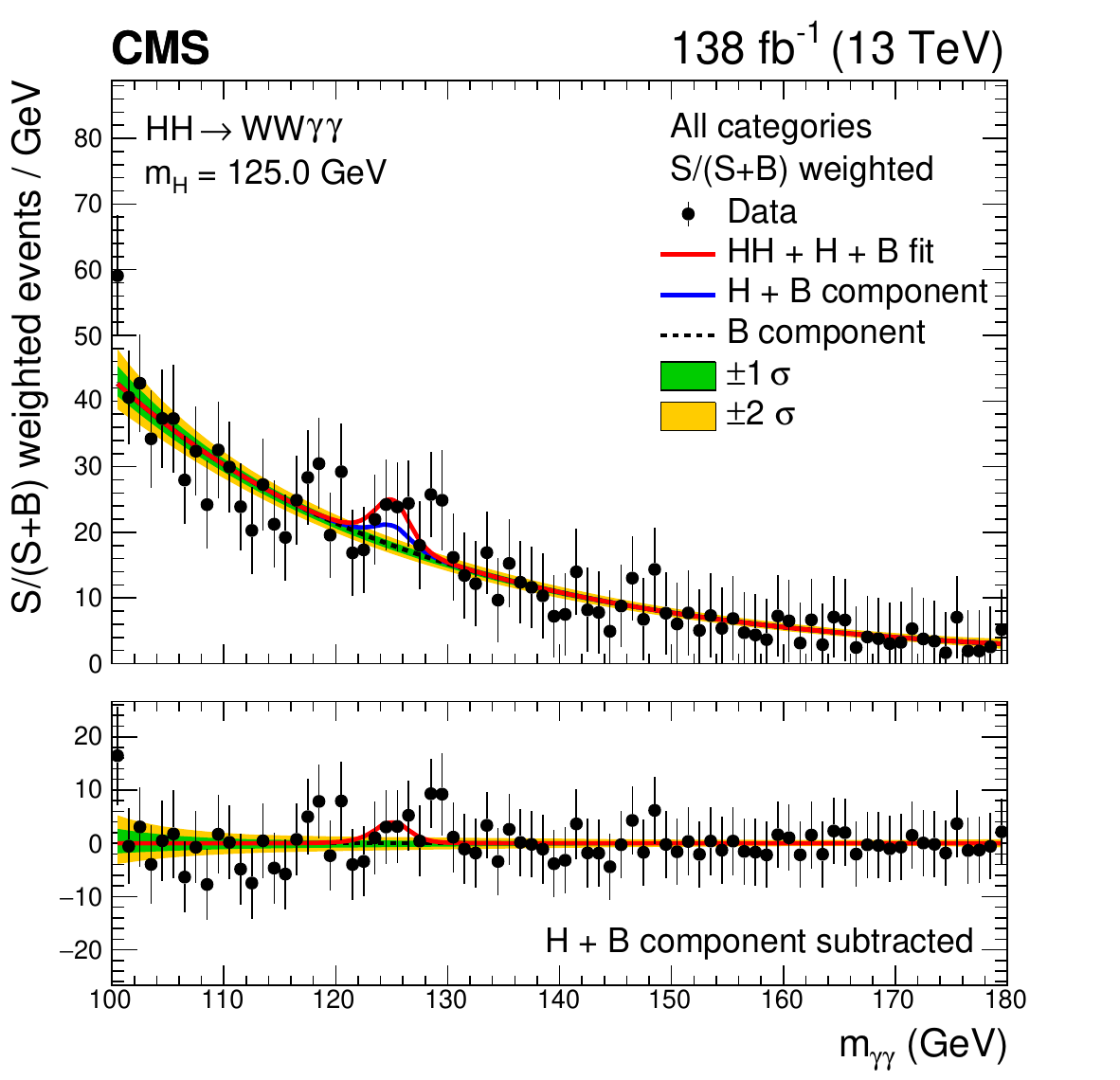}
    \caption{
        The distributions of DNN scores for the signal and main backgrounds in the 1\Pell (upper left) and fully hadronic (upper right) channels of the \WWgg analysis.
        The signal-plus-background (red), single \PH plus continuum background (blue), and continuum background (dashed black) fits for all channels weighted by $S/(S+B)$ (lower).
    }
    \label{fig:WWgg_plots}
\end{figure*}

A BDT classifier is trained for photon-jet separation, resulting in a photon identification score.
The jets from the hadronisation of bottom quarks are identified using the \textsc{DeepJet} algorithm.
The number of light leptons is used to maintain orthogonality in all three analyses.
We assign an event to the fully hadronic channel if it has no leptons, to the 1\Pell channel if it has exactly one lepton, and to the 2\Pell channel if it has two leptons.
Each channel requires a diphoton candidate, which is defined as a pair of two photons.

The 1\Pell channel uses a multiclassifier DNN to differentiate between signal, single Higgs boson, and background events, as shown in Fig.~\ref{fig:WWgg_plots}, upper left.

The \bbgg signal contaminates the \WWgg phase space in the fully hadronic channel.
Therefore, two binary classifiers are used for the fully hadronic channel, one acting as a \bbgg discriminator and the other distinguishing between signal and background events, the latter shown in Fig.~\ref{fig:WWgg_plots}, upper right.
A specific challenge in the fully hadronic channel is differentiating between the $\PZ\PZ\PGg\PGg$ and the \WWgg signal processes, due to the identical experimental signatures of the hadronic $\PZ$ and $\PW$ decays.
There are residual \bbgg events that pass the \WWgg event selection but are not favoured by the \bbgg analysis because of low \bbgg discriminator scores.
Therefore, the combination of \WWgg, $\PZ\PZ\PGg\PGg$, and residual \bbgg processes is considered signal in this analysis.
Finally, because of the limited size of the samples available for training and validating a multivariate event classifier, a selection based on traditional variables is adopted for the 2\Pell channel.
The diphoton invariant mass distribution is used for the signal extraction across all channels.
The signal-plus-background and background-only fits for all channels weighted by $S/(S+B)$ are shown in Fig.~\ref{fig:WWgg_plots}, lower.

\section{Systematic uncertainties}
\label{sec:syst}

Hundreds of systematic uncertainties are considered, affecting \HH signal and background processes' yields (\ie normalisation uncertainties) and/or shapes (\ie shape uncertainties).
The systematic uncertainties are introduced as nuisance parameters in the maximum likelihood fit used to extract the results.
Common uncertainties among different analyses are fully correlated, while others specific to each analysis are treated as uncorrelated.
The uncertainties with the biggest impact (20--23\%) are the ones affecting the background in \bbbb and \bbtt analyses as well as the uncertainty in the theoretical prediction of the \HH production cross section.

\subsection{Theory uncertainties}
\label{sec:syst_th}
Theoretical uncertainties on the renormalisation and factorisation scale, strong coupling constant \alpS, and PDFs affecting the cross section of all the simulated processes are included.
Other uncertainties considered concern EW corrections for the $\ttbar\PZ$, $\ttbar\PW$, $\qqbar\PZ\PZ$, and $\Pg\Pg\PZ\PZ$ processes.
Theoretical uncertainties in the Higgs boson branching fractions~\cite{deFlorian:2016spz} are applied to both the \HH signal and single-Higgs backgrounds.
The assigned uncertainties are 1.2\% for $\PHiggs\to\bb$, 1.5\% for $\PHiggs\to\WW$ and $\PHiggs\to\ZZ$, 2.1\% for $\PHiggs\to\PGg\PGg$, and 1.6\% for $\PHiggs\to\tau\tau$.
These uncertainties are treated as correlated across all channels.
The uncertainty in the rate of Higgs boson production associated with heavy-flavour jets is assumed to be 50\%~\cite{manzoni2023}.

Theoretical uncertainties on the nonresonant \HH cross section via \GGF is applied as a function of \klambda and include a combination of uncertainties in the renormalisation and factorisation scales, as well as the uncertainty in the top quark mass.
In the SM case when $\klambda=1$, this combined uncertainty corresponds to $-23$\%/$+6$\%.
The PDF uncertainty is 3\%~\cite{Baglio:2020wgt}.
Uncertainties for \VBF production include $+0.03$\%/$-0.04$\% (scales) and 2.1\% (PDF+\alpS)~\cite{Ling:2014sne, Dreyer:2018rfu}.
An additional scale uncertainty is applied to the \bbtt, \bbWW and multilepton analyses related to \PYTHIA and the colour-correlated recoil effect.
This effect was found to be negligible in other analyses.

Some analyses suffer from a significant \ttbar background contribution (\bbWW, multilepton, and \bbtt) and a shape uncertainty that corresponds to the NNLO correction on the top quark \pt is applied to the \ttbar simulated samples~\cite{Czakon_2017}.
The \bbWW and multilepton analyses include additional uncertainties on the parton shower initial- and final-state radiation for background processes, on the top quark mass for the \ttbar background, as well as renormalisation and factorisation scale uncertainties for \ttbar and single top quark production.

\subsection{Experimental uncertainties}
\label{sec:syst_ex}

The integrated luminosities for the 2016, 2017, and 2018 data-taking years have 1.2--2.5\% uncertainty per year~\cite{LUM-17-003,LUM-17-004,LUM-18-002}, while the overall uncertainty for the 2016--2018 period is 1.6\%.
The number of $\Pp\Pp$ interactions is calculated from the instantaneous luminosity and an estimated inelastic $\Pp\Pp$ collision cross section with 4.6\% uncertainty~\cite{CMS:2018mlc}.
An uncertainty in the shape of the distribution of the mean number of $\Pp\Pp$ interactions per bunch crossing (pileup) is also applied to all simulated samples.
The trigger selection efficiency is corrected in each analysis to account for the differences between the data and simulation, and a corresponding uncertainty is applied.

When identification criteria are applied to muons, electrons, and \tauh candidates, the identification efficiencies in the simulated samples are corrected to match those in data.
The uncertainties related to these corrections affect the shapes of the kinematic distributions.
For analyses that use the same selection criteria, these uncertainties are treated as correlated.
For analyses that use photons, the shape uncertainties related to the photon identification efficiency are also considered.

Similarly, the uncertainties in the efficiency of the jet selection criteria in simulation, are considered as uncertainties on the shape of the kinematic distributions.
Nuisance parameters that affect the shape of the jet flavour discriminant are also included.
Different types of flavour contamination are treated with separate parameters.
Uncertainties in the jet energy scale and resolution are used as shape uncertainties for all simulated samples.
For analyses where \ptmiss is relevant, uncertainties related to the unclustered energy reconstruction efficiency are taken into account as correlated uncertainties in \ptmiss distribution shapes in simulated samples.

During data collection in 2016 and 2017, a gradual change in the timing of the ECAL first-level trigger inputs in the region with $\abs{\eta} > 2.0$ caused a specific trigger inefficiency.
For events containing an electron (jet) with $\pt \gtrsim 50$ (100)\GeV in the region $2.5 < \abs{\eta} < 3.0$ the efficiency loss is approximately 10--20\%, depending on \pt, $\eta$, and time.
Correction factors are computed from data and applied to the acceptance evaluated from simulation.
In addition, a normalisation uncertainty is included in the global maximum likelihood fit, described in Section~\ref{sec:results}.

A correction factor is applied to simulated data for the 2018 era to account for two HCAL modules being switched off, and an associated shape uncertainty is introduced which is treated as correlated among most analyses.
Analyses that perform data-driven background estimation methods include dedicated uncertainties, which are treated as uncorrelated.

Finally, uncertainties from the limited number of simulated events are taken into account using the Barlow--Beeston approach~\cite{Barlow:249779}.
These uncertainties are not considered in the case of the HL-LHC projections, as it is assumed that we will have sufficient simulated data at the time.

\section{Results}
\label{sec:results}

Upper limits on \HH production cross sections and constraints on coupling modifiers that contribute to SM \HH production are derived based on the asymptotic formulae for the profile likelihood ratio test statistic $-2\Delta\ln(L)$~\cite{ATLAS:2011tau, Cowan:2010js} and the \CLs~\cite{Junk:1999kv,Read:2002hq} criterion.
Generally, the test statistic is defined as
\begin{equation}
    -2\Delta\ln(L) \equiv -2\ln\left(\frac{L(\vmu,\hat{\vtheta}({\vmu}))}{L(\hat{\vmu},\hat{\vtheta})}\right),
\end{equation}
where $L$ is the combined likelihood function, $\vmu$ are the parameters of interest, such as signal strengths or coupling modifiers, $\hat{\vmu}$ are the best fit values of $\vmu$, $\hat{\vtheta}({\vmu})$ is the maximum likelihood estimator of the nuisance parameters for a given set of values of $\vmu$, and $\hat{\vtheta}$ is the global maximum likelihood estimator of the nuisance parameters for the best fit values of $\vmu$.
Generally, confidence intervals at 68\% (95\%) \CL are derived by finding the values of the parameters of interest for which the test statistic is less than 1 (3.84), following Wilks' theorem~\cite{Wilks}.
However, because of the quadratic dependence of the expected yields on the values of the coupling modifiers, the corresponding intervals may undercover or overcover~\cite{Bernlochner:2022oiw,CMS:2025ugn}.
For comparison, constraints on coupling modifiers are also computed by performing a scan of the 95\% \CL upper limit on the signal strength of \HH production while varying the coupling modifier of interest, as reported in Table~\ref{tab:ranges_results}.
The results of the different analyses described in this paper have been statistically combined in a global maximum likelihood fit.
The HEFT parametrisation is used to provide constraints on a number of different BSM scenarios.
In all cases, expected limits are derived under the background-only hypothesis.
The results have been determined using the CMS statistical analysis tool \textsc{Combine}~\cite{CAT-23-001}, which is based on the \textsc{RooFit}~\cite{Verkerke:2003ir} and \textsc{RooStats}~\cite{Moneta:2010pm} frameworks.

\begin{figure}[tbh]
    \centering
    \includegraphics[width=0.45\textwidth]{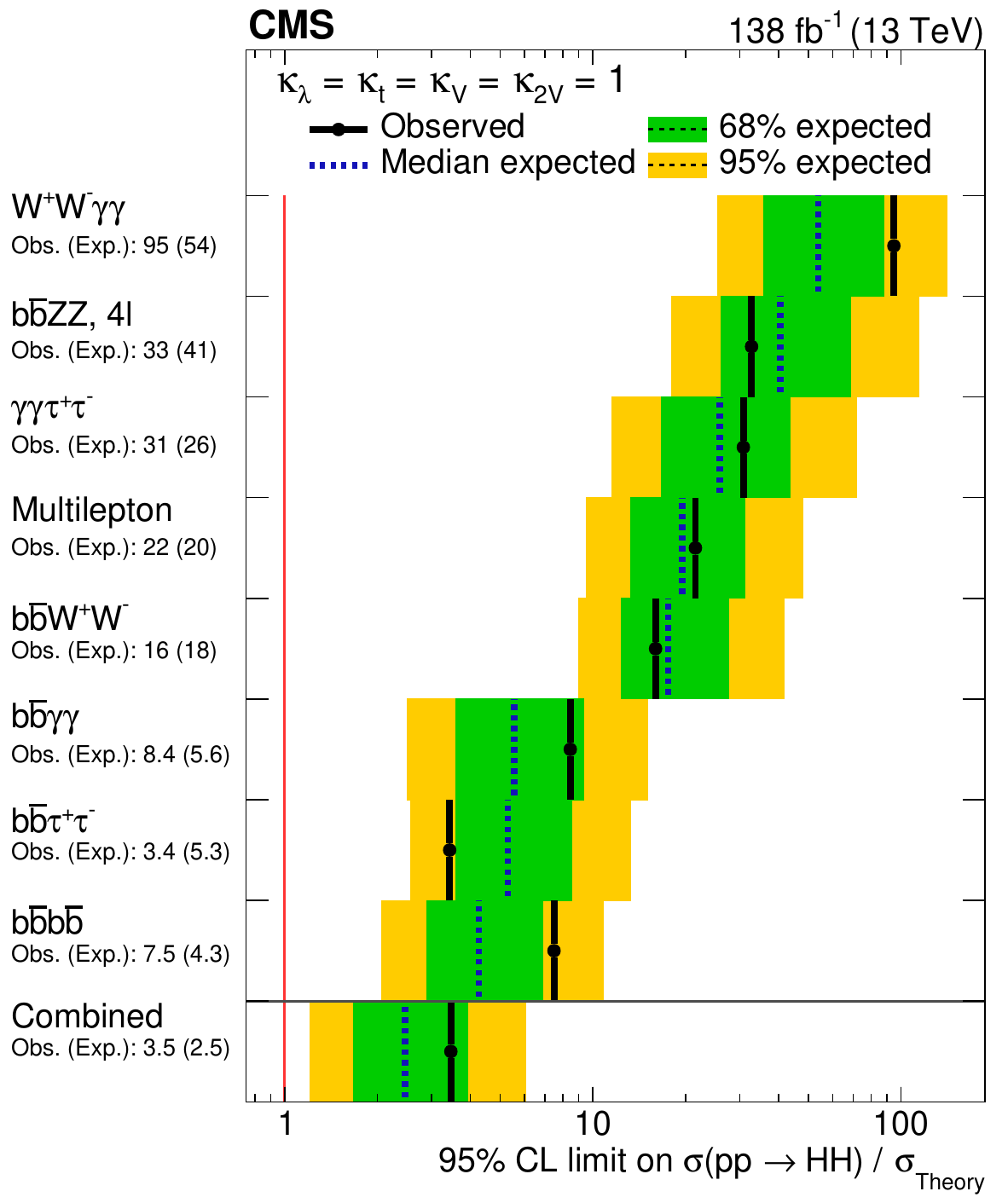}
    \caption{
        The upper limits at 95\% \CL on the inclusive signal strength $\mu = \sigmaHH/\sigmaHH^\text{SM}$ for each channel and their combination.
        The inner (green) and outer (yellow) bands indicate the 68 and 95\% \CL intervals, respectively, under the background-only hypothesis.
        The individual contributions within the \bbbb and \bbWW channels have been combined in order to simplify the presentation of results.
    }
    \label{fig:limits_cross_section_r}
\end{figure}
\begin{figure}[tbh]
    \centering
    \includegraphics[width=0.45\textwidth]{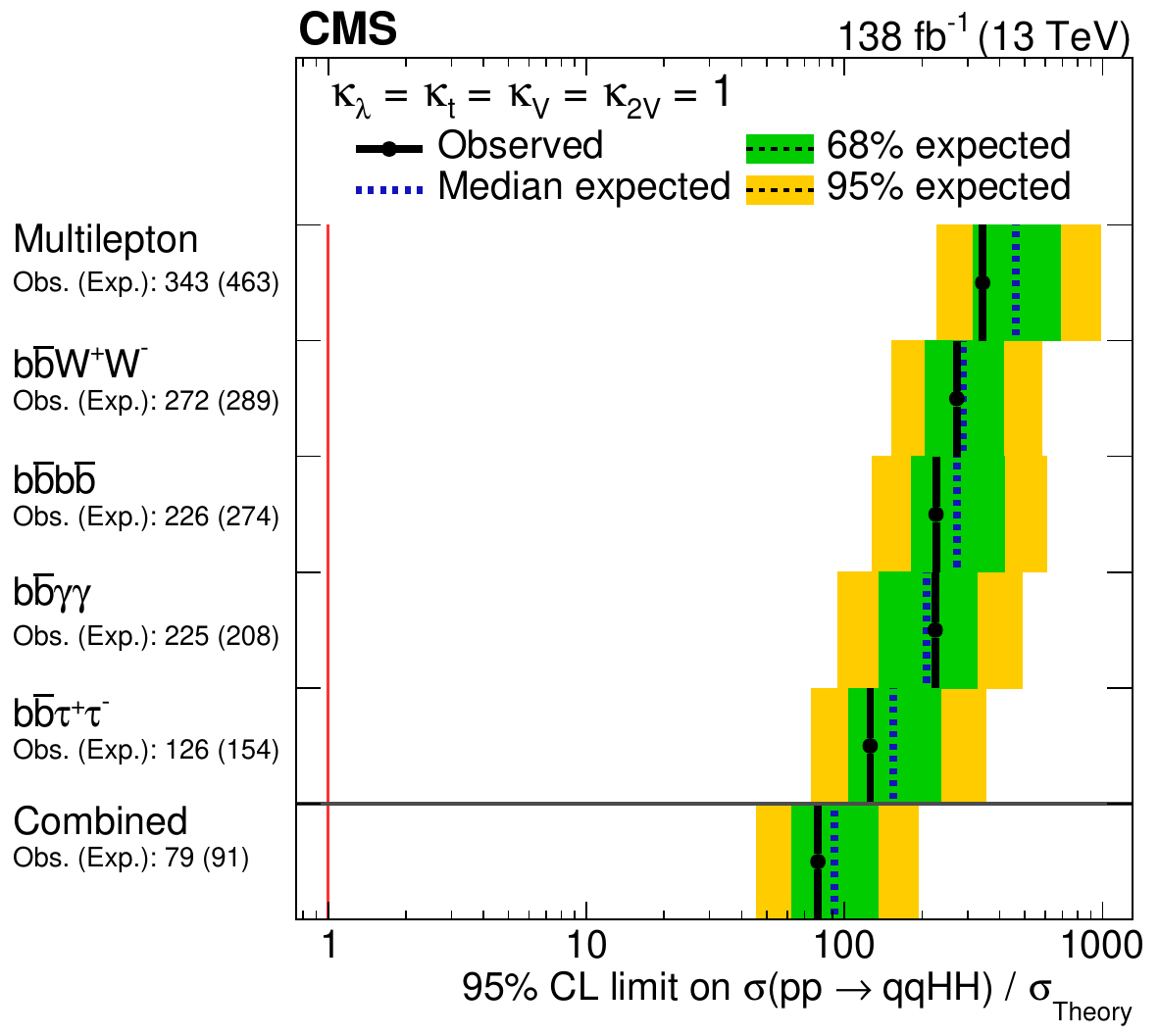}
    \caption{
        The upper limits at 95\% \CL on the \VBF signal strength $\mu_{\VBF~\HH} = \sigmaVBF/\sigmaVBF^\text{SM}$ for each channel and their combination.
        The inner (green) and outer (yellow) bands indicate the 68 and 95\% \CL intervals, respectively, under the background-only hypothesis.
        The five contributing channels are indicated in the figure.
        The individual contributions within the \bbbb and \bbWW channels have been combined in order to simplify the presentation of results.
    }
    \label{fig:limits_cross_section_rqq}
\end{figure}

The upper limit at 95\% \CL on the inclusive \HH production cross section is observed (expected) to be 3.5 (2.5) times the SM prediction, while the equivalent upper limit for the \VBF alone is 79 (91) times the SM prediction.
The limits on the \HH cross section divided by the SM prediction, \ie the signal strength $\mu$, for each contributing channel and their combination are shown in Figs.~\ref{fig:limits_cross_section_r} and~\ref{fig:limits_cross_section_rqq} for the inclusive and \VBF cases, respectively.

\begin{figure}[thbp!]
    \centering
    \includegraphics[width=0.45\textwidth]{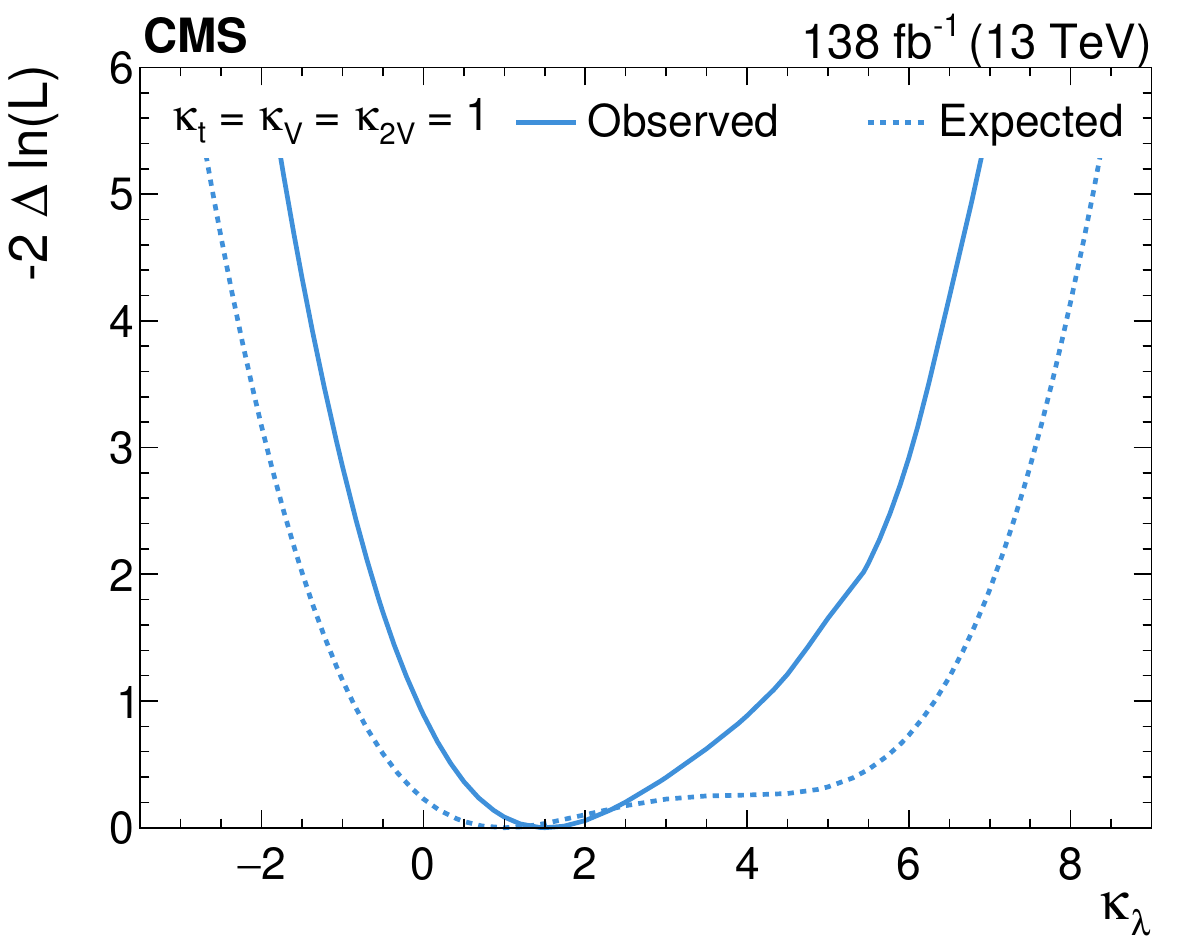}
    \includegraphics[width=0.45\textwidth]{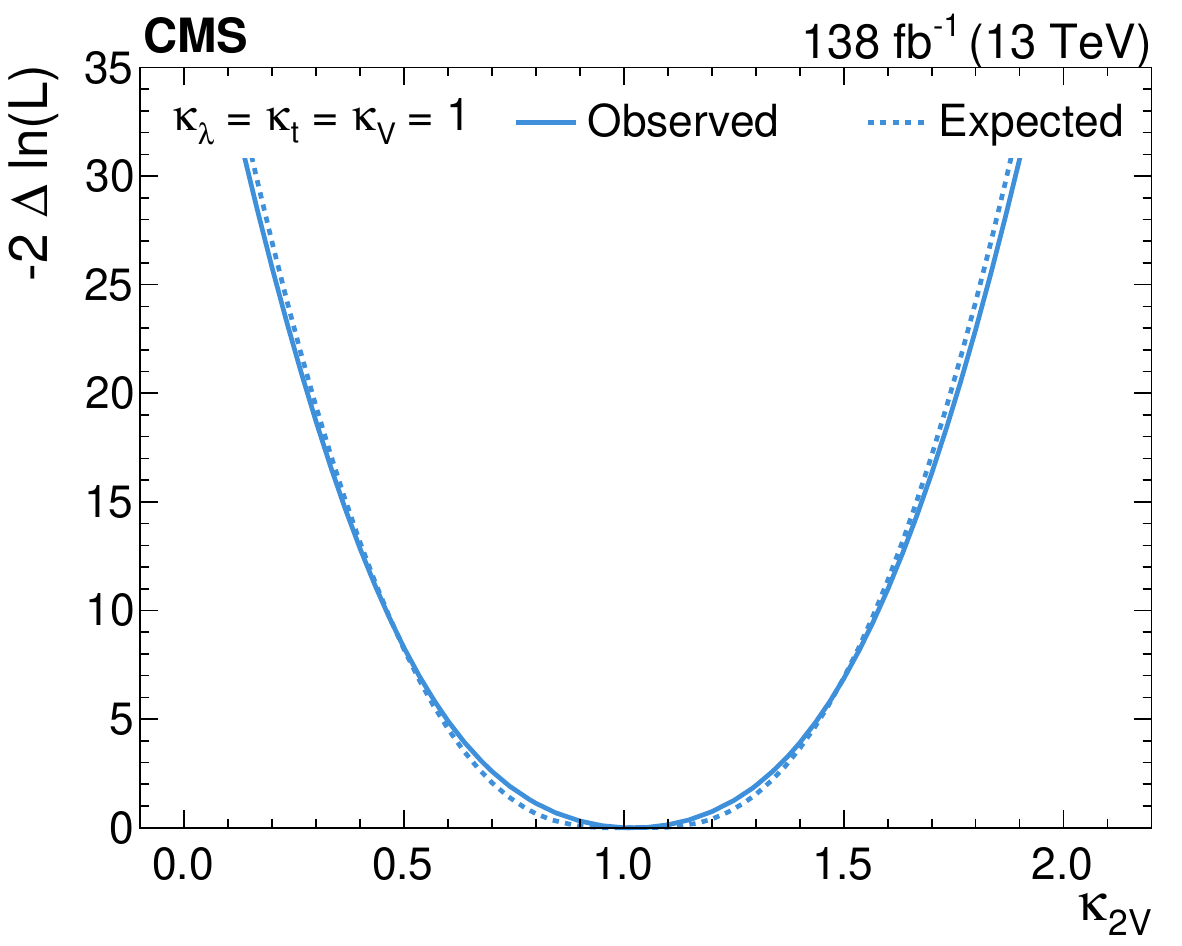}
    \caption{
        The $-2\Delta\ln(L)$ scan as functions of coupling modifiers \klambda (\cmsLeft) and \CVV (\cmsRight) for the combination of all channels when all the other parameters are fixed to their SM values.
    }
    \label{fig:likelihood_kl_k2v_kv_combination_linear}
\end{figure}

The $-2\Delta\ln(L)$ scans as a function of \klambda and \CVV is shown in Fig.~\ref{fig:likelihood_kl_k2v_kv_combination_linear}.
Besides the ones varied as shown, all other couplings are set to the values expected by the SM.
All eight channels are included in the combination.
The best fit value of \klambda is found to be 1.51 and \klambda is constrained at 68\% \CL to be within $-0.07$ and 4.18, while the expected constraint is from $-0.87$ to 6.31.
At 95\% \CL, \klambda is constrained to be within $-1.35$ to 6.37, with an expected constraint of $-2.24$ to 7.89.

For \CVV, the best fit value is at 1.02, the 68\% \CL interval is from 0.81 to 1.23 (0.77 to 1.26 expected), and the 95\% \CL interval is from 0.64 to 1.40 (0.62 to 1.41 expected).
The value of $\CVV=0$ is excluded with a significance of 6.6 standard deviations.
This level of significance marks a decisive step in establishing the presence of quartic Higgs-vector boson interactions in nature.

\begin{figure*}[thbp!]
    \centering
    \includegraphics[width=0.45\textwidth]{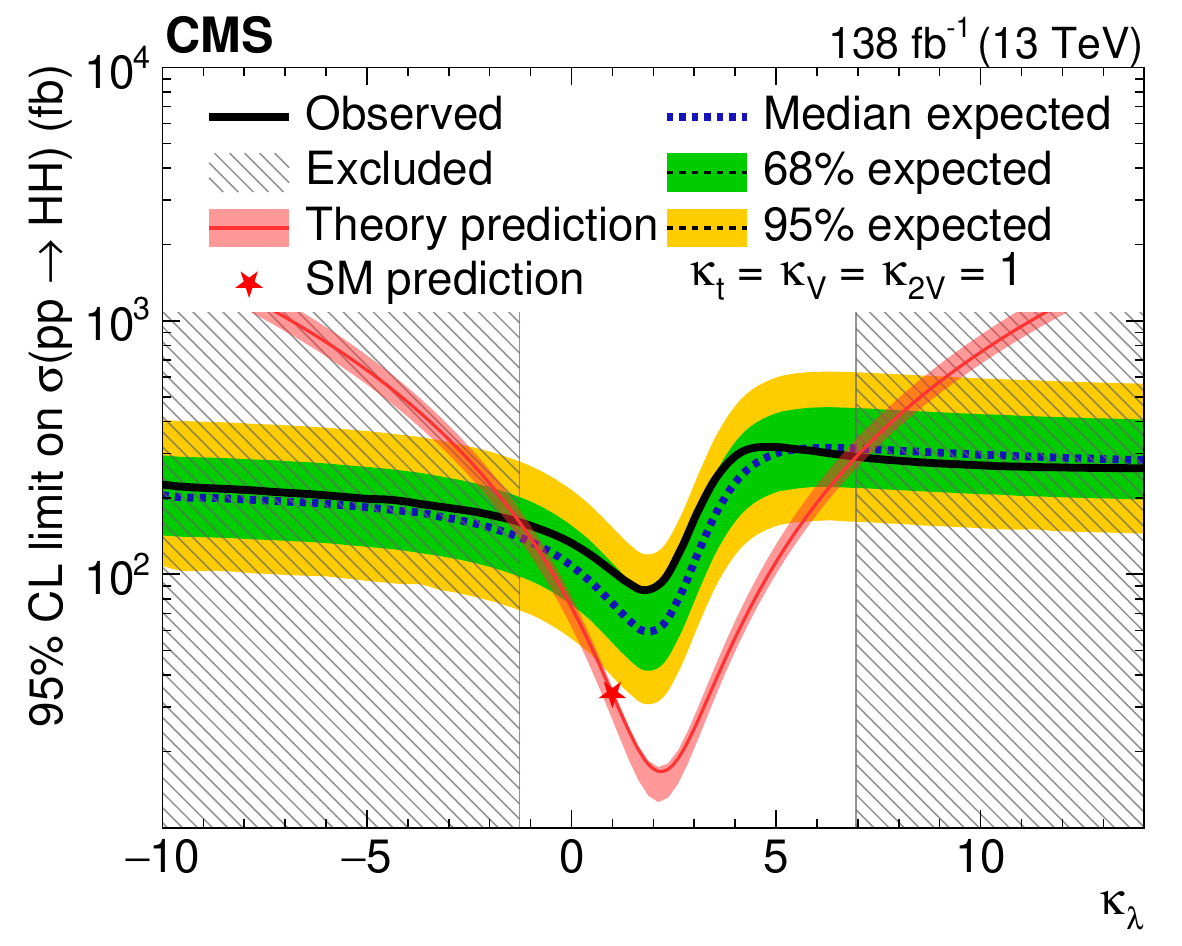}
    \includegraphics[width=0.45\textwidth]{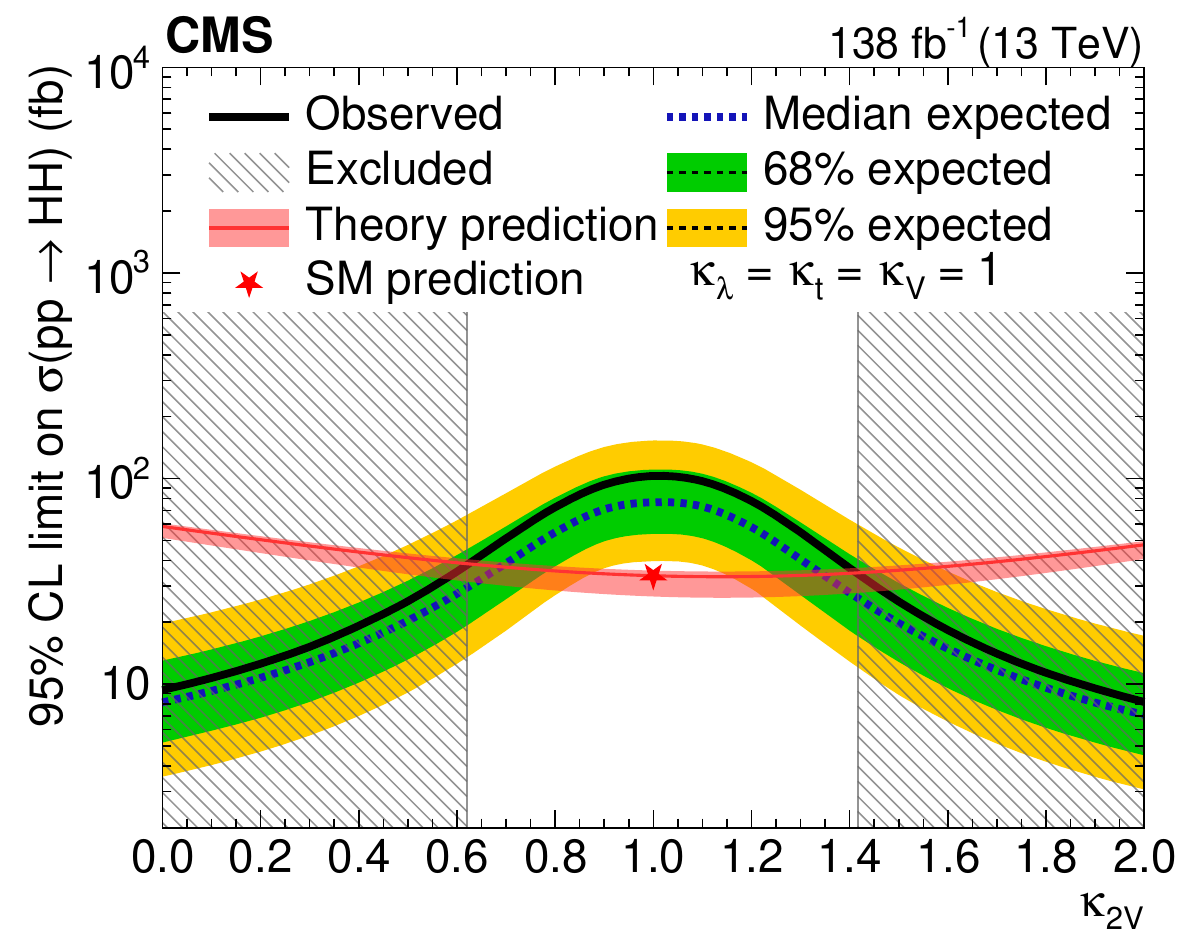}
    \caption{
        The 95\% \CL upper limits on the inclusive \HH cross section as functions of \klambda (left) and \CVV (right).
        All other couplings are set to the values predicted by the SM.
        The theoretical uncertainties in the \GGF and \VBF \HH signal cross sections are not considered because here we directly constrain the measured cross section.
        The inner (green) band and the outer (yellow) band indicate the 68 and 95\% \CL intervals, respectively, under the background-only hypothesis.
        The star shows the limit at the SM value for \klambda and \CVV.
    }
    \label{fig:limits_cross_section_kl_k2V}
\end{figure*}

In Fig.~\ref{fig:limits_cross_section_kl_k2V}, the upper limits at 95\% \CL on the \HH cross section are shown as functions of \klambda (left) and \CVV (right), respectively.
We exclude \HH production at 95\% \CL, meaning the 95\% \CL upper limit on the inclusive \HH signal strength $\mu < 1$, for values of \klambda outside the range from $-1.39$ to 7.02.
Similarly, we exclude \HH production at 95\% \CL for values of the \CVV coupling modifier outside the range from 0.62 to 1.42, with the expected range from 0.69 to 1.35.

The summary of the constraints on the coupling modifiers \klambda and \CVV is given in Table~\ref{tab:ranges_results}.

\begin{figure*}[tbh!]
    \centering
    \includegraphics[width=0.45\textwidth]{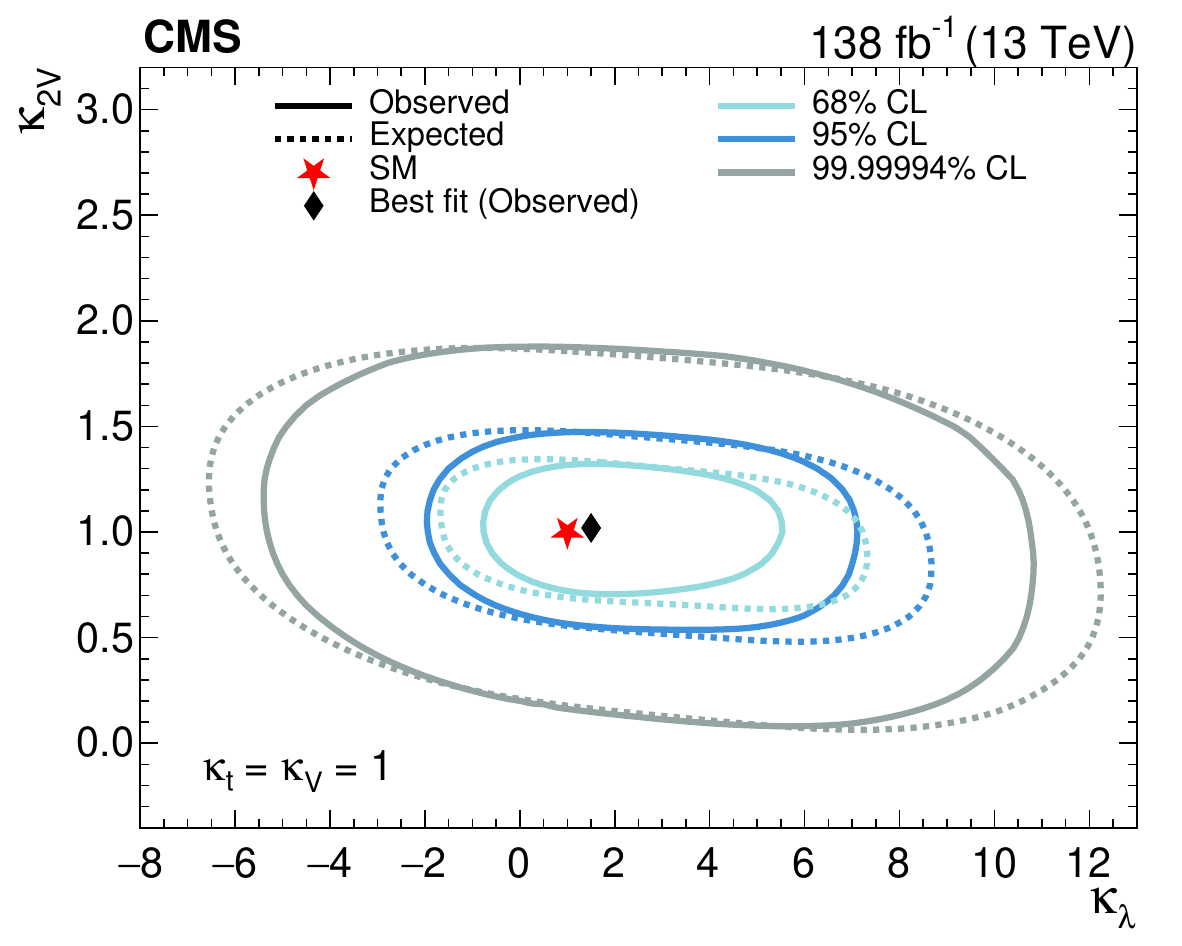}
    \includegraphics[width=0.45\textwidth]{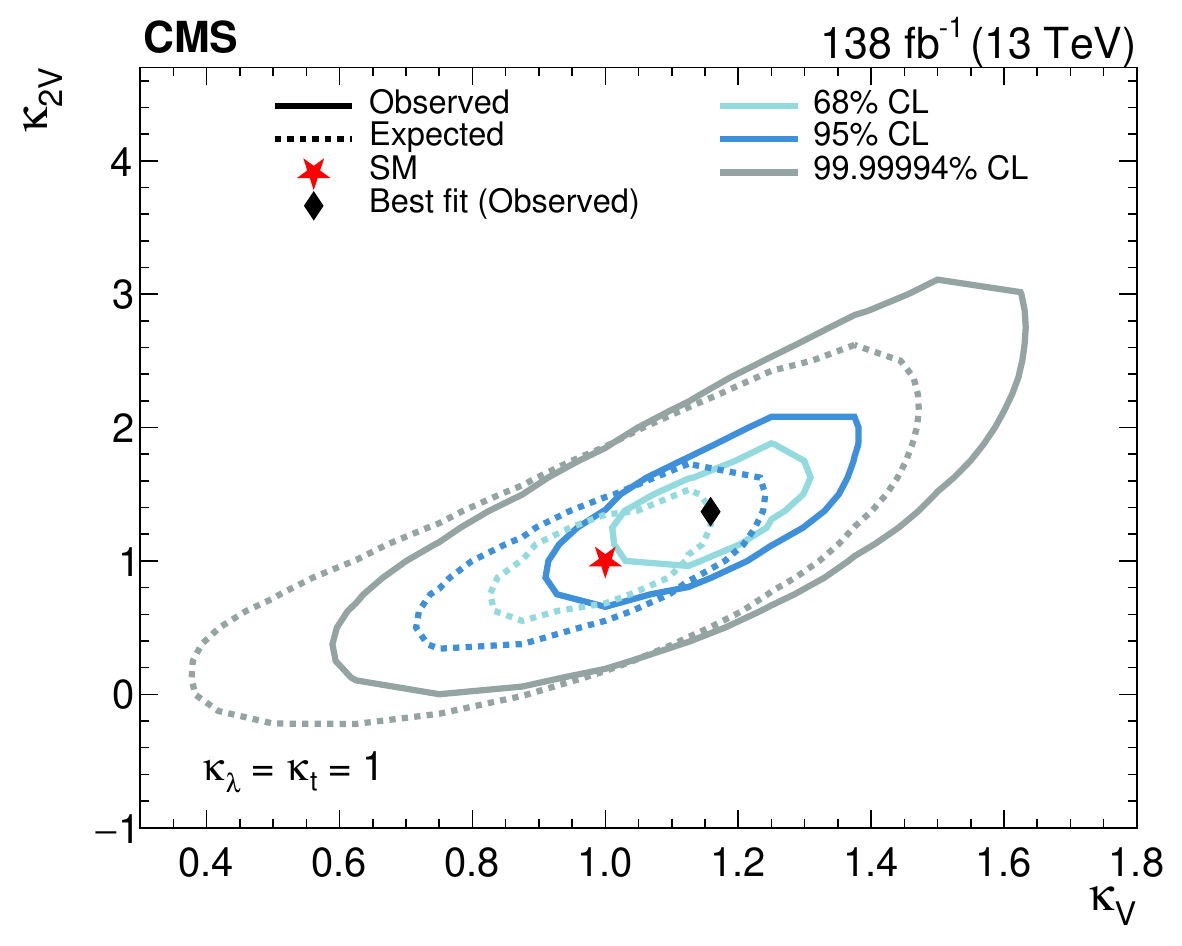}\\
    \includegraphics[width=0.45\textwidth]{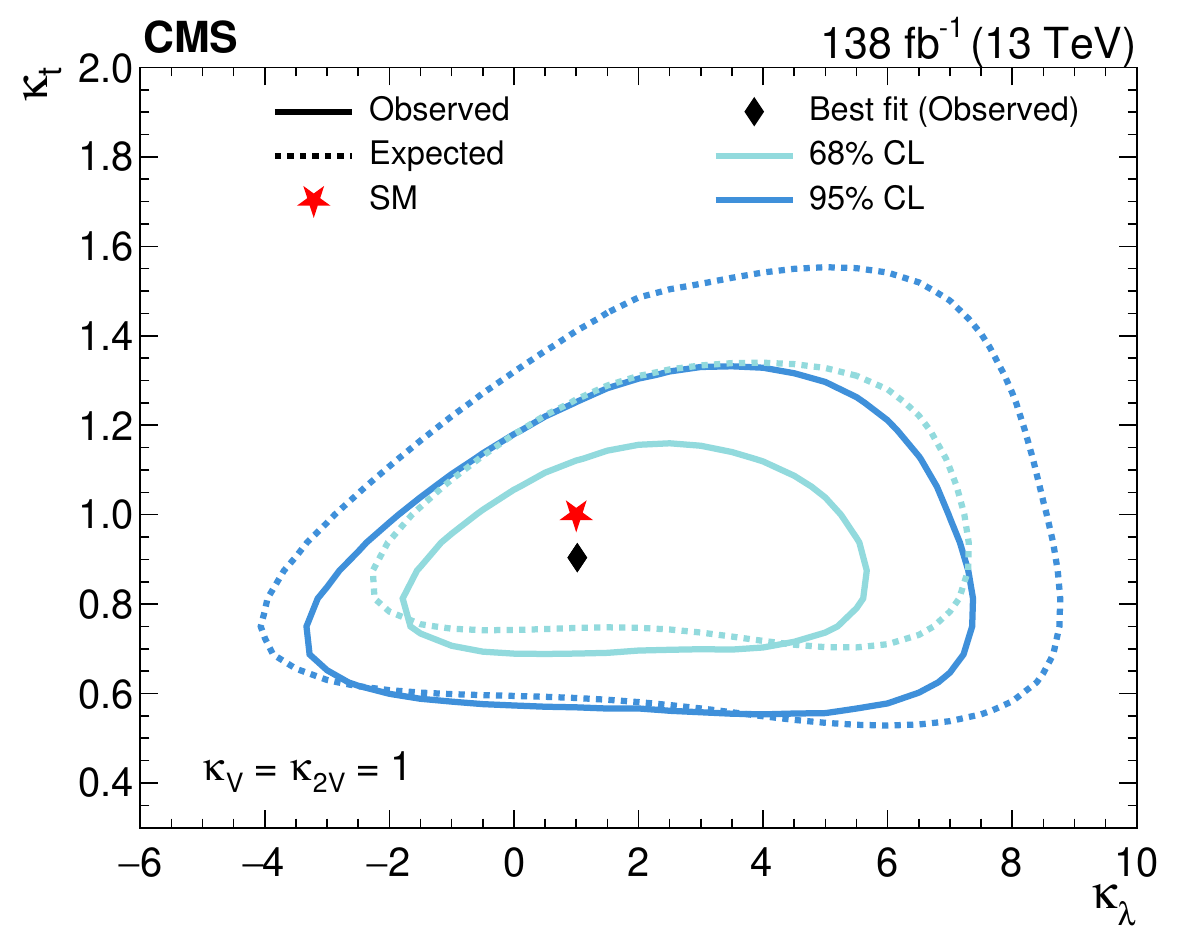}
    \caption{
        The 68 and 95\% \CL contours of $-2\Delta\ln(L)$ in the (\klambda, \CVV) (upper left), (\CV, \CVV) (upper right), and (\klambda, \ktop) (lower) planes for the combination of all channels when all the other parameters are fixed to their SM values.
        In the (\klambda, \CVV) and (\CV, \CVV) planes, the 5 standard deviation \CL contours are also shown.
    }
    \label{fig:likelihoods_2D}
\end{figure*}

Figure~\ref{fig:likelihoods_2D} shows the 68 and 95\% \CL contours of $-2\Delta\ln(L)$ in the (\klambda, \CVV), (\CV, \CVV), and (\klambda, \ktop) planes for the combination of all contributing channels.
In the (\klambda, \CVV) and (\CV, \CVV) planes, the 5 standard deviation \CL contours are also shown.
All the other parameters, besides the ones scanned, are set to the values expected by the SM.
The value of $\CVV=0$ is excluded at a \CL of more than 5 standard deviations, for every value of \klambda, adding to the robustness of this exclusion.
No significant deviation from the SM is observed.

\begin{table*}[htbp!]
    \centering
    \topcaption{
        Summary of results on constraints to the coupling modifiers \klambda and \CVV.
        The ranges are either extracted using either the $-2\Delta\ln(L)$ scan at 68\% and 95\% \CL, or upper limits on the signal strength $\mu$ at 95\% \CL.
        The theoretical uncertainties in the \GGF and \VBF \HH signal cross sections are considered in all results tabulated here.
    }
    \label{tab:ranges_results}
    \cmsTable{
        \begin{tabular}{ l  c  c }
            \hline
            {Method}                        & \klambda observed (expected)      & \CVV observed (expected)    \\
            \hline
            $-2\Delta\ln(L)$ scan, 68\% \CL & [$-0.07$, 4.18] ([$-0.87$, 6.31]) & [0.81, 1.23] ([0.77, 1.26]) \\
            $-2\Delta\ln(L)$ scan, 95\% \CL & [$-1.35$, 6.35] ([$-2.24$, 7.89]) & [0.64, 1.40] ([0.62, 1.41]) \\
            Limits on $\mu$, 95\% \CL       & [$-1.39$, 7.02] ([$-1.02$, 7.19]) & [0.62, 1.42] ([0.69, 1.35]) \\
            \hline
        \end{tabular}}
\end{table*}

Beyond varying \klambda, \CVV, \CV, and \ktop, we interpret the data in terms of couplings that are not predicted in the SM, namely \ctwo, \cg, and \cgg.
All BSM interpretations studied in this paper only alter the \GGF production, while the \VBF production is assumed to be as predicted by the SM.
The channels contributing to the BSM interpretations are \bbgg, \bbbb boosted and resolved, \bbtt, \bbWW, multilepton, \WWgg, and \ttgg.

\begin{figure*}[tbh!]
    \centering
    \includegraphics[width=0.9\textwidth]{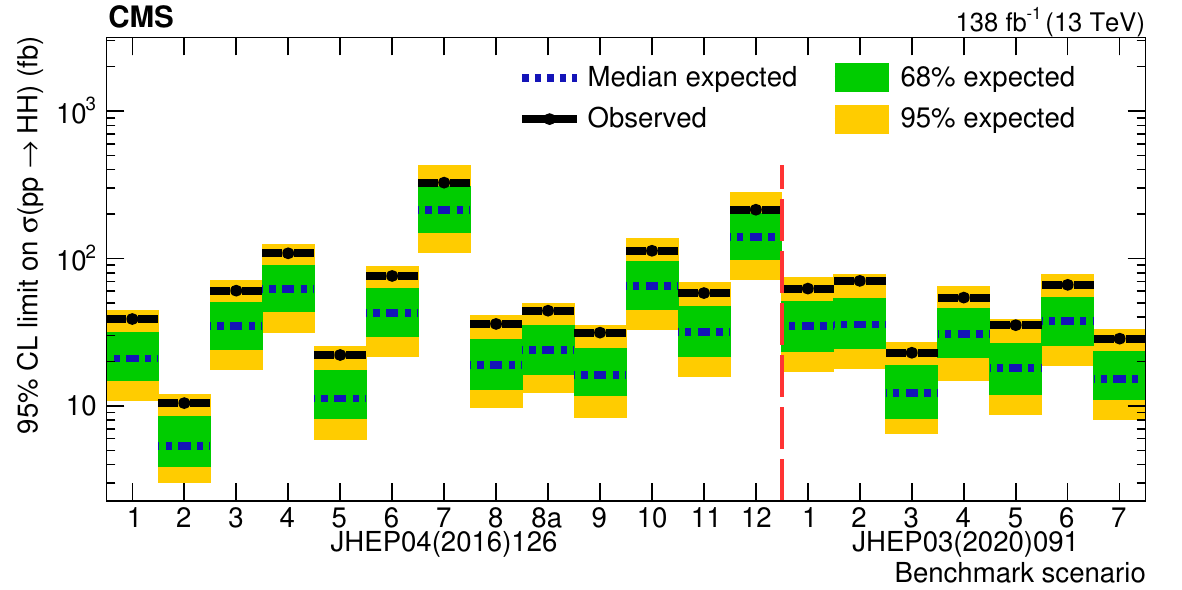}
    \caption{
        The upper limits at 95\% \CL on the \HH production cross section for the two sets of HEFT benchmarks.
        The theoretical uncertainties in the \GGF \HH signal cross section are not considered because we directly constrain the measured cross section.
    }
    \label{fig:limits_EFT_BM}
\end{figure*}

First, we interpret the results in the context of two sets of benchmarks, JHEP04(2016)126~\cite{Carvalho_2016} and JHEP03(2020)091~\cite{Capozi_2020}, combinations of the coupling modifiers ($\klambda, \ktop, \ctwo, \cg, \cgg$) as described in Section~\ref{sec:MC}.
The \ttgg channel only contributes to the results for the benchmarks JHEP04(2016)126, while the rest of the channels contribute to both sets.
The upper limits on the \HH cross section at 95\% \CL are shown in Fig.~\ref{fig:limits_EFT_BM}.
No significant deviations from expectations are observed, but there is an overall excess in all benchmarks, between 68\% and 95\% \CL, consistent with excesses observed in the most sensitive channels, namely \bbgg and \bbbb especially in the boosted topology, visible in Fig.~\ref{fig:limits_cross_section_r}.

\begin{figure*}[tbh!]
    \centering
    \includegraphics[width=0.45\textwidth]{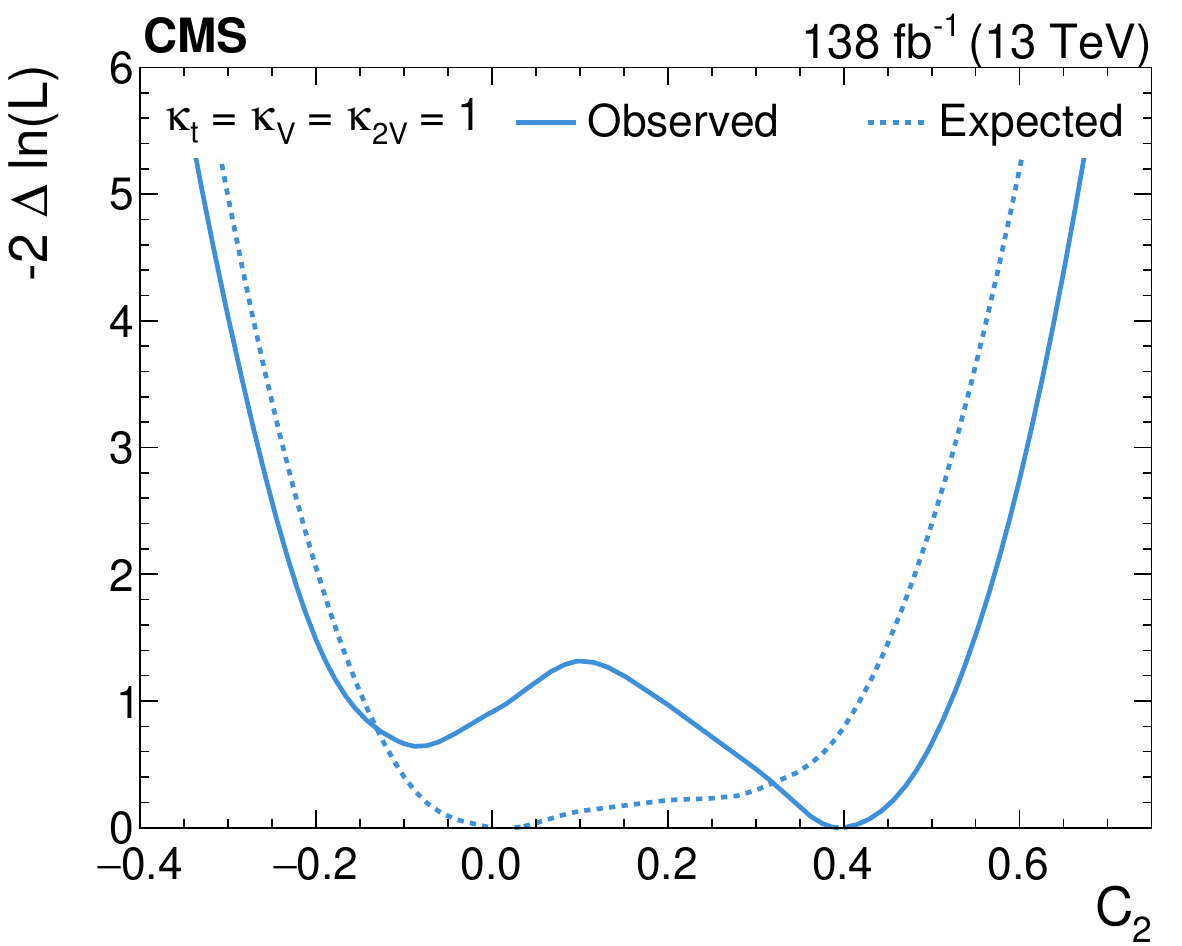}
    \includegraphics[width=0.45\textwidth]{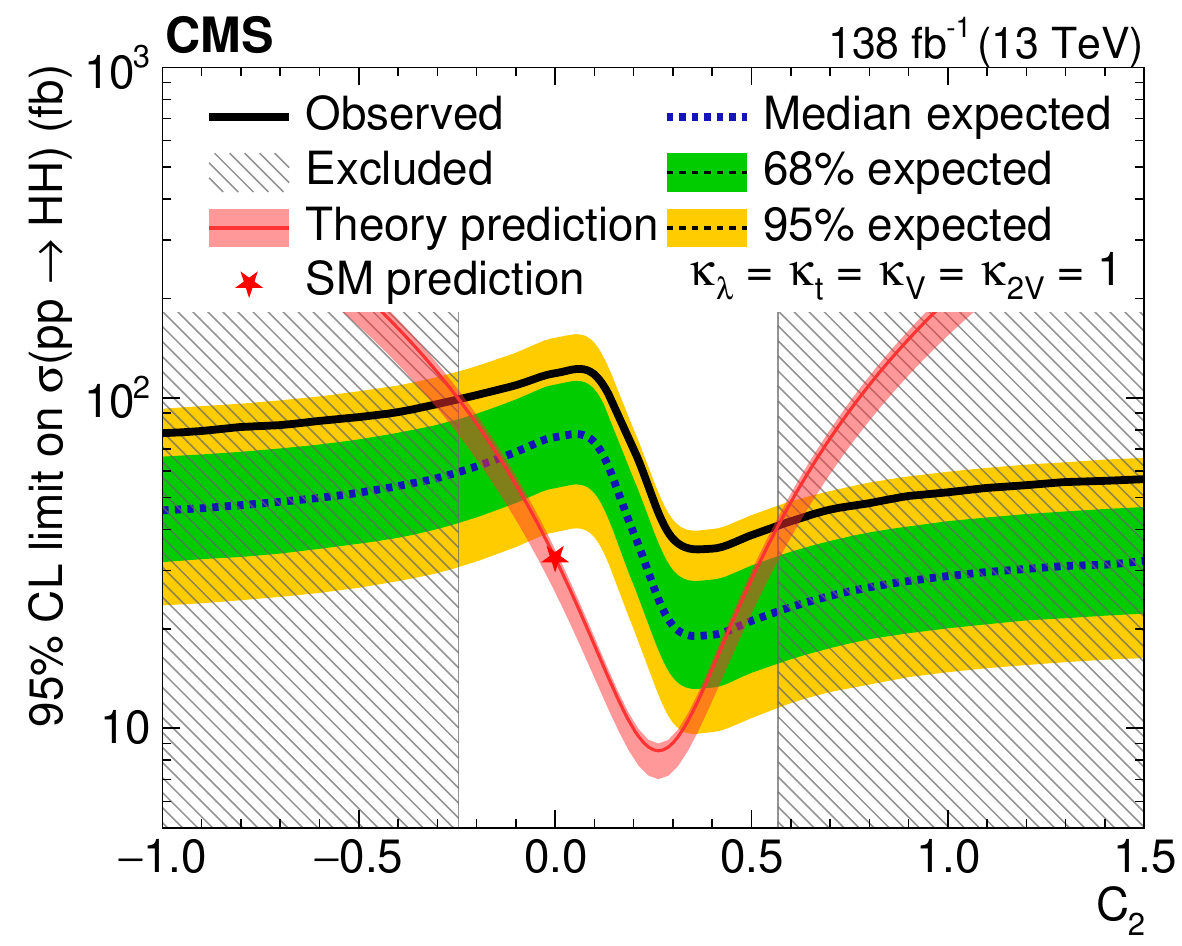}

    \caption{
        The upper limits at 95\% \CL on the \HH cross section as a function of the \ctwo coupling modifier (left).
        The theoretical uncertainties in the \GGF \HH signal cross section are not considered because we directly constrain the measured cross section.
        The $-2\Delta\ln(L)$ scan as a function of the \ctwo coupling modifier (right).
    }
    \label{fig:c2_limits_likelihood}
\end{figure*}

\begin{figure*}[tbh!]
    \centering
    \includegraphics[width=0.45\textwidth]{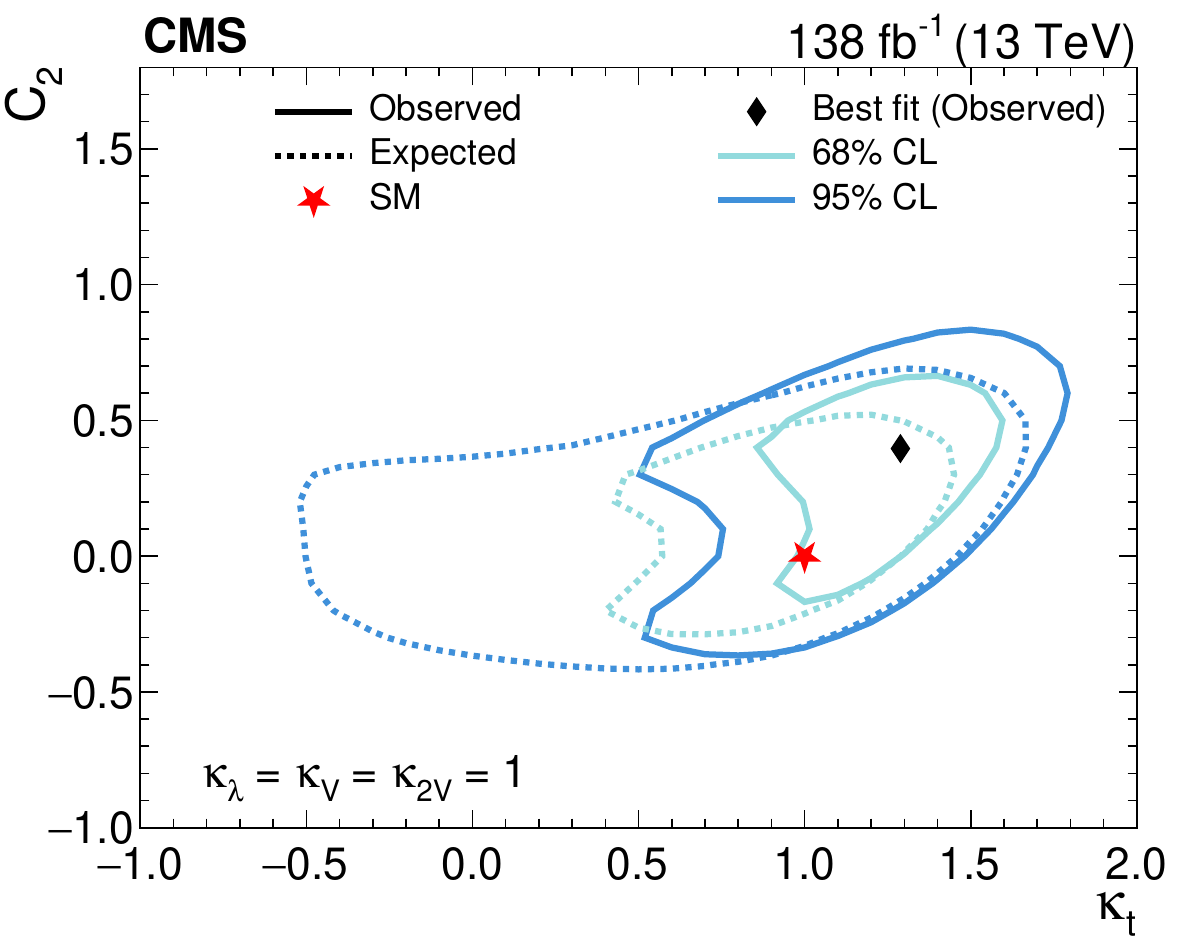}
    \includegraphics[width=0.45\textwidth]{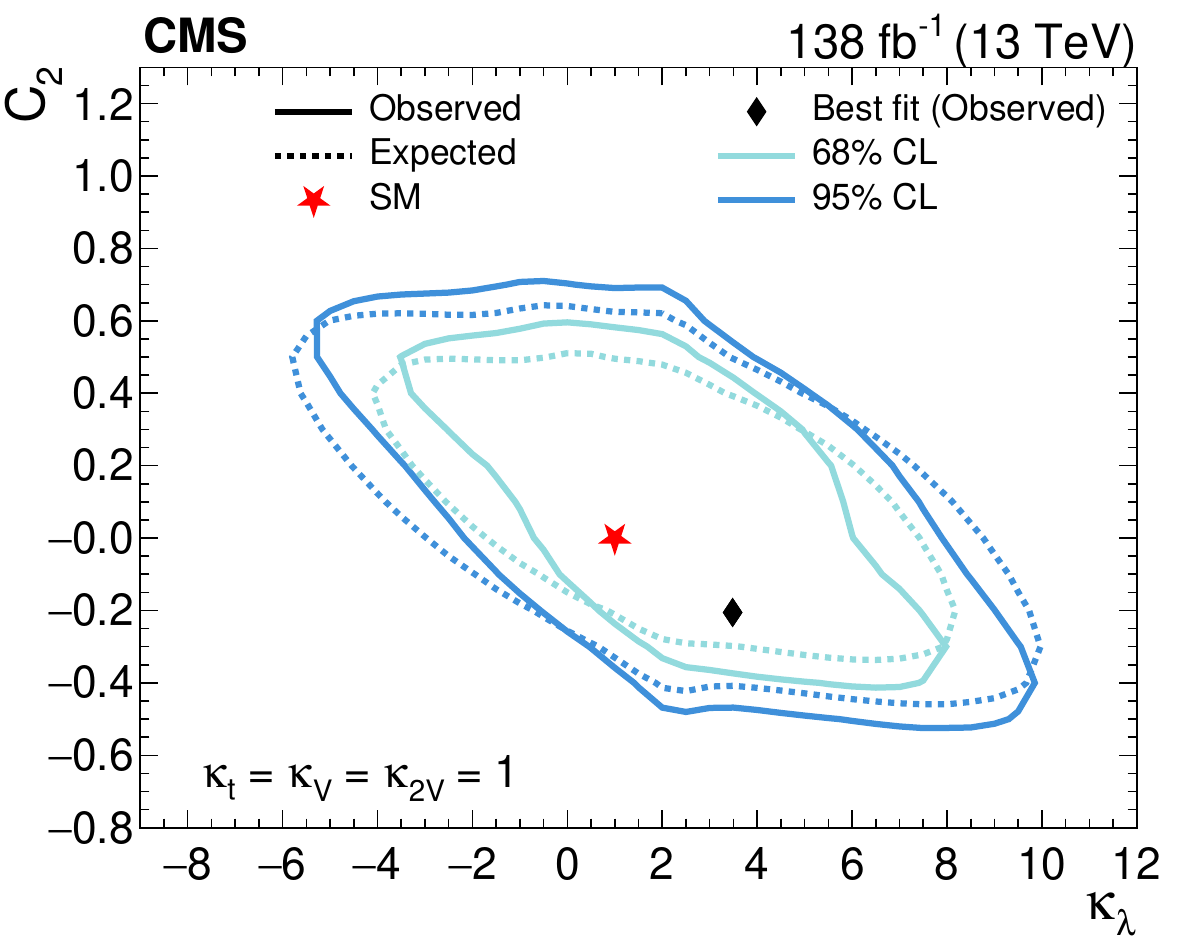}
    \caption{
        The 68 and 95\% \CL contours of $-2\Delta\ln(L)$ in the $(\ctwo, \ktop)$ (left) and $(\ctwo, \klambda)$ (right) planes for the combination of all channels when all the other parameters are fixed to their SM values.
    }
    \label{c2kt_c2kl_likelihood_2D}
\end{figure*}

In the HEFT Lagrangian, the term containing \ktop is correlated with the coupling modifier \ctwo, which corresponds to the BSM coupling between two top quarks and two Higgs bosons.
Figure~\ref{fig:c2_limits_likelihood} shows the $-2\Delta\ln(L)$ scan for \ctwo on the left and the upper limits on the \HH cross section as a function of \ctwo on the right.
For \ctwo, the best fit to the data is found to be 0.40, the 68\% \CL interval is from 0.20 to 0.52 ($-0.15$ to 0.42 expected), and the 95\% \CL interval is from $-0.29$ to 0.63 ($-0.27$ to 0.56 expected).
Taking into account the theoretical uncertainties in the \GGF and \VBF \HH signal cross sections, we exclude \HH production at 95\% \CL for \ctwo values outside the range from $-0.28$ to 0.59.
The corresponding expected range is from $-0.17$ to 0.47.
Figure~\ref{c2kt_c2kl_likelihood_2D} shows the two-dimensional contours of $-2\Delta\ln(L)$ in the (\ctwo, \klambda) and (\ctwo, \ktop) planes for the combination of all contributing channels.

The data are further interpreted in the context of several UV complete models, exploiting the HEFT parametrisation described in Section~\ref{sec:MC}.
Experimental constraints on the parameters that characterise each model are extracted, using a mapping between \klambda, \ktop, and \ctwo and the specific model parameters.

\begin{figure*}[tbh!]
    \centering
    \includegraphics[width=0.45\textwidth]{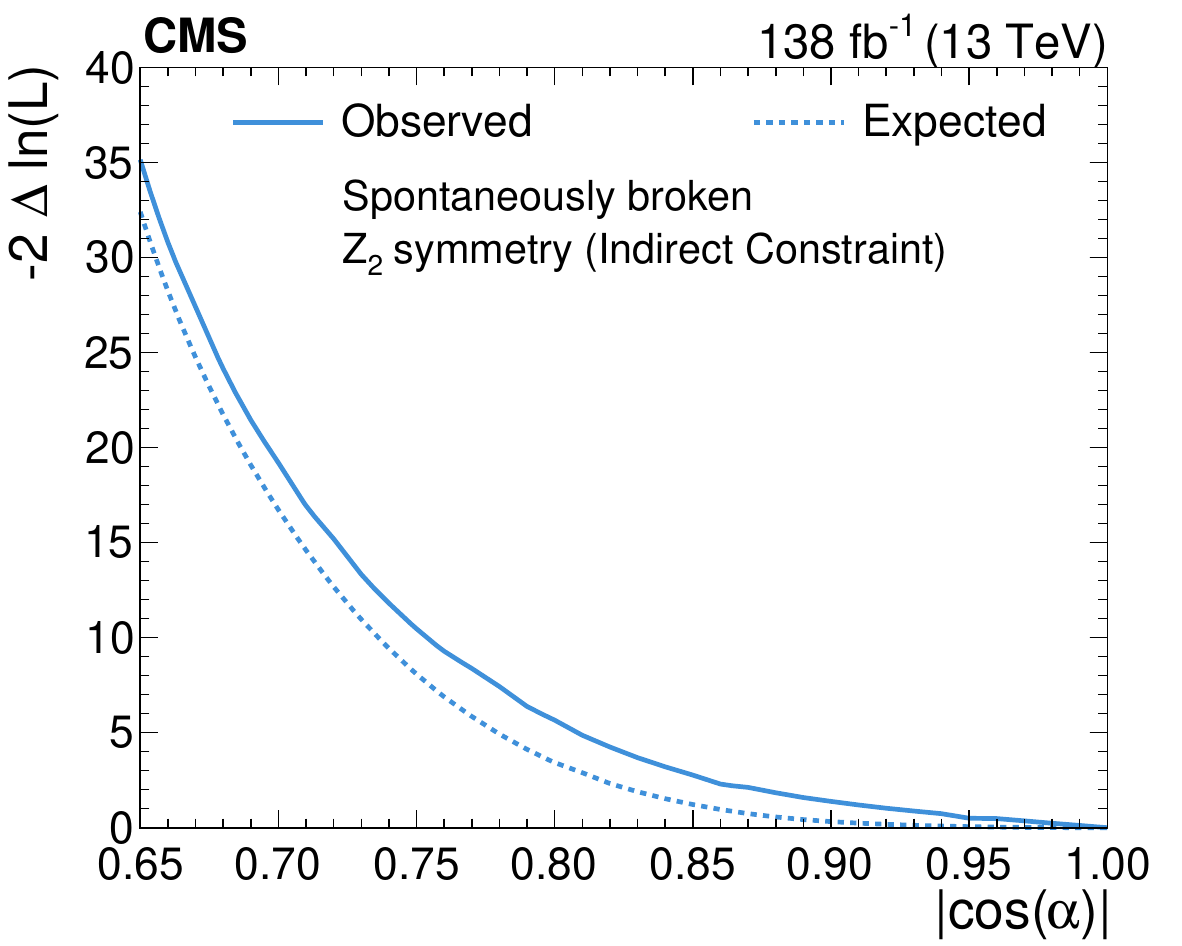}
    \includegraphics[width=0.45\textwidth]{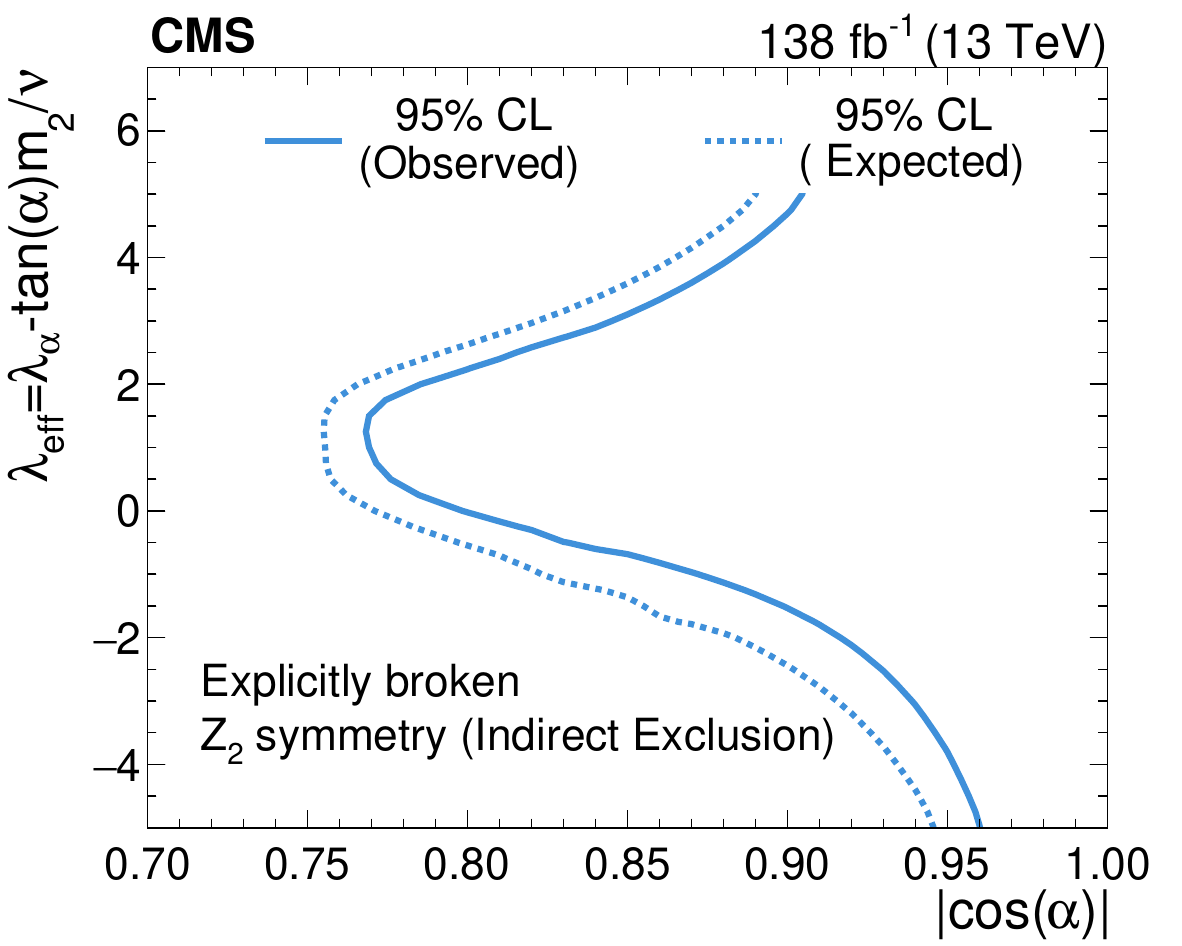}
    \caption{
        The $-2\Delta\ln(L)$ scan for the combination of all channels as a function of $\cos{\alpha}$ for singlet extensions of the SM with spontaneous $Z_2$ symmetry breaking (left).
        The $\cos{\alpha}$ is constrained at 95\% \CL from 0.79 to 1.
        The 95\% \CL contour of $-2\Delta\ln(L)$ in the ($\cos{\alpha}$, $\lambda_\text{eff}$) plane (right), where $\lambda_\text{eff}=\lambda_\alpha-\tan{(\alpha)\frac{m_{2}}{\nu}}$
        The ranges of $\cos{\alpha}$ and $\lambda_\text{eff}$ are chosen to guarantee the validity of the models.
        The excluded regions are below and to the left of the curves shown.
    }
    \label{fig:res_singletlikelihood}
\end{figure*}

The simplest extensions of the SM Higgs sector are models with the addition of a new scalar singlet $\Phi$ with couplings to the Higgs doublet~\cite{Dawson:2017vgm,deBlas:2014mba}.
For the following we consider both the case where the $Z_{2}$ symmetry of the resulting potential $V(\Phi, \PH)$ is either explicitly or spontaneously broken.
In the explicit case the modifications of Higgs boson interactions can be written in terms of $\alpha$, the mixing angle of the new scalar singlet with the Higgs doublet, $m_{2}$, the coefficient of the triple scalar singlet coupling and $\lambda_\alpha$ the coefficient of the biquadratic term between scalar singlet and Higgs potentials.
In the spontaneously broken symmetry case, the modifications depend only on $\alpha$.
Figure~\ref{fig:res_singletlikelihood} shows the $-2\Delta\ln(L)$ scan as a function of $\cos{\alpha}$ and the 95\% \CL contour in the ($\cos{\alpha}$, $\lambda_\text{eff}$) plane, where $\lambda_\text{eff}=\lambda_\alpha-\tan{(\alpha)\frac{m_{2}}{\nu}}$, for singlet extensions of the SM with both explicit and spontaneous $Z_{2}$ symmetry breaking.
For the spontaneous case, the allowed range of $\cos{\alpha}$ at 68\% \CL is from 0.92 to 1 (expected 0.86 to 1) and at 95\% \CL is from 0.79 to 1 (expected 0.82 to 1).
The strongest constraints come from the \bbtt and \bbbb boosted channels.

In the 2HDM additional bosons arise from the introduction of a second complex Higgs doublet that mixes with the SM Higgs doublet~\cite{PhysRevD.8.1226,Branco:2011iw,Haber:2015pua,Kling:2016opi}.
Four cases are considered, based on which of the two doublets the SM fermions couple to.
For `Type-I', the charged fermions only couple to the second doublet, for `Type-II', up- and down- type fermions couple to different doublets, in `Type-X', charged leptons couple to the first doublet and both up- and down- type quarks couple to the second doublet, and finally for `Type-Y' up-type quarks and charged leptons couple to the second doublet, while down type quarks couple to the first.
Based on the mixing of the two doublets, five scalar bosons arise: a CP even scalar $\PH$, a CP odd scalar $\PA$, both of which are usually heavy; a light SM-like Higgs boson $\Phlight$, and two charged Higgs bosons $\PH^\pm$.
The model is characterised by the masses of these five bosons, the mixing angle of the CP-even Higgs fields $\alpha$, the ratio of the vacuum expectation values in the two associated Higgs potentials $\tan{\beta}=\frac{\nu_2}{\nu_1}$ and the mass mixing parameter $m_{12}$.
The masses of the heavy scalars $\PH$ and $\PA$ are considered to be equal and noted as $m_\PA$.

\begin{figure*}[tbh!]
    \centering
    \includegraphics[width=0.45\textwidth]{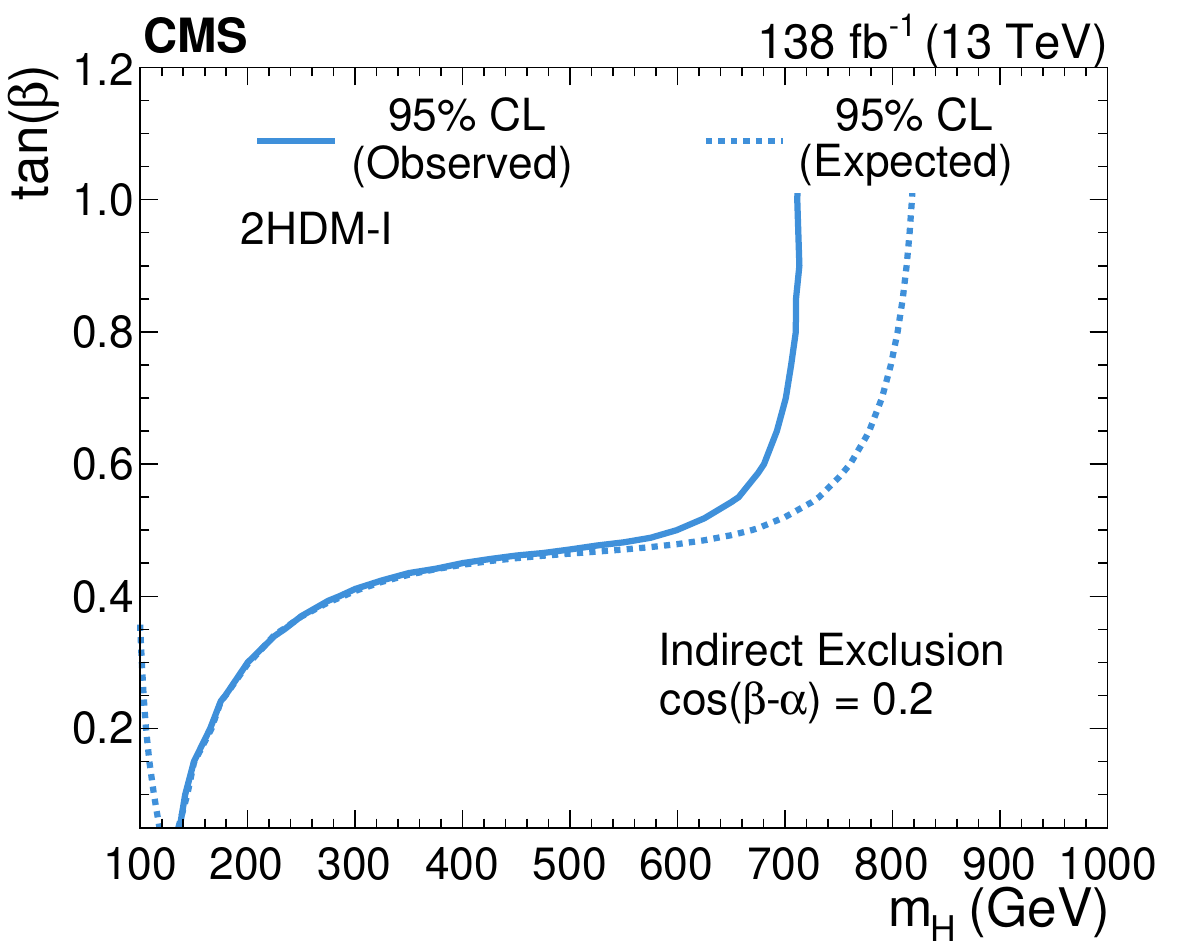}
    \includegraphics[width=0.45\textwidth]{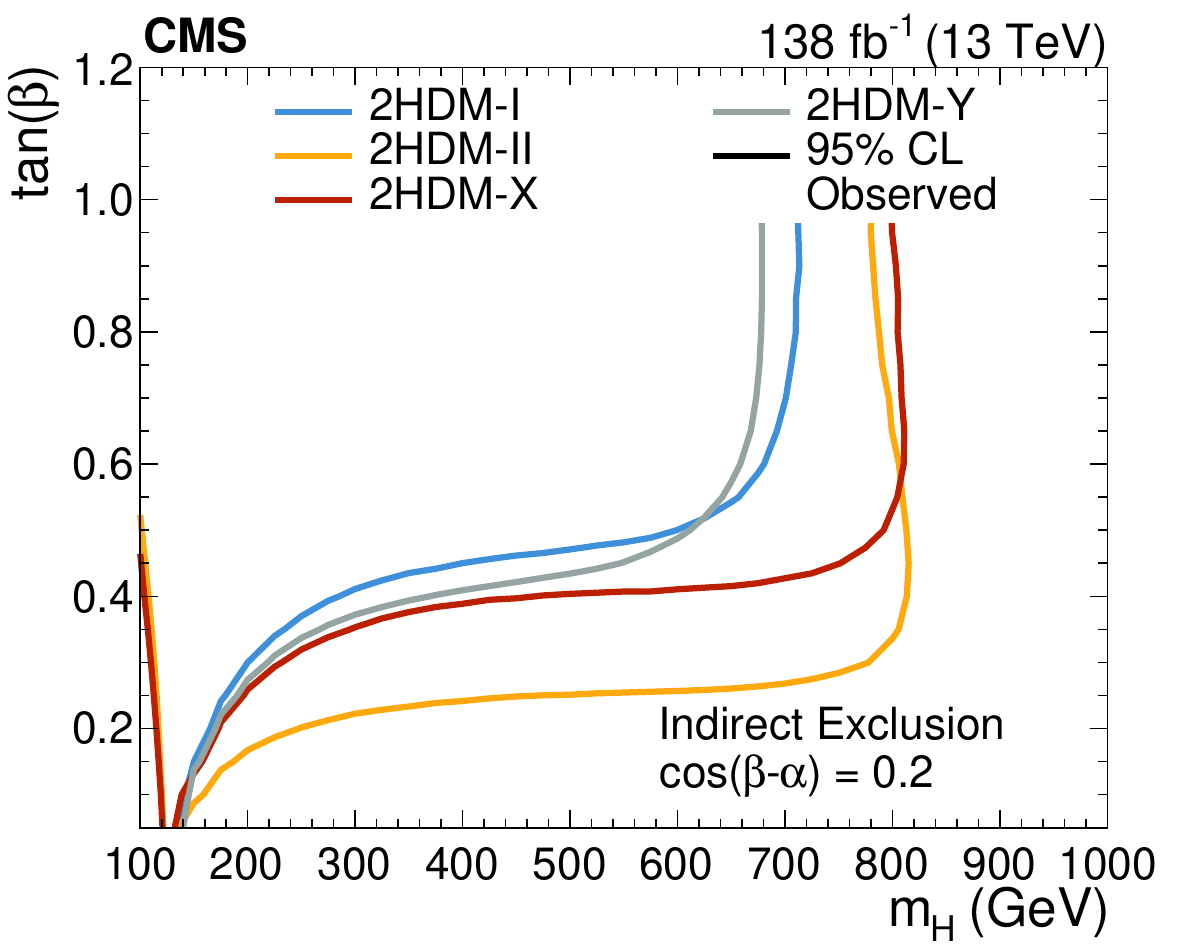}
    \caption{
    The 95\% \CL contour of $-2\Delta\ln(L)$ for the combination of all channels in the ($m_{\PH}$, $\tan\beta$) plane for a fixed value of $\abs{\cos(\beta-\alpha)}=0.2$ in the 2HDM-I model (left) and the four considered 2HDM models (right).
    In these and all other cases considered in this paper, $m_{\PH}=m_{\PA}$.
    The ranges of $m_{\PH}$ and $\tan\beta$ are chosen to guarantee the validity of the models.
    The excluded regions are below the curves shown.
    The value $\tan\beta=0.5$ is excluded for $m_{\PH}>800\GeV$ for all models considered.
    }
    \label{fig:res_2HDM_MHTB}
\end{figure*}

\begin{figure*}[tbh!]
    \centering
    \includegraphics[width=0.45\textwidth]{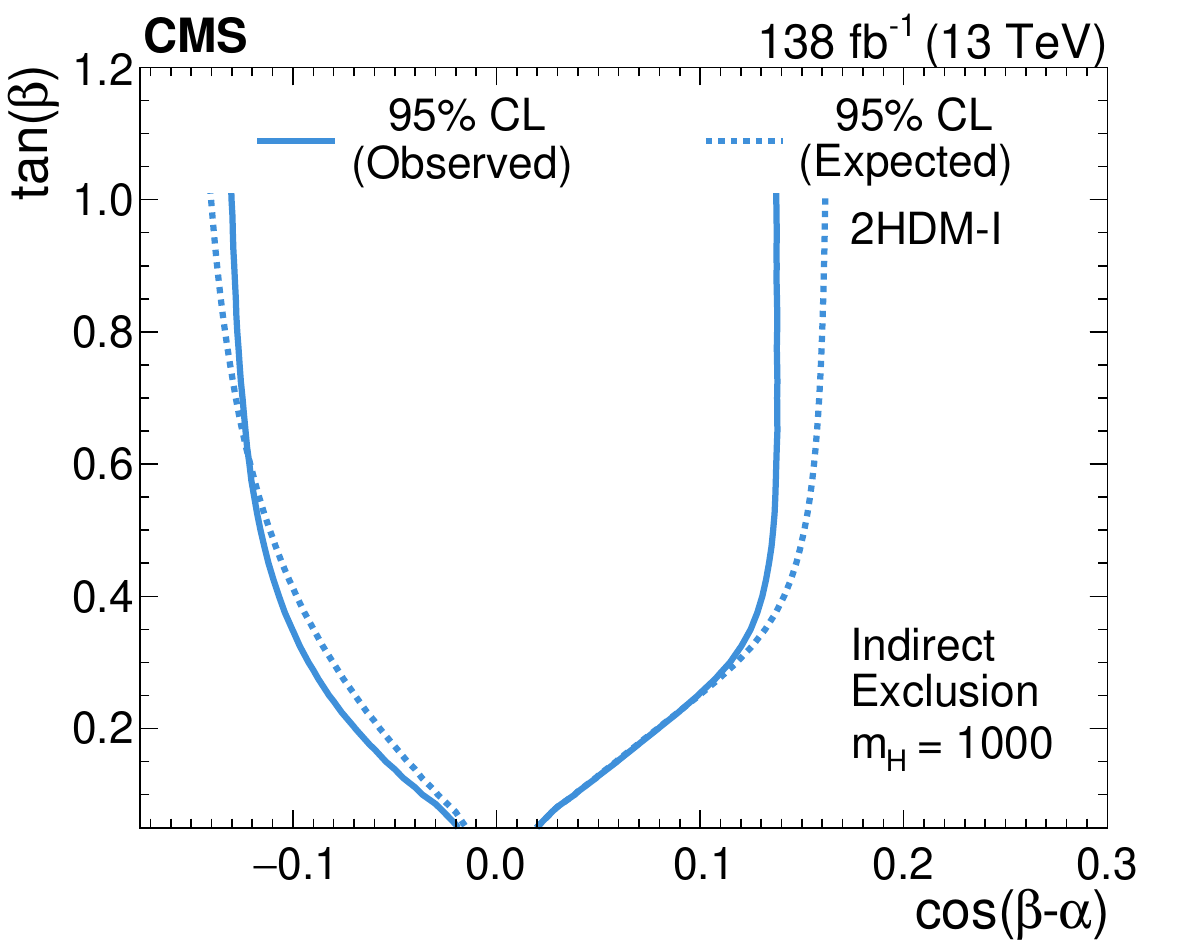}
    \includegraphics[width=0.45\textwidth]{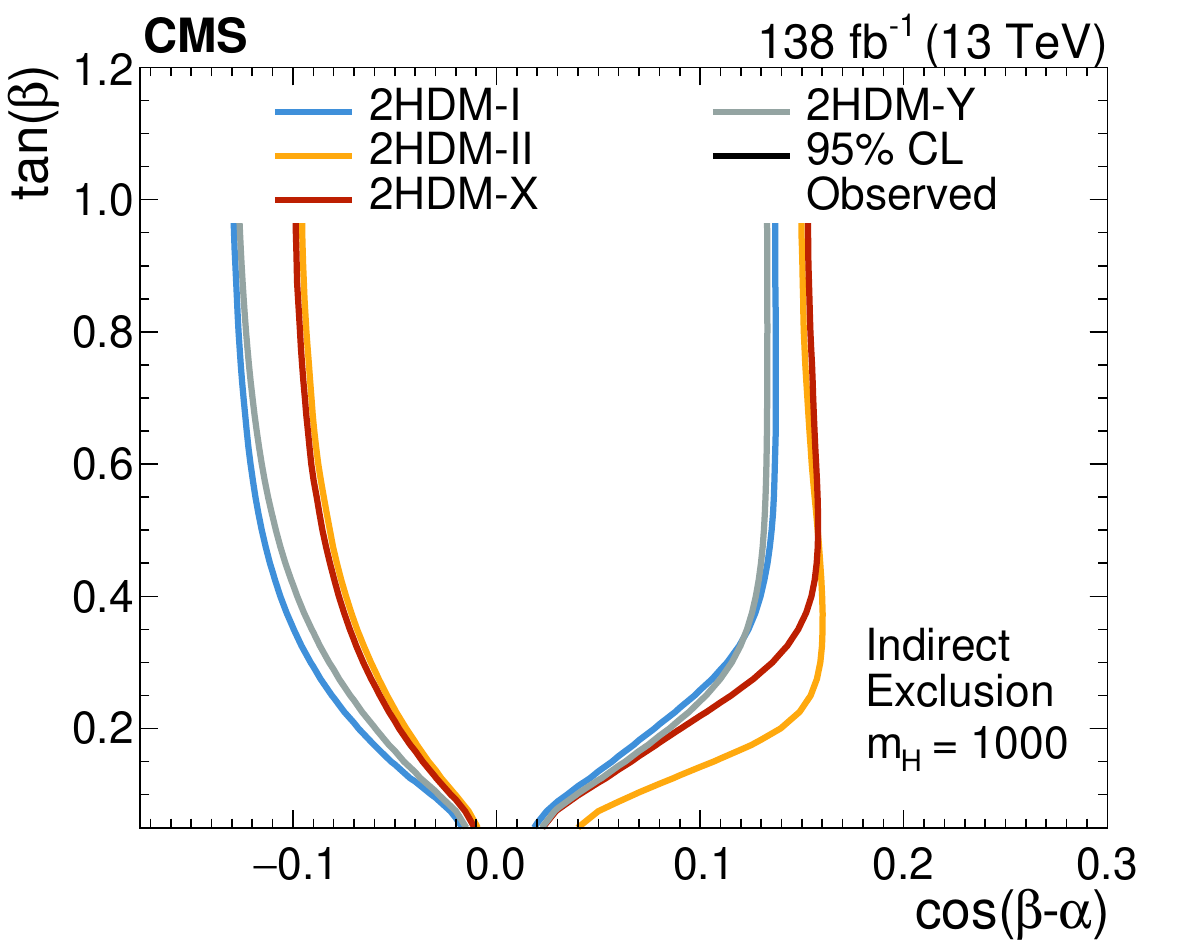}
    \caption{
    The 95\% \CL contour of $-2\Delta\ln(L)$ for the combination of all channels in the ($\cos(\beta-\alpha)$, $\tan\beta$) plane for a fixed value of $m_{\PH}=1000\GeV$ in the 2HDM-I model (left) and the four considered 2HDM models (right).
    In these and all other cases considered in this paper, $m_{\PH}=m_{\PA}$.
    The ranges of $\cos(\beta-\alpha)$ and $\tan\beta$ are chosen to guarantee the validity of the model.
    The excluded regions are below the curves shown.
    The value $\tan\beta=0.5$ is excluded for $\cos(\beta-\alpha)>0.16$ and $\cos(\beta-\alpha)<-0.13$ for all models considered.
    }
    \label{fig:res_2HDM_CBATB}
\end{figure*}

Figures~\ref{fig:res_2HDM_MHTB} and~\ref{fig:res_2HDM_CBATB} show the results of the combination in the ($m_{\PH}$, $\tan\beta$) and ($\abs{\cos(\beta-\alpha)}$, $\tan\beta$) planes, respectively.
On the left are the observed and expected 95\% \CL contours for the 2HDM-I case, while the right shows the observed contours in all 2HDM cases.
The excluded regions are below the curves shown.
The results are derived under the assumption that we are not sensitive to the direct features of the resonant signal and that the interference effects between the resonant and nonresonant signal are negligible.
Therefore, no interference effects are taken into account in this study.

Another simple extension of the SM is the minimal composite Higgs model (MCHM)~\cite{Giudice:2007fh,Grober:2010yv,Contino:2013kra,Contino:2010rs}.
Two cases are studied, MCHM$_4$ and MCHM$_5$, where the SM fermions either transform as spinor or fundamental representations of the underlying SO(5) symmetry.
These models are characterised by the compactness parameter $\xi=(\frac{\nu}{f})^{2}$ with $\nu$ being the vacuum expectation value and $f$ the strong dynamics scale.

\begin{figure*}[tbh!]
    \centering
    \includegraphics[width=0.45\textwidth]{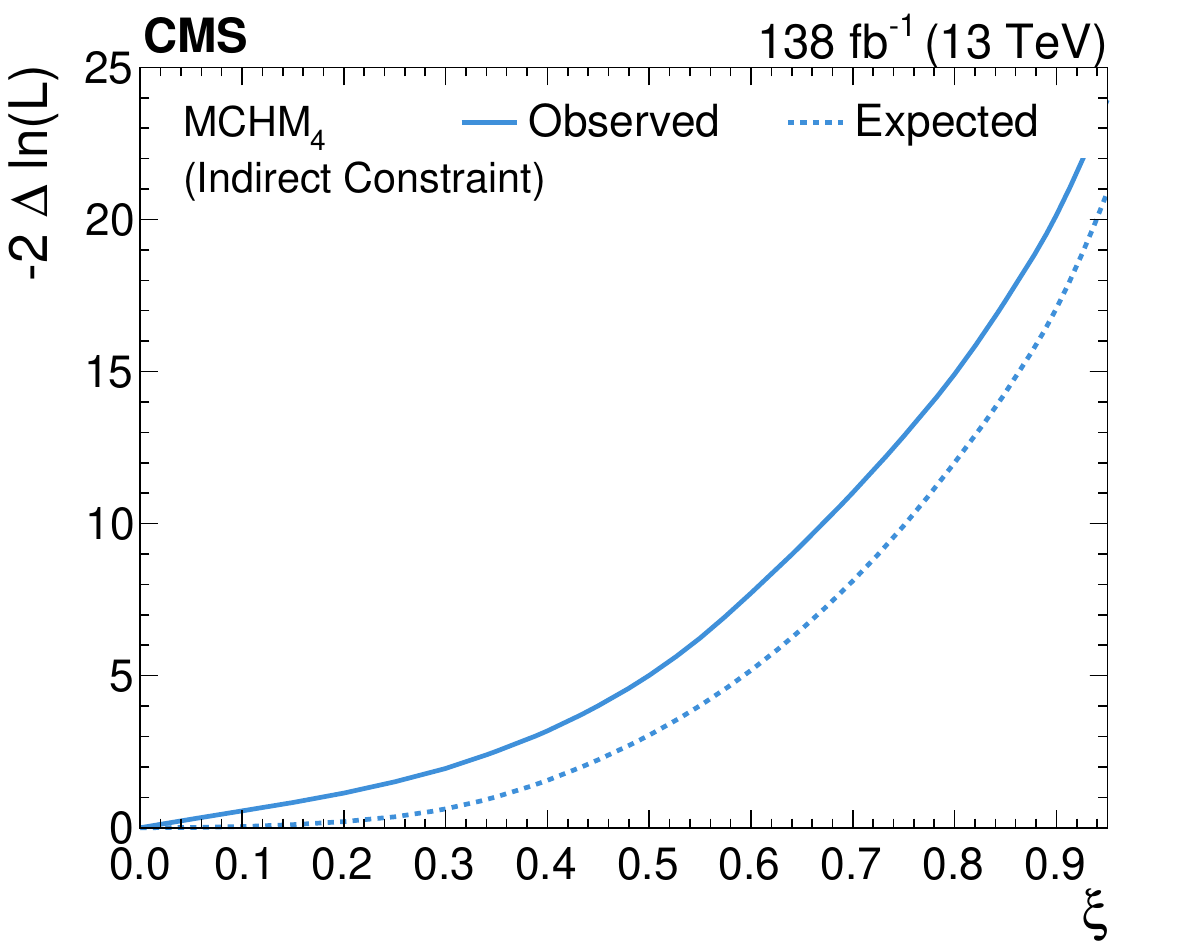}
    \includegraphics[width=0.45\textwidth]{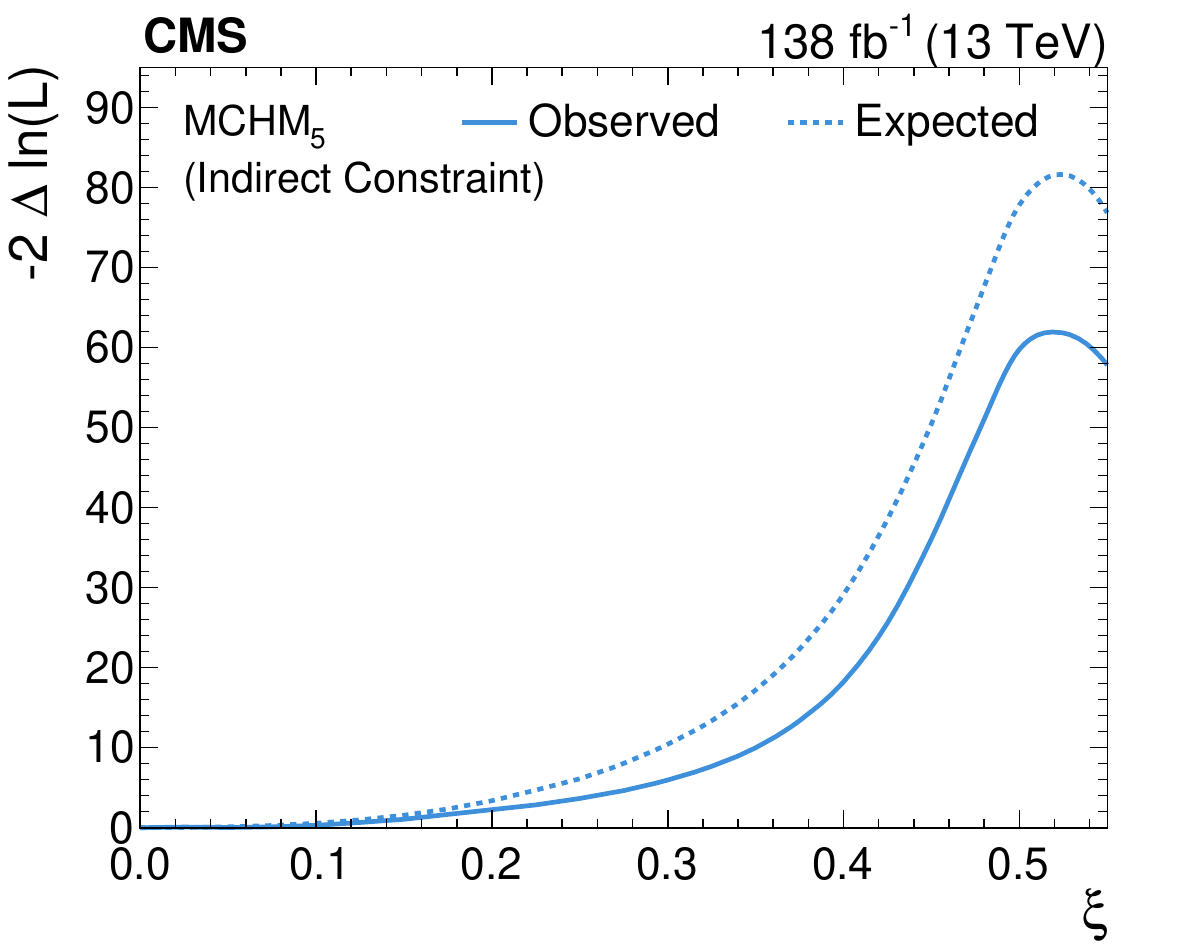}
    \caption{
        The $-2\Delta\ln(L)$ scan for the combination of all channels as a function of $\xi$ for MCHM$_4$ (left) and MCHM$_5$ (right).
        The range of $\xi$ is chosen to guarantee the validity of the model.
        At 95\% \CL, $\xi$ is constrained to be between 0 and 0.45 in MCHM$_4$ and 0 and 0.26 in MCHM$_5$.
    }
    \label{fig:res_mchmlihood}
\end{figure*}

Figure~\ref{fig:res_mchmlihood} shows the observed and expected $-2\Delta\ln(L)$ scan as a function of $\xi$ for MCHM$_4$ and MCHM$_5$.
For MCHM$_4$, the best fit is $\xi=0.0$, while the observed (expected) 68 and 95\% \CL ranges are found to be from 0 to 0.18 (0 to 0.35) and 0 to 0.45 (0 to 0.55), respectively.
The strongest constraints come from the \WWgg and \bbWW channels.
Likewise, for MCHM$_5$, the best fit is $\xi=0.05$, while the observed (expected) 68 and 95\% \CL ranges are found to be from 0 to 0.15 (0 to 0.13) and 0 to 0.26 (0 to 0.21), respectively.
In this case, the strongest constraints come from the multilepton and \WWgg channels.

\clearpage
\section{Projections for the HL-LHC}
\label{sec:results_projections}

To further probe the SM, with particular interest in the Higgs boson self-coupling, the High-Luminosity LHC (HL-LHC)~\cite{Aberle:2749422} operation is currently scheduled to begin in 2030.
An integrated luminosity up to 3000\fbinv is expected to be collected over the anticipated ten years of data taking.
This unparalleled data set will open a unique window on the weak-scale nature of the universe, and the study of the Higgs boson self-coupling represents one of the most important targets of the HL-LHC.
It is therefore of interest to extrapolate the current results to predict the sensitivity that can be achieved at the HL-LHC.

The results are projected to different integrated luminosity values, \ie 300, 1000, 2000, and 3000\fbinv to track the evolution during the data taking.
In the projection studies, only the \bbgg, \bbtt, \bbbb, multilepton, and \bbWW channels are included as they are the ones with the greatest sensitivity.
The procedure for the statistical combination is the same as described in Sections~\ref{sec:syst} and~\ref{sec:results}.

The extrapolation of the Run-2 results to the HL-LHC follows the same strategy used in previous projections of Higgs boson searches and measurements~\cite{Cepeda:2019klc}.
For the projections to a certain integrated luminosity $L$, the Run-2 signal and background yields, as well as the selected data events, are scaled up by a factor $k_L$ equal to the increase of integrated luminosity with respect to Run~2.
The scaling of the data events is necessary for those analyses that rely on the data to properly model the backgrounds in the fit.
The efficiency of the physics object reconstruction and identification is assumed to be the same as in Run~2.
The same assumption is made regarding the experimental energy or momentum resolution of the physics objects.
This is based on the premise that the upgraded CMS detector will ensure performance comparable to Run~2, despite the larger pileup and radiation damage to the detector components.

The exact nature and sizes of the systematic uncertainties in CMS during the HL-LHC are unknown.
Therefore, we derive the projections in three different scenarios of systematic uncertainties:
\begin{itemize}
    \item ``S1": The systematic uncertainties are assumed to be at the same level of Run~2.
    \item ``S2": The systematic uncertainties with a statistical origin, \eg statistical uncertainties in data/MC scale factors, are reduced by a factor $\sqrt{k_L}$, until ``floor" values are reached.
          The ``floor" values prevent uncertainties from becoming unreasonably small and are based on studies in Ref.~\cite{Cepeda:2019klc}.
          The theoretical uncertainties in the signal and background cross sections are halved to account for the expected progress in the theoretical calculations throughout the next years.
          The uncertainties originating from the limited size of the MC samples are also removed under the assumption of very large MC data sets.
    \item ``stat. only": No systematic uncertainties are considered in the fit.
\end{itemize}
It should be noted that the systematic uncertainties related to specific detector issues encountered in Run~2 have been removed from all three scenarios, including S1.
The nominal scenario for the HL-LHC conditions is S2.

The treatment of the uncertainties common to multiple analysis channels is summarised in Table~\ref{tab:CommonHLSyst}.
The analysis-specific uncertainties are treated case by case.

\begin{table*}[ht]
    \centering
    \topcaption{
        Treatment of most important common systematic uncertainties in the S2 scenario.
    }
    \label{tab:CommonHLSyst}
    \begin{tabular}{ c  c }
        \hline
        {Uncertainty}                                & {Scaling with respect to Run~2} \\
        \hline
        Theoretical uncertainties                    &
        $1/2$                                                                          \\
        Stat. uncertainties in MC simulation         &
        Removed                                                                        \\
        \PQb-tagging efficiency stat. component      &
        $1/\sqrt{k_L}$                                                                 \\
        \PQb-tagging efficiency (nonstat. component) &
        Unchanged                                                                      \\
        AK4 jet absolute energy scale                &
        $\max(0.3, 1/\sqrt{k_L})$                                                      \\
        AK4 jet energy scale dependence on flavour   &
        $\max(0.5, 1/\sqrt{k_L})$                                                      \\
        AK4 jet relative energy scale                &
        $\max(0.2, 1/\sqrt{k_L})$                                                      \\
        AK4 jet energy scale method                  &
        $1/\sqrt{k_L}$                                                                 \\
        AK4 jet energy resolution                    &
        $\max(0.5, 1/\sqrt{k_L})$                                                      \\
        \MET                                         &
        $\max(0.5, 1/\sqrt{k_L})$                                                      \\
        Luminosity                                   &
        0.6                                                                            \\
        $\tauh$ ID                                   &
        Unchanged                                                                      \\
        $\tauh$ energy scale                         &
        Unchanged                                                                      \\
        Pileup                                       &
        Unchanged                                                                      \\
        Run-2 issues                                 &
        Removed                                                                        \\
        \hline
    \end{tabular}
\end{table*}

\begin{figure}[ht]
    \centering
    \includegraphics[width=.45\textwidth]{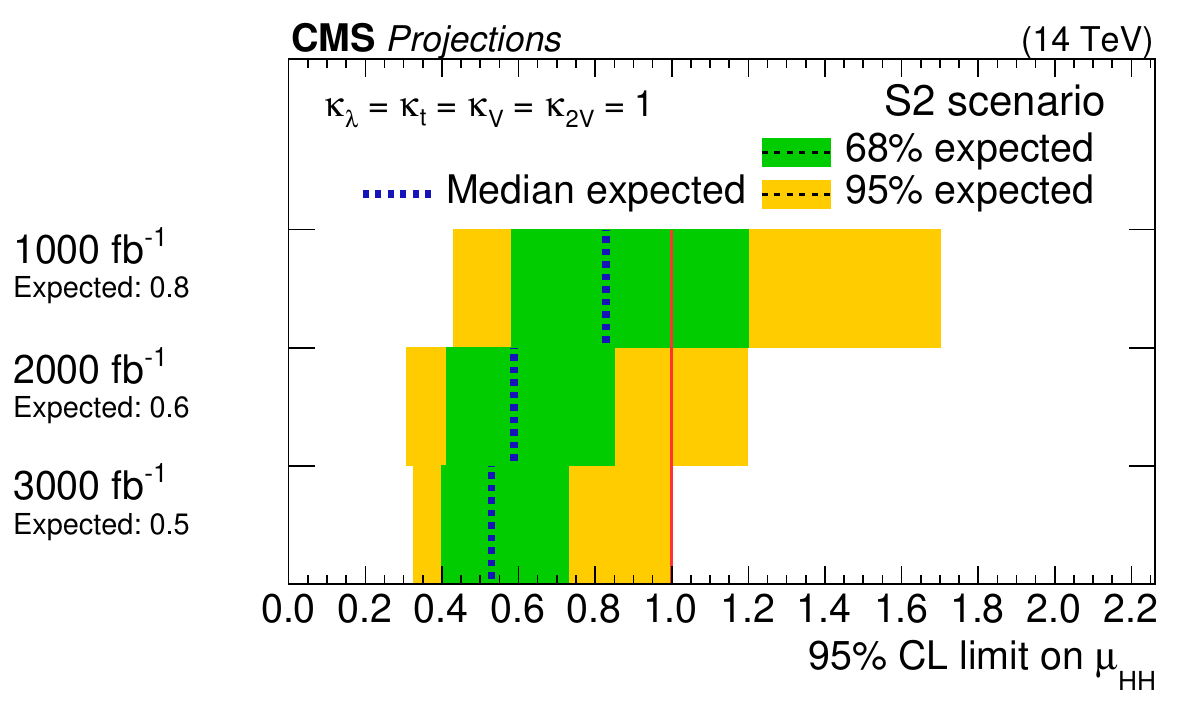}
    \includegraphics[width=.45\textwidth]{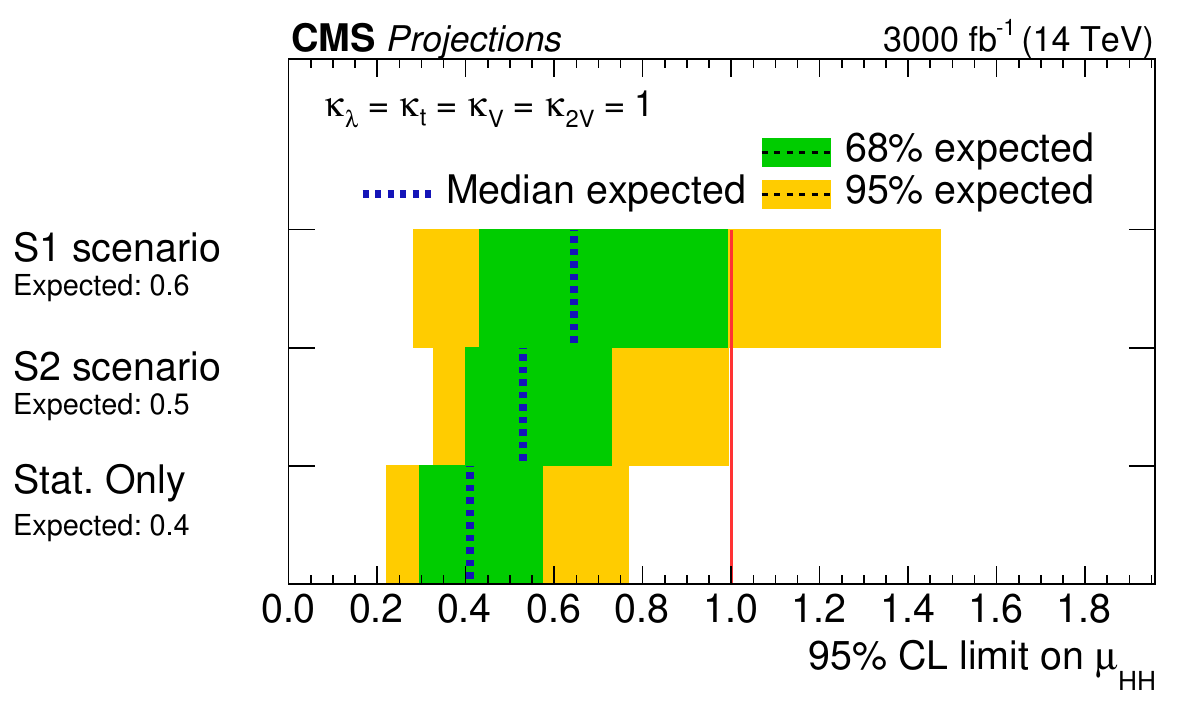}
    \caption{
        The expected upper limits at 95\% \CL on the \HH signal strength from the combination of all the considered channels projected to different integrated luminosities (\cmsLeft), and under different assumptions on the systematic uncertainties for an integrated luminosity of 3000\fbinv (\cmsRight).
    }
    \label{fig:HHcomb_projected_limits}
\end{figure}

The projected upper limits on the \HH signal strength for the statistical combination of the considered \HH channels at different integrated luminosities are shown in the left panel of Fig.~\ref{fig:HHcomb_projected_limits}.
We will become sensitive to SM \HH production at a 95\% CL for integrated luminosities larger than 1000\fbinv, when the expected upper limit drops below one.
The projections at 3000\fbinv under the different systematic uncertainty scenarios are shown in the right panel of Fig.~\ref{fig:HHcomb_projected_limits}.

\begin{figure*}[ht]
    \centering
    \includegraphics[width=.45\textwidth]{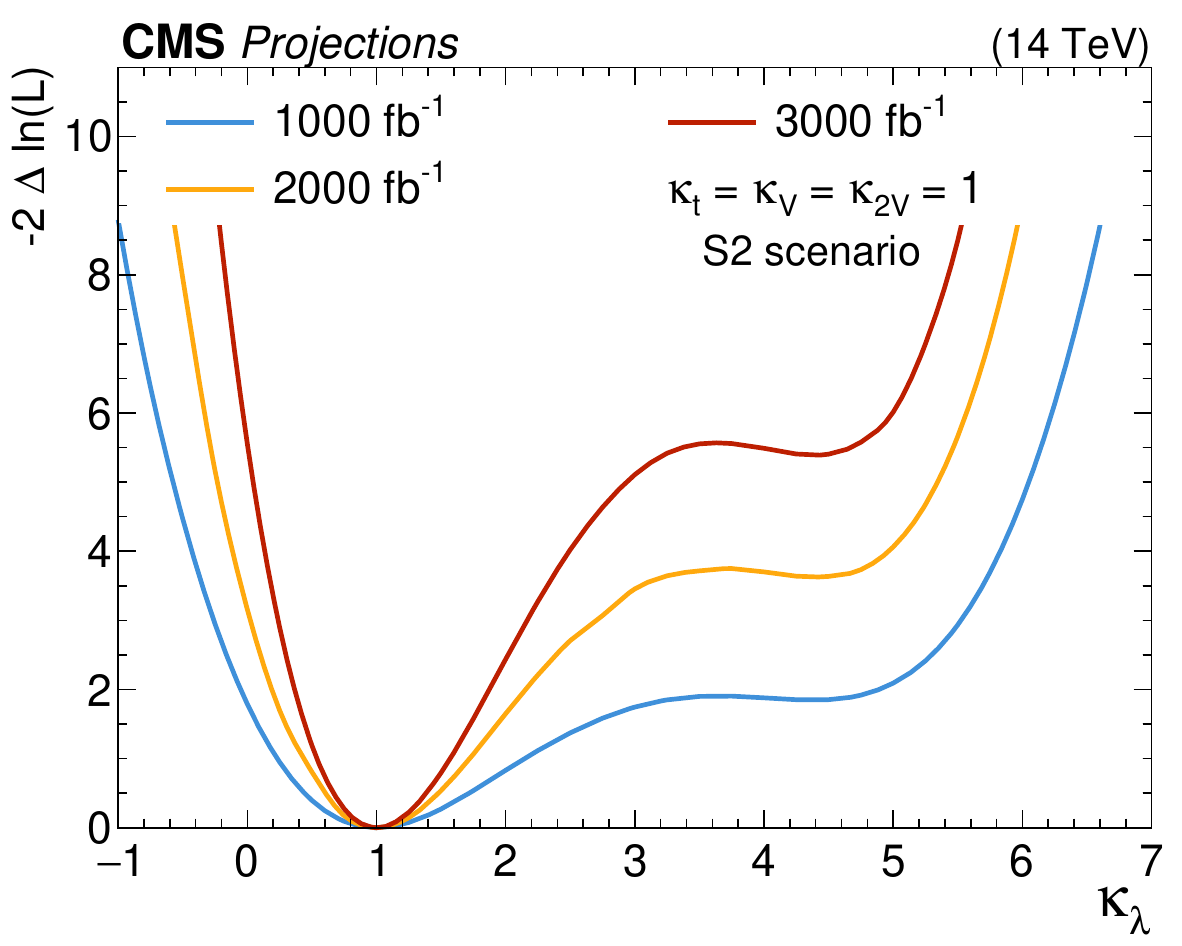}
    \includegraphics[width=.45\textwidth]{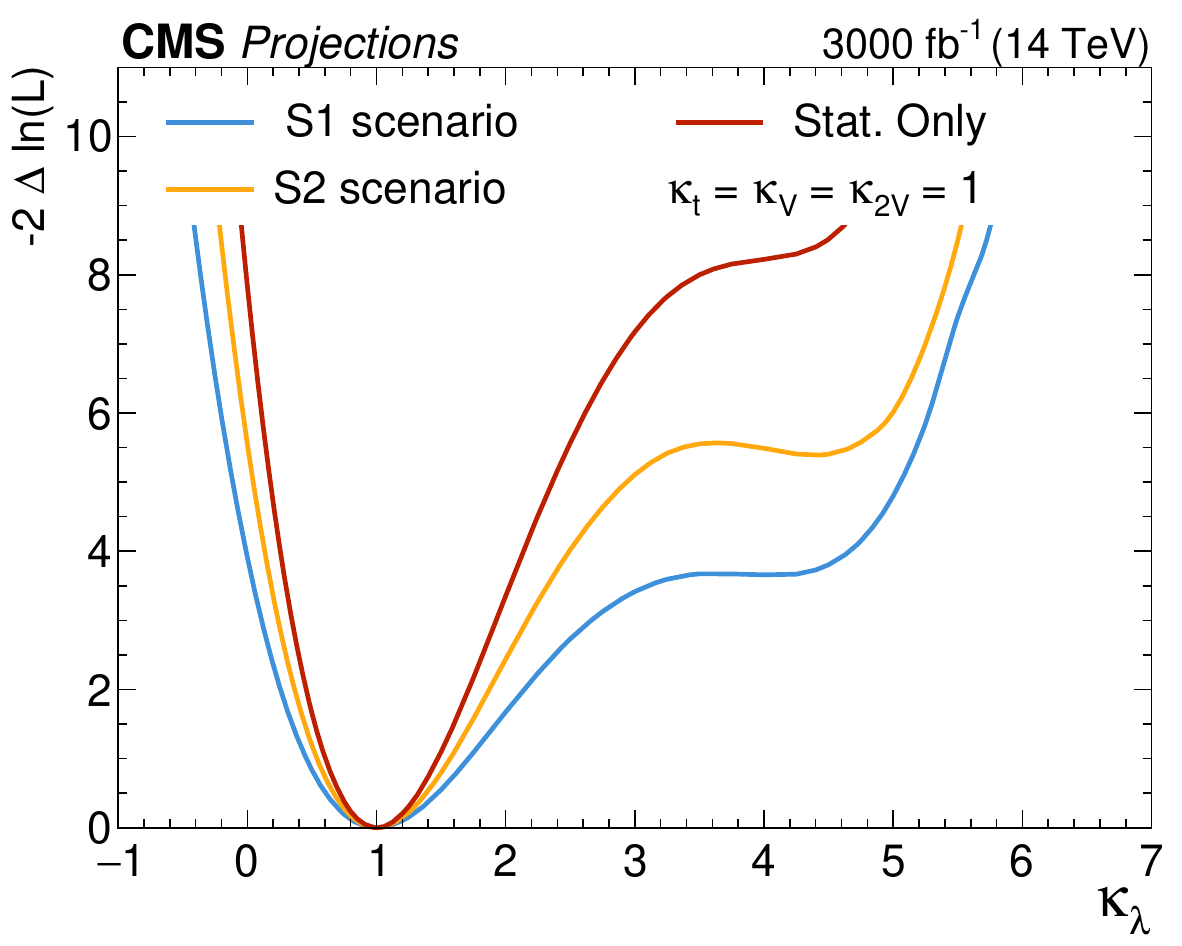}
    \caption{
        The expected $-2\Delta\ln(L)$ scan as a function of coupling modifier \klambda for the combination of all contributing channels, projected to different integrated luminosities (left), and under different assumptions on the systematic uncertainties for an integrated luminosity of 3000\fbinv (right).
        All other parameters are fixed to their SM values.
    }
    \label{fig:HHcomb_projected_klscans}
\end{figure*}

The projected \klambda likelihood scans are shown in Fig.~\ref{fig:HHcomb_projected_klscans}.
In the S2 scenario, the expected 1 standard deviation uncertainty on \klambda is $+80\%$/$-60\%$ and $+60\%$/$-50\%$ for an integrated luminosity of 2000 and 3000\fbinv, respectively.

\begin{figure*}[ht]
    \centering
    \includegraphics[width=.45\textwidth]{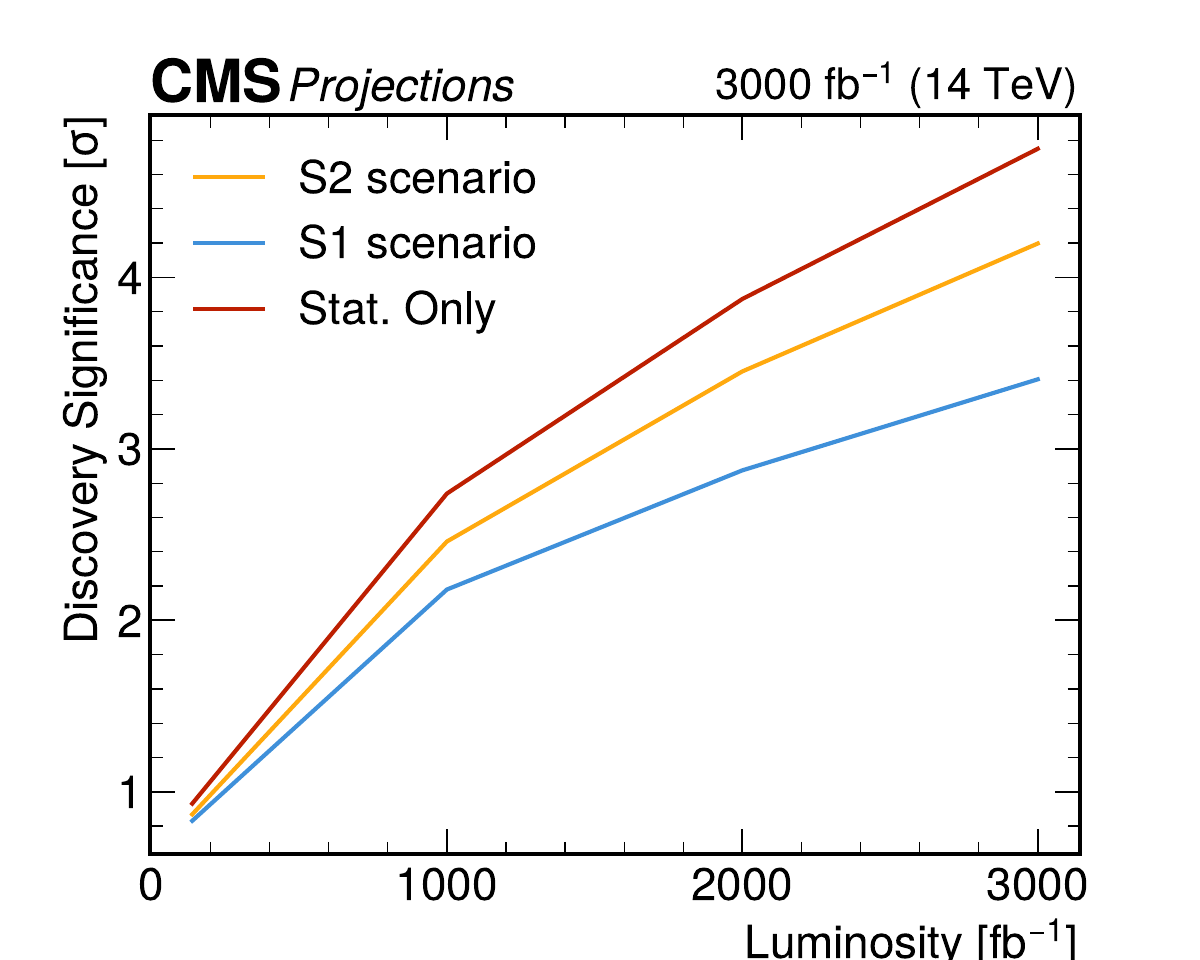}
    \includegraphics[width=.45\textwidth]{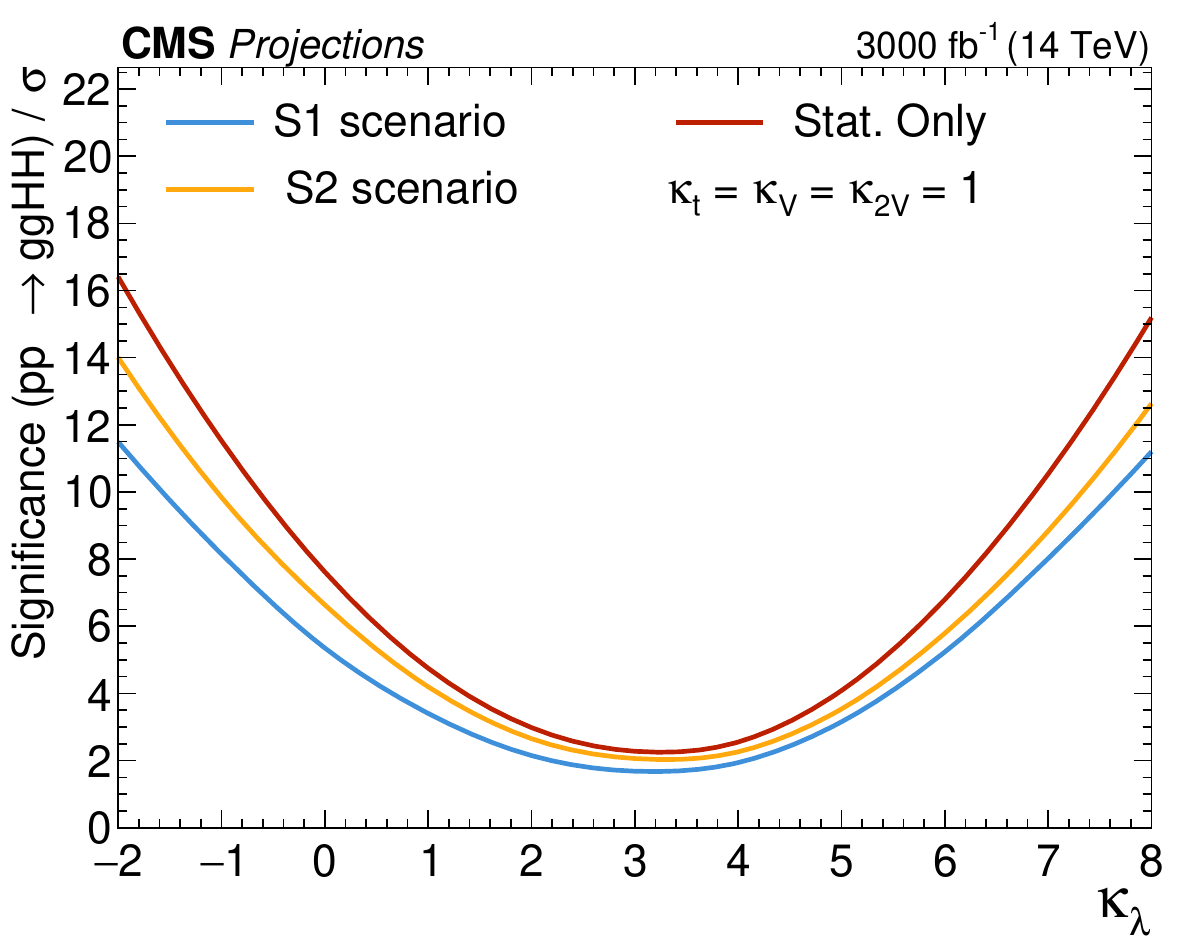}
    \caption{
        The expected signal significance as a function of integrated luminosity under different assumptions on the systematic uncertainties (left).
        The expected signal significance as a function of \klambda under different assumptions on the systematic uncertainties for an integrated luminosity of 3000\fbinv (right).
    }
    \label{fig:significanceVSkl}
\end{figure*}

The expected significance for the \HH signal strength assuming SM values for all parameters is summarised in Table~\ref{tab:HHcomb_significance}.
Before the end of the HL-LHC, the CMS experiment will have evidence for \HH production with a significance of 3.5 standard deviations for 2000\fbinv in the nominal systematic uncertainty scenario and assuming that the signal is SM-like.
Figure~\ref{fig:significanceVSkl} on the left shows the significance that can be achieved for the SM signal as a function of the integrated luminosity.
On the right, the significance is shown as a function of \klambda.
The sensitivity to the \HH signal varies with \klambda due to the effects of the interference between the box and triangle diagrams, discussed in Section~\ref{sec:intro}.
The two diagrams have different kinematic properties, therefore, the interference varies not only the \HH cross section but the \HH signal acceptance rate as well.
In particular, for $\klambda<1$ and $\klambda>5$, the destructive interference is reduced, leading to an enhancement of the cross section relative to the SM expectation.
In these regimes, evidence for the \HH process could be established with smaller data sets compared to the SM case.
The \HH signal is suppressed for $1<\klambda<5$, and the significance will be lower than expected, with minimum significance at $\klambda=3.4$.
The projections presented here do not take into account potential improvements in triggering~\cite{CERN-LHCC-2020-004,CERN-LHCC-2021-007}, object reconstruction and selection such as \PQb tagging, or analysis techniques.
Historically, advancements in software and analysis methods have allowed us to exceed expectations, therefore the projections shown in this paper are most likely conservative.
Results from Run~3 will also add valuable information regarding the potential improvements for measuring the \HH process.
If the trend of improvement continues, an observation of \HH production by the end of the HL-LHC is within reach.

\begin{table*}[ht]
    \centering
    \topcaption{
        Expected significance for the \HH signal projected to 2000 or 3000\fbinv under different assumptions of systematic uncertainties.
    }
    \label{tab:HHcomb_significance}
    \begin{tabular}{ l  c  c  c  c }
        \hline
                            & \multicolumn{2}{ c}{Significance ($\sigma$) at 2000\fbinv}
                            & \multicolumn{2}{ c }{Significance ($\sigma$) at 3000\fbinv}                                   \\
                            & S2                                                          & Stat. only & S2    & Stat. only \\
        \hline
        \bbbb resolved jets & 1.0                                                         & 1.4        & 1.2   & 1.7        \\
        \bbbb merged jets   & 1.8                                                         & 1.9        & 2.2   & 2.3        \\
        \bbtt               & 1.9                                                         & 2.1        & 2.4   & 2.6        \\
        \bbgg               & 2.0                                                         & 2.0        & 2.4   & 2.5        \\
        \bbWW               & 0.3                                                         & 0.8        & 0.3   & 1.0        \\
        \HH multilepton     & 0.4                                                         & 0.6        & 0.5   & 0.8        \\
        Combination         & {3.5}                                                       & {3.9}      & {4.2} & {4.8}      \\
        \hline
    \end{tabular}
\end{table*}

\clearpage

\section{Summary}
\label{sec:summary}
A combined search for nonresonant Higgs boson pair (\HH) production was performed using the proton-proton collision data set produced by the LHC at $\sqrt{s} = 13\TeV$, and collected by the CMS experiment from 2016 to 2018 (Run~2), which corresponds to an integrated luminosity of 138\fbinv.
Searches for \HH production via gluon-gluon (\GGF) and vector boson fusion (\VBF) production, were carried out in the \bbgg, \bbtt, \bbbb, \bbWW, and multilepton channels.
Additionally, searches for \GGF \HH production were conducted in the \bbZZ (with both $\PZ$ bosons decaying to leptons), \WWgg, and \ttgg final states, which have clean signatures but relatively small branching fractions.
We searched for the associated production mechanism with a vector boson in the \bbbb final state, which has the largest branching fraction.
The analyses of these channels were combined to probe the Higgs boson trilinear self-coupling and the quartic coupling between two vector bosons and two Higgs bosons (\VVHH), and to search for beyond the standard model (SM) physics scenarios in the Higgs effective field theory (HEFT) approach.

The observed and expected upper limits at 95\% confidence level (\CL) on the cross section of \GGF \HH production were found to be 3.5 and 2.5 times the SM expectation, respectively.
For \VBF production, the observed and expected 95\% \CL upper limits are 79 and 91 times the SM expectation, respectively.
When all other parameters are set to their SM values, we constrain the Higgs boson trilinear self-coupling modifier \klambda in the range from $-1.35$ to 6.37 at 95\% \CL (expected 95\% \CL range is $-2.24$ to 7.89).
Likewise, the \VVHH coupling modifier \CVV is constrained in the range from 0.64 to 1.40 (0.62 to 1.41 expected).

Two-dimensional measurements were also performed, including simultaneous measurements of \klambda and \CVV, \klambda and the modifier of the Higgs boson coupling to the top quark (\ktop), and \CVV and the modifier of the Higgs boson coupling to vector bosons (\CV).
The results are in agreement with the SM predictions.

Under the HEFT framework, the cross section of the nonresonant \GGF \HH pair production was parametrized as a function of anomalous couplings of the Higgs boson, involving the contact interactions between two Higgs and two top quarks, between two gluons and two Higgs bosons, and between two gluons and a Higgs boson.
We performed searches for benchmark signals under different anomalous coupling scenarios and set upper limits on their cross sections at 95\% \CL.
We exclude \HH production at 95\% \CL when the coupling modifier of the contact interaction between two Higgs bosons and two top quarks is outside the range from $-0.28$ to 0.59 (expected 95\% \CL range is $-0.17$ to 0.47).
The HEFT parametrisation is also exploited to study various ultraviolet complete models with an extended Higgs sector and set constraints on specific parameters.

These results constitute the most stringent limits and constraints obtained from the searches for nonresonant \HH production using the LHC Run-2 data set collected by the CMS experiment.
Extrapolating our current results to the integrated luminosity anticipated of the High-Luminosity LHC, it is expected to see first evidence for \HH production with $\approx$2000\fbinv of data.

\begin{acknowledgments}
    We congratulate our colleagues in the CERN accelerator departments for the excellent performance of the LHC and thank the technical and administrative staffs at CERN and at other CMS institutes for their contributions to the success of the CMS effort. In addition, we gratefully acknowledge the computing centres and personnel of the Worldwide LHC Computing Grid and other centres for delivering so effectively the computing infrastructure essential to our analyses. Finally, we acknowledge the enduring support for the construction and operation of the LHC, the CMS detector, and the supporting computing infrastructure provided by the following funding agencies: SC (Armenia), BMBWF and FWF (Austria); FNRS and FWO (Belgium); CNPq, CAPES, FAPERJ, FAPERGS, and FAPESP (Brazil); MES and BNSF (Bulgaria); CERN; CAS, MoST, and NSFC (China); MINCIENCIAS (Colombia); MSES and CSF (Croatia); RIF (Cyprus); SENESCYT (Ecuador); ERC PRG, TARISTU24-TK10 and MoER TK202 (Estonia); Academy of Finland, MEC, and HIP (Finland); CEA and CNRS/IN2P3 (France); SRNSF (Georgia); BMFTR, DFG, and HGF (Germany); GSRI (Greece); NKFIH (Hungary); DAE and DST (India); IPM (Iran); SFI (Ireland); INFN (Italy); MSIT and NRF (Republic of Korea); MES (Latvia); LMTLT (Lithuania); MOE and UM (Malaysia); BUAP, CINVESTAV, CONACYT, LNS, SEP, and UASLP-FAI (Mexico); MOS (Montenegro); MBIE (New Zealand); PAEC (Pakistan); MES, NSC, and NAWA (Poland); FCT (Portugal); MESTD (Serbia); MICIU/AEI and PCTI (Spain); MOSTR (Sri Lanka); Swiss Funding Agencies (Switzerland); MST (Taipei); MHESI (Thailand); TUBITAK and TENMAK (T\"{u}rkiye); NASU (Ukraine); STFC (United Kingdom); DOE and NSF (USA).

    \hyphenation{Rachada-pisek} Individuals have received support from the Marie-Curie programme and the European Research Council and Horizon 2020 Grant, contract Nos.\ 675440, 724704, 752730, 758316, 765710, 824093, 101115353, 101002207, 101001205, and COST Action CA16108 (European Union); the Leventis Foundation; the Alfred P.\ Sloan Foundation; the Alexander von Humboldt Foundation; the Science Committee, project no. 22rl-037 (Armenia); the Fonds pour la Formation \`a la Recherche dans l'Industrie et dans l'Agriculture (FRIA) and Fonds voor Wetenschappelijk Onderzoek contract No. 1228724N (Belgium); the Beijing Municipal Science \& Technology Commission, No. Z191100007219010, the Fundamental Research Funds for the Central Universities, the Ministry of Science and Technology of China under Grant No. 2023YFA1605804, the Natural Science Foundation of China under Grant No. 12061141002 and USTC Research Funds of the Double First-Class Initiative No.\ YD2030002017 (China); the Ministry of Education, Youth and Sports (MEYS) of the Czech Republic; the Shota Rustaveli National Science Foundation, grant FR-22-985 (Georgia); the Deutsche Forschungsgemeinschaft (DFG), among others, under Germany's Excellence Strategy -- EXC 2121 ``Quantum Universe" -- 390833306, and under project number 400140256 - GRK2497; the Hellenic Foundation for Research and Innovation (HFRI), Project Number 2288 (Greece); the Hungarian Academy of Sciences, the New National Excellence Program - \'UNKP, the NKFIH research grants K 131991, K 133046, K 138136, K 143460, K 143477, K 146913, K 146914, K 147048, 2020-2.2.1-ED-2021-00181, TKP2021-NKTA-64, and 2021-4.1.2-NEMZ\_KI-2024-00036 (Hungary); the Council of Science and Industrial Research, India; ICSC -- National Research Centre for High Performance Computing, Big Data and Quantum Computing, FAIR -- Future Artificial Intelligence Research, and CUP I53D23001070006 (Mission 4 Component 1), funded by the NextGenerationEU program (Italy); the Latvian Council of Science; the Ministry of Education and Science, project no. 2022/WK/14, and the National Science Center, contracts Opus 2021/41/B/ST2/01369, 2021/43/B/ST2/01552, 2023/49/B/ST2/03273, and the NAWA contract BPN/PPO/2021/1/00011 (Poland); the Funda\c{c}\~ao para a Ci\^encia e a Tecnologia, grant CEECIND/01334/2018 (Portugal); the National Priorities Research Program by Qatar National Research Fund; MICIU/AEI/10.13039/501100011033, ERDF/EU, "European Union NextGenerationEU/PRTR", and Programa Severo Ochoa del Principado de Asturias (Spain); the Chulalongkorn Academic into Its 2nd Century Project Advancement Project, the National Science, Research and Innovation Fund program IND\_FF\_68\_369\_2300\_097, and the Program Management Unit for Human Resources \& Institutional Development, Research and Innovation, grant B39G680009 (Thailand); the Kavli Foundation; the Nvidia Corporation; the SuperMicro Corporation; the Welch Foundation, contract C-1845; and the Weston Havens Foundation (USA).

\end{acknowledgments}\section*{Data availability} Release and preservation of data used by the CMS Collaboration as the basis for publications is guided by the  \href{https://doi.org/10.7483/OPENDATA.CMS.1BNU.8V1W}{CMS data preservation, re-use and open access policy}.
\bibliography{auto_generated}

\ifthenelse{\boolean{cms@external}}{}{
    \clearpage
    \numberwithin{table}{section}
    \numberwithin{figure}{section}
    \appendix
    \section{Supplementary material}
    \label{app:suppMat}
    \begin{figure*}[tbh!]
  \centering
  \includegraphics[width=0.45\textwidth]{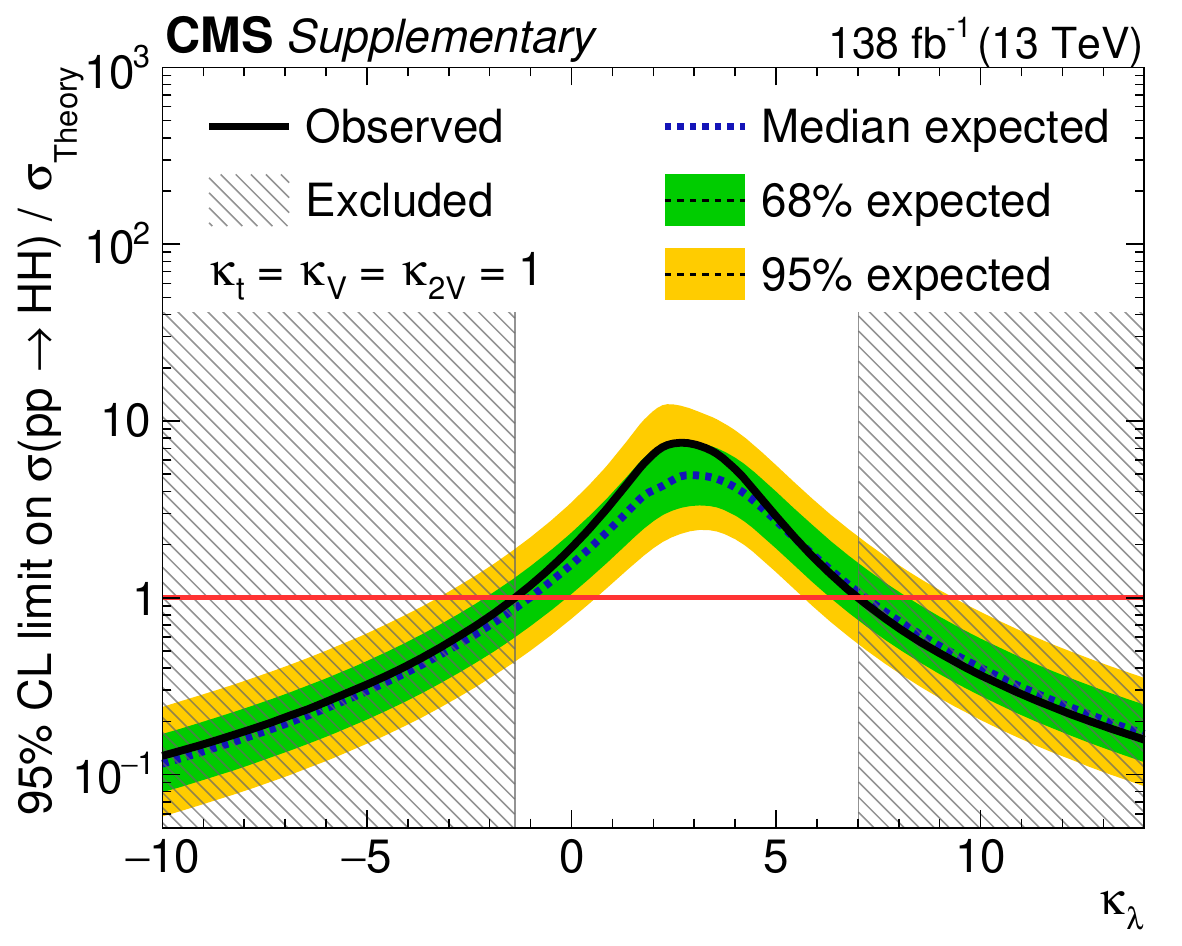}
  \includegraphics[width=0.45\textwidth]{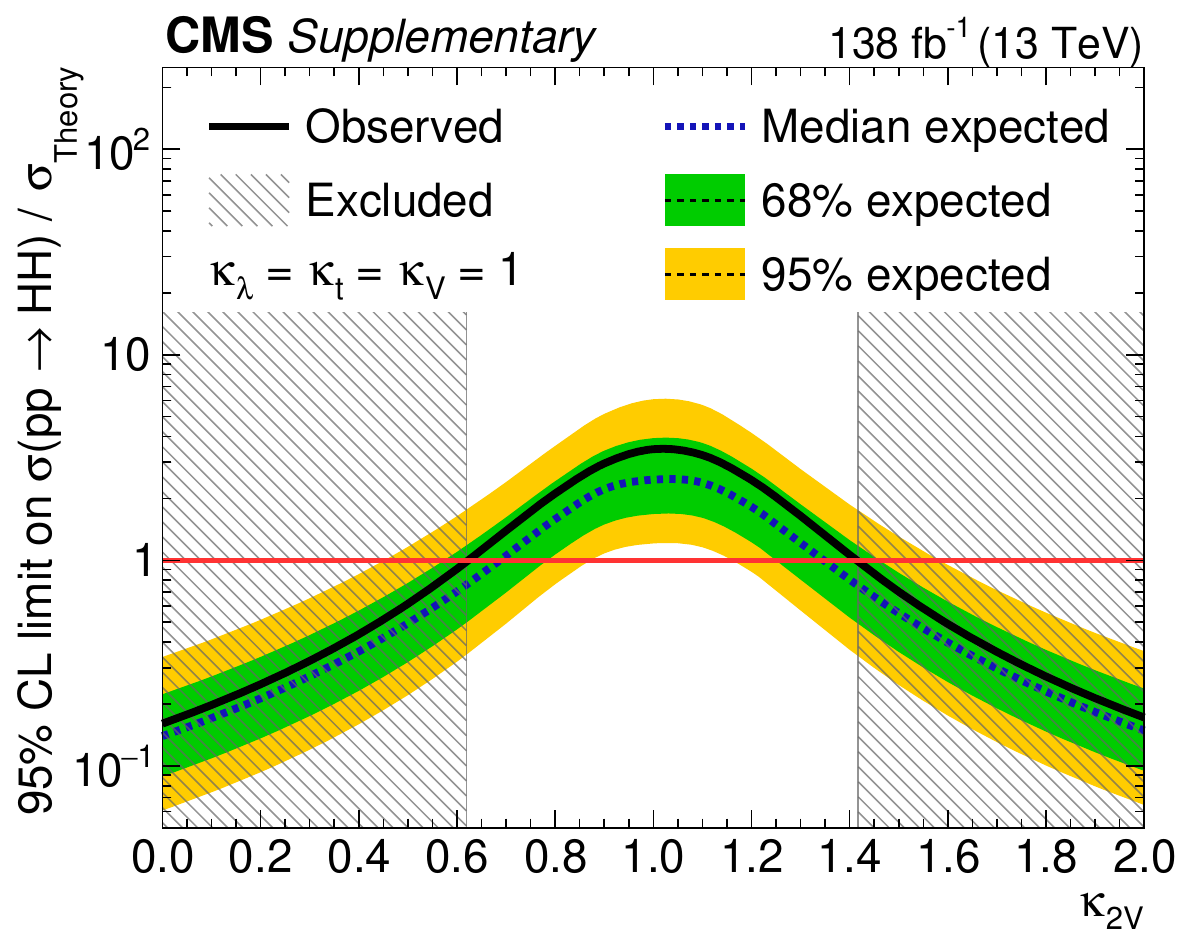}
  \caption{
    The 95\% \CL upper limits on the inclusive \HH signal strength as functions of \klambda (left) and \CVV (right).
    All other couplings are set to the values predicted by the SM.
    The theoretical uncertainties in the \HH \GGF and \VBF signal cross sections are considered in this case.
    The inner (green) band and the outer (yellow) band indicate the 68 and 95\% \CL intervals, respectively, under the background-only hypothesis.
  }
  \label{fig:limits_r_kl_cvv}
\end{figure*}

\begin{figure*}[tbh!]
  \centering
  \includegraphics[width=0.45\textwidth]{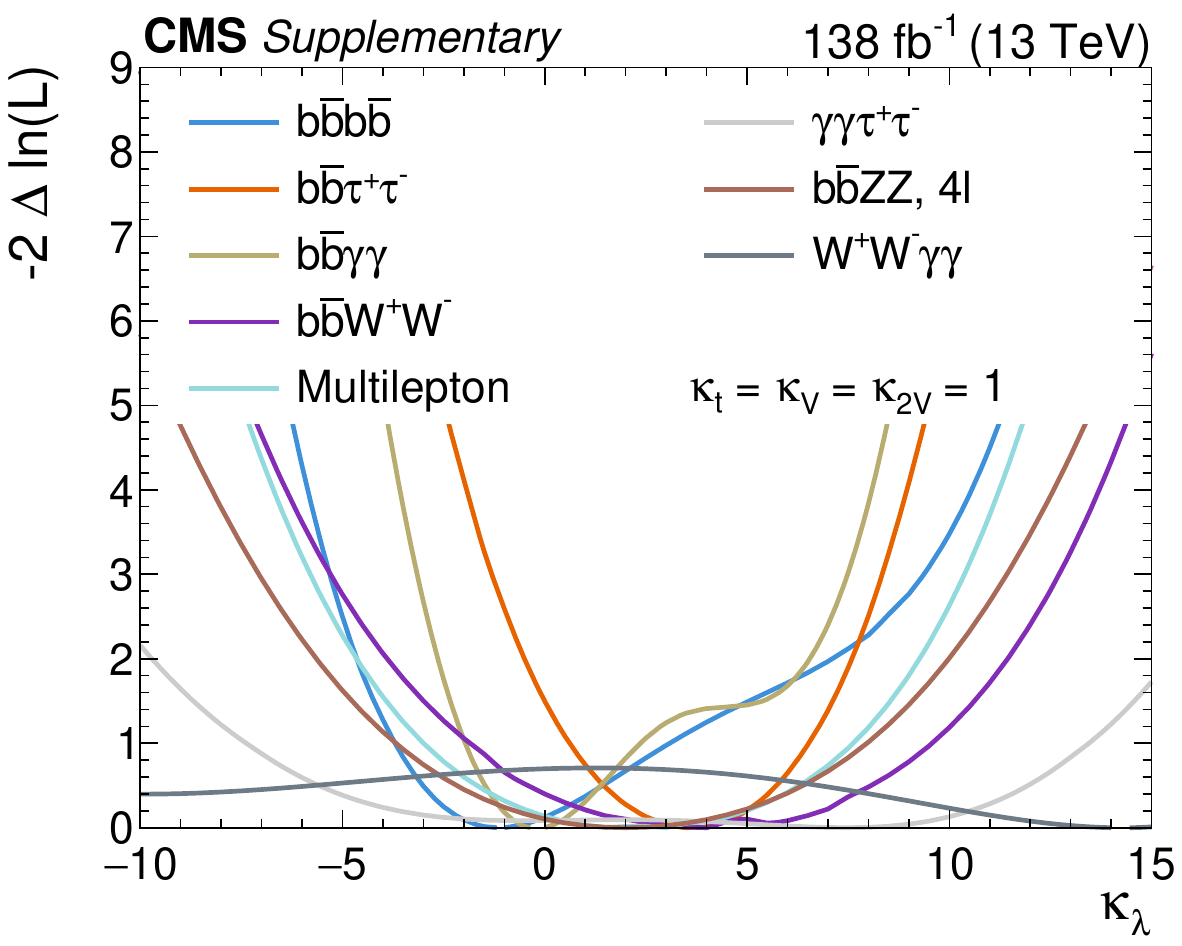}
  \includegraphics[width=0.45\textwidth]{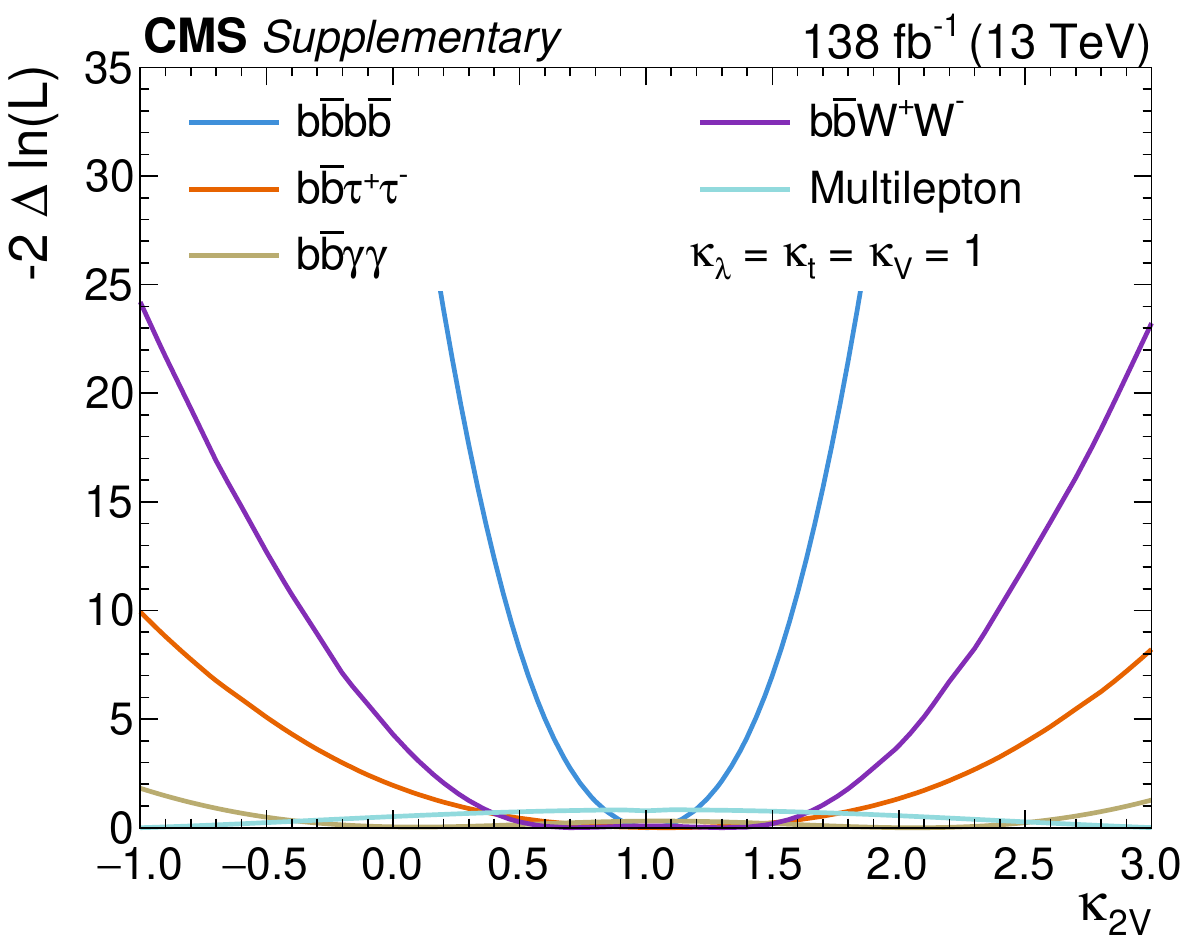}
  \caption{
    The $-2\Delta\ln(L)$ scan as functions of \klambda (left) and \CVV (right) for all channels, when all the other parameters are fixed to their SM value.
  }
  \label{fig:likelihood_kl_cvv_channels_linear}
\end{figure*}

\begin{figure}[tbh!]
  \centering
  \includegraphics[width=0.45\textwidth]{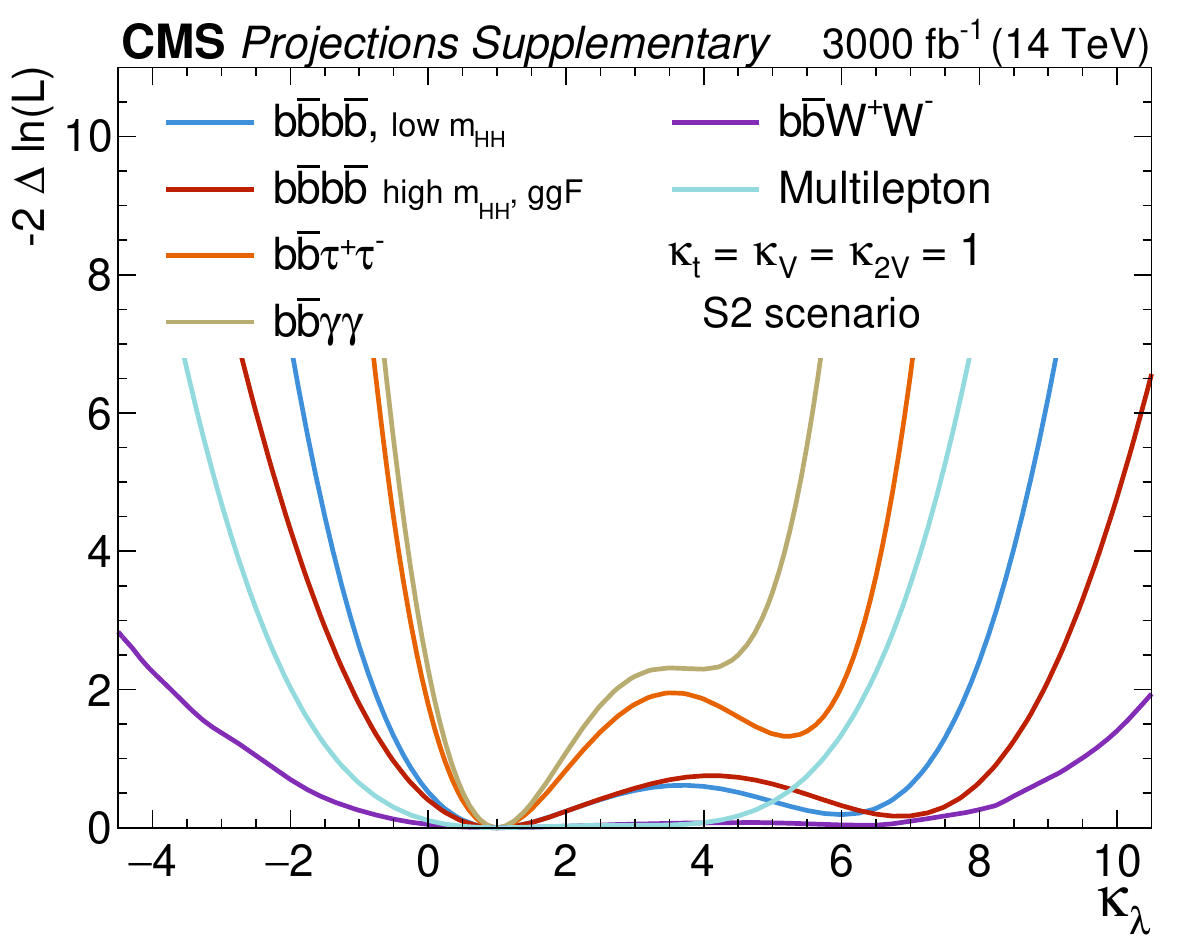}
  \caption{
    The expected $-2\Delta\ln(L)$ scan as a function of coupling modifier \klambda for all channels and an integrated luminosity of 3000\fbinv.
    All the other parameters are fixed to their SM values.
  }
  \label{fig:likelihood_kl_channels_linear_projections}
\end{figure}

\begin{figure*}[tbh!]
  \centering
  \includegraphics[width=0.45\textwidth]{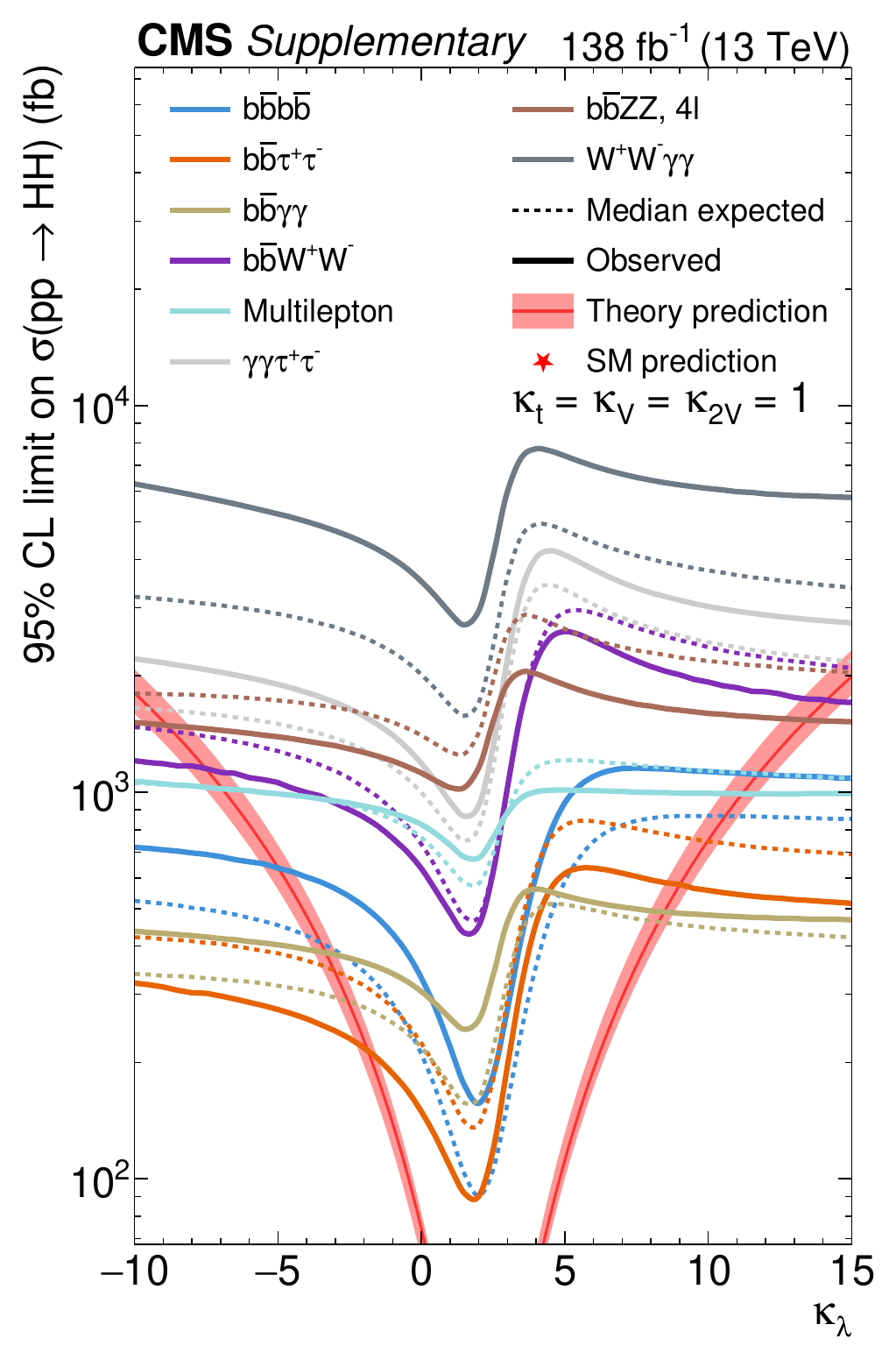}
  \includegraphics[width=0.45\textwidth]{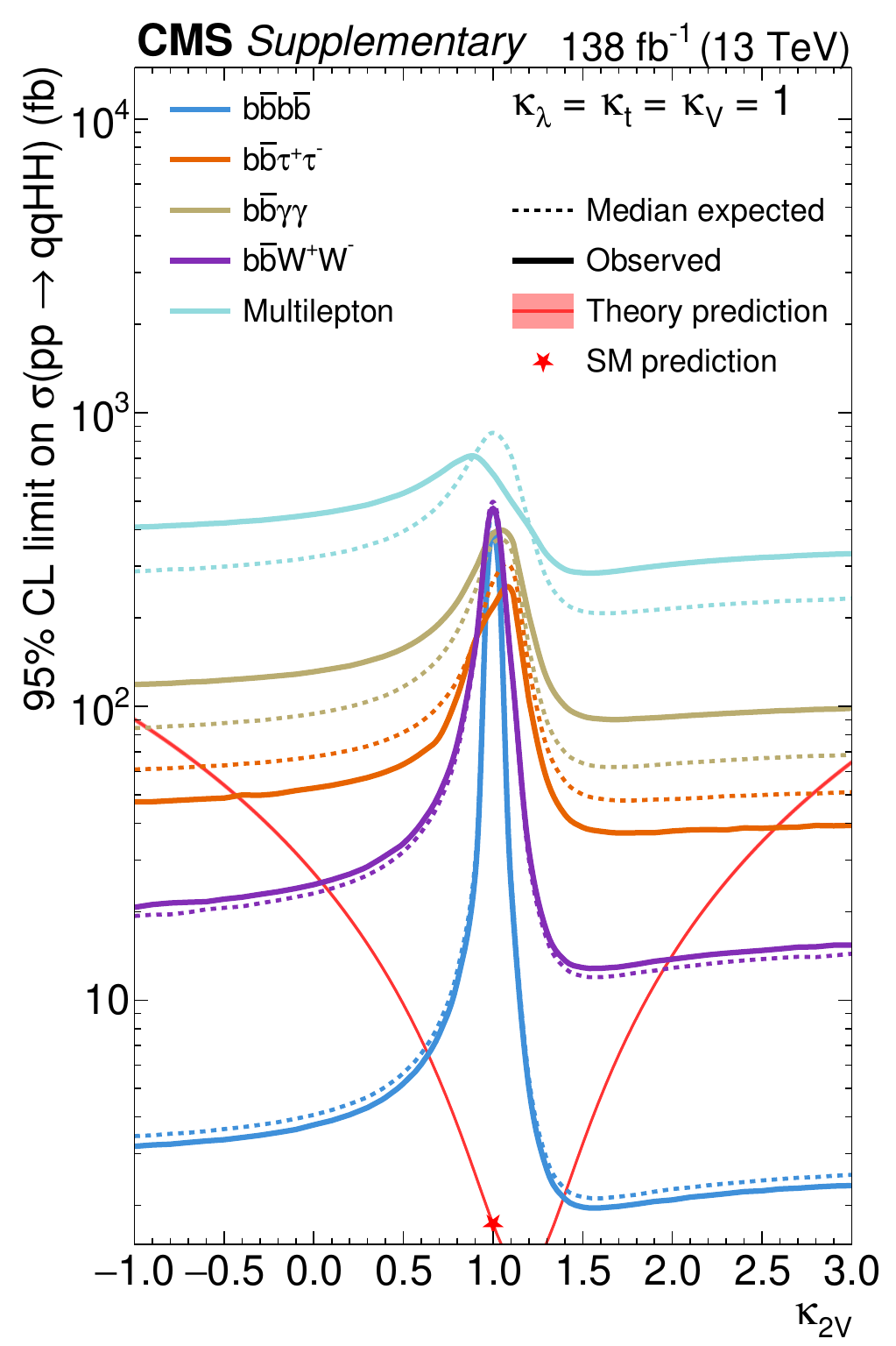}
  \caption{
    The 95\% \CL upper limits on the inclusive (left) and \VBF (right) \HH cross section as functions of \klambda and \CVV, respectively, for all channels.
    All other couplings are set to the values predicted by the SM.
    The theoretical uncertainties in the \HH \GGF and \VBF signal cross sections are not considered because we directly constrain the measured cross section.
  }
  \label{fig:limits_xs_kl_cvv_channels_linear}
\end{figure*}

\begin{figure*}[tbh!]
  \centering
  \includegraphics[width=0.45\textwidth]{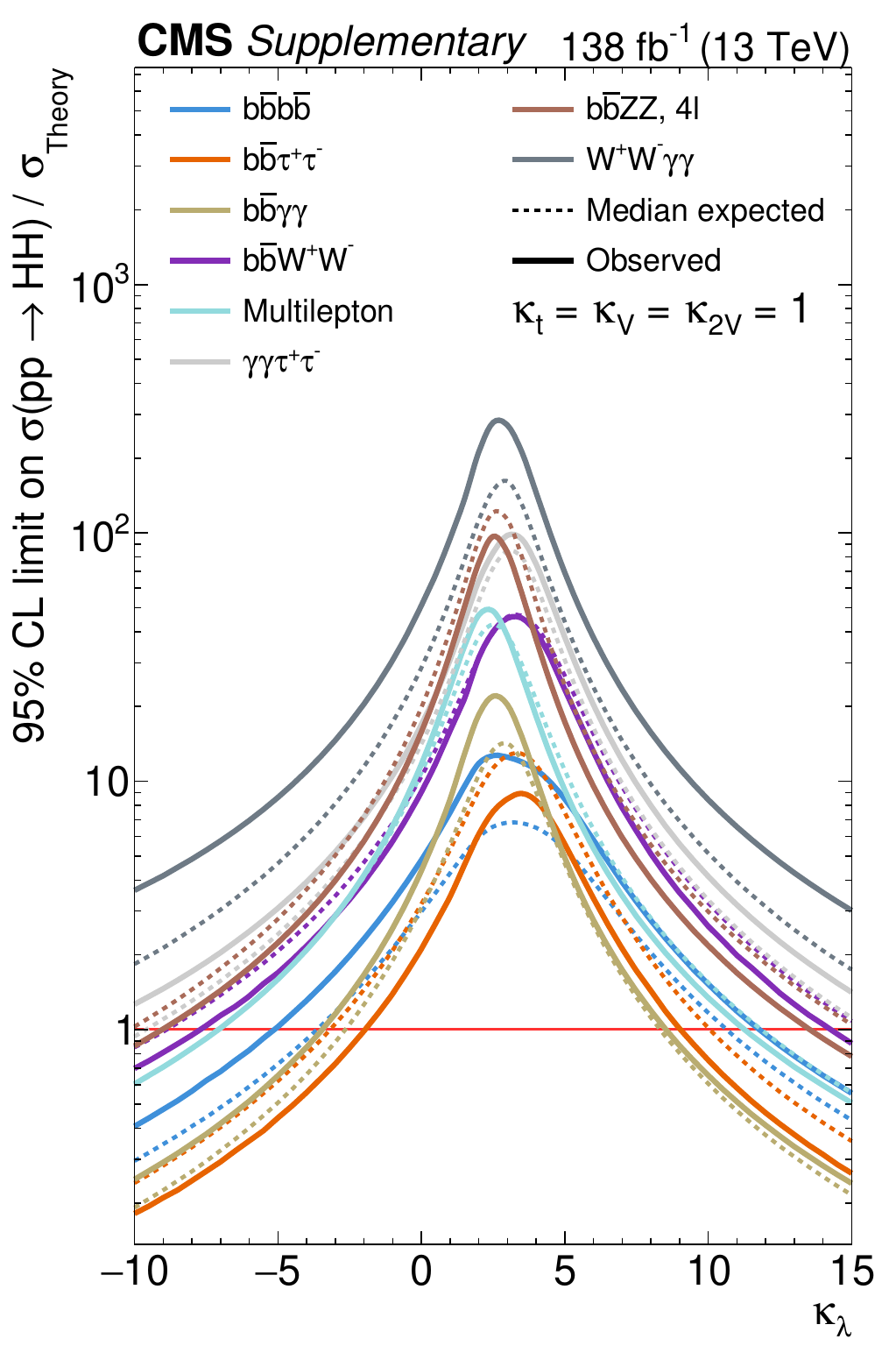}
  \includegraphics[width=0.45\textwidth]{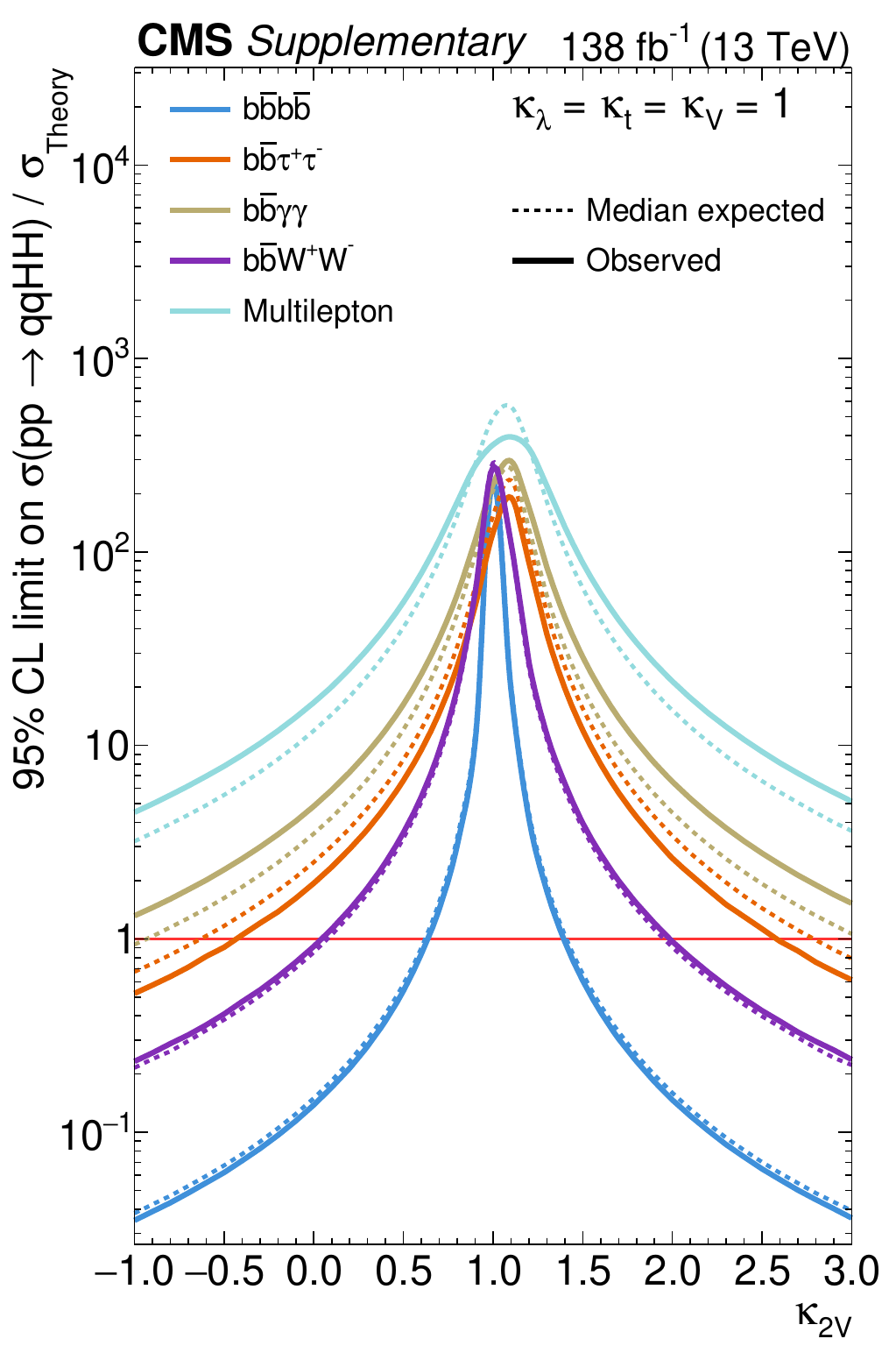}
  \caption{
    The 95\% \CL upper limits on the inclusive (left) and \VBF (right) \HH signal strength as functions of \klambda and \CVV, respectively, for all channels.
    All other couplings are set to the values predicted by the SM. The theoretical uncertainties in the \HH \GGF and \VBF signal cross sections are considered in this case.
  }
  \label{fig:limits_r_kl_cvv_channels_linear}
\end{figure*}

\begin{figure*}[tbh!]
  \centering
  \includegraphics[width=0.45\textwidth]{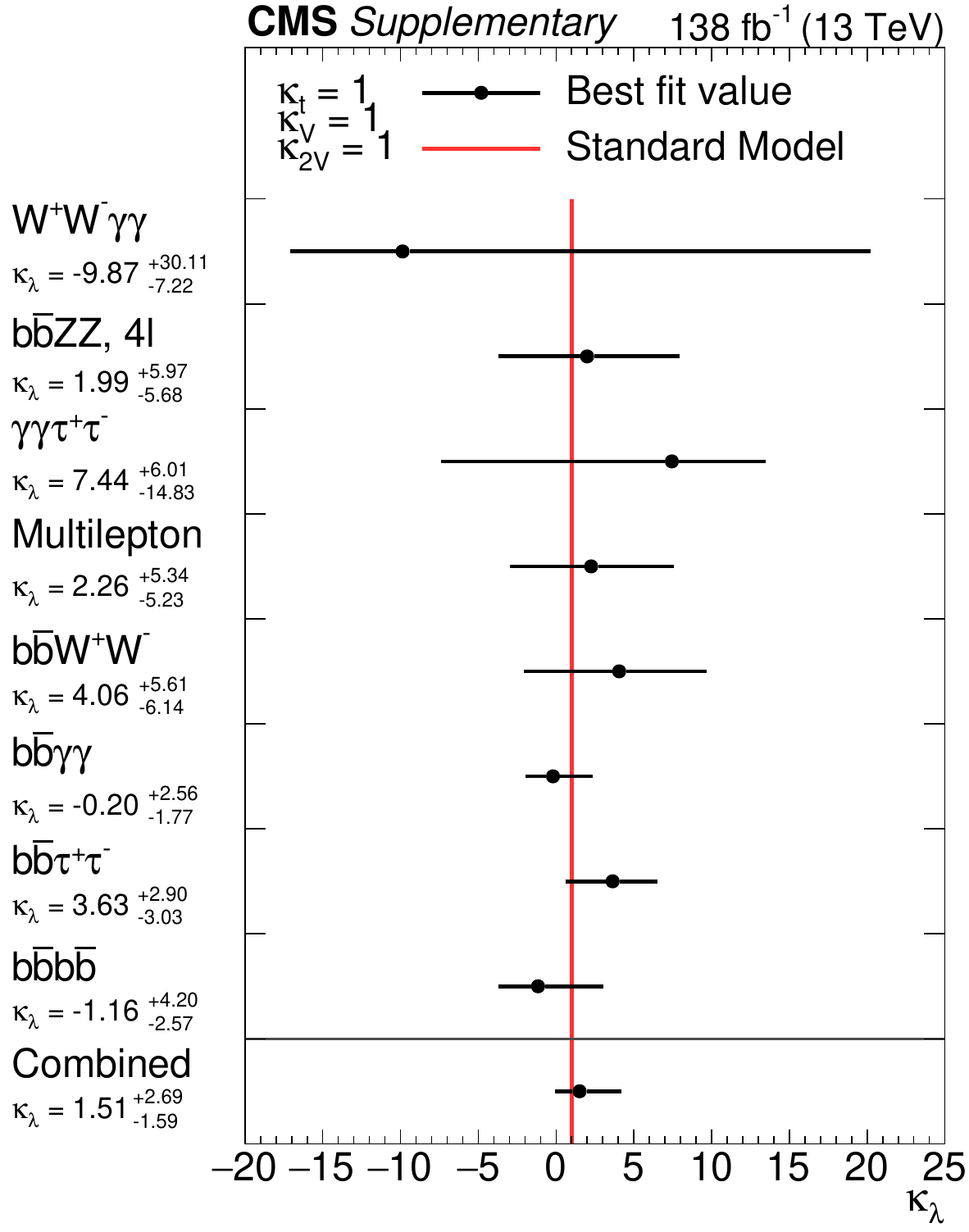}
  \includegraphics[width=0.45\textwidth]{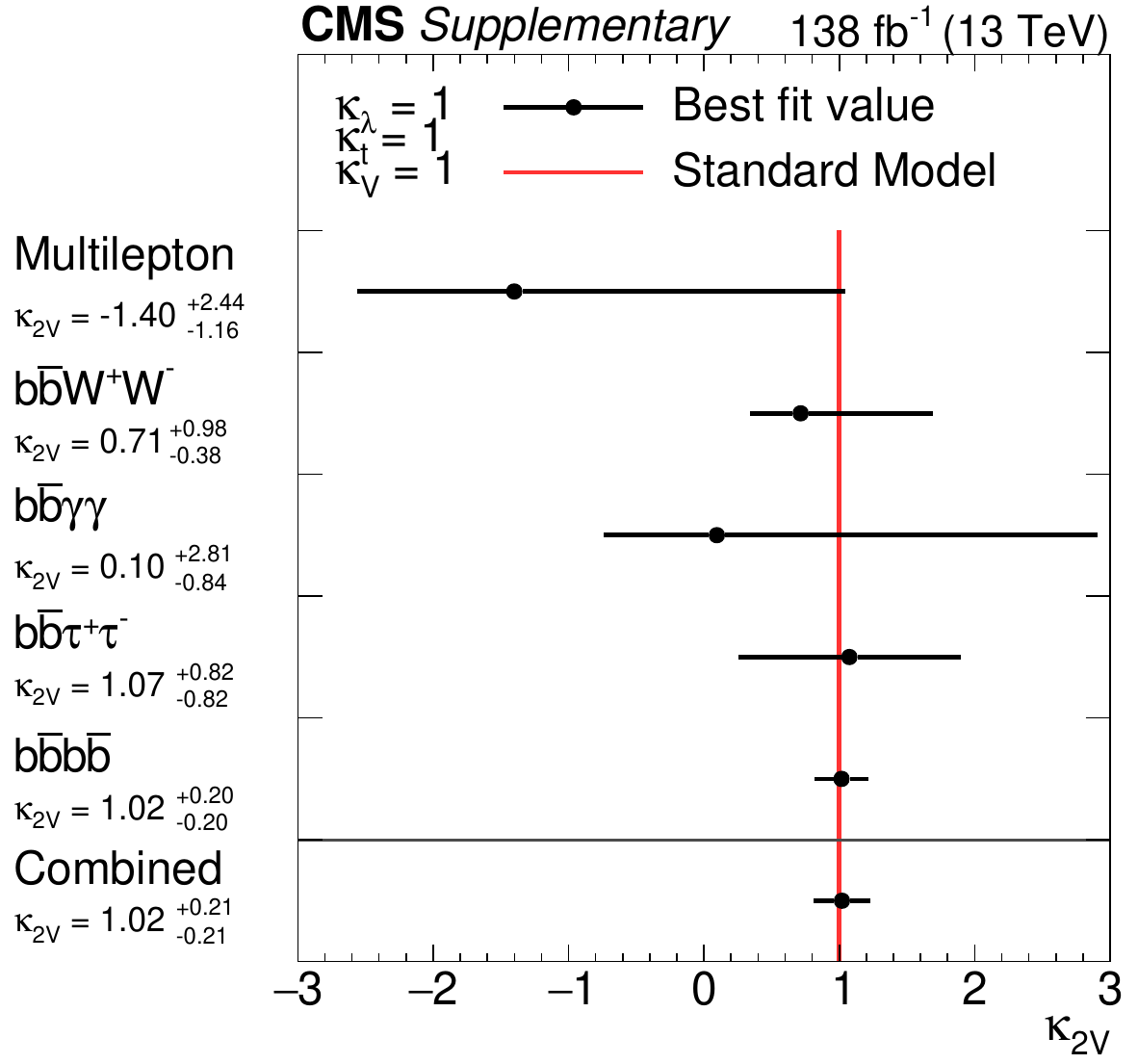}
  \caption{
    The best fit value for \klambda (left) and \CVV (right) compared to the SM expectation for all channels and their combination, when all the other parameters are fixed to their SM values.
  }
  \label{fig:bestfit_kl_cvv_channels_linear}
\end{figure*}

\begin{figure}[tbh!]
  \centering
  \includegraphics[width=0.45\textwidth]{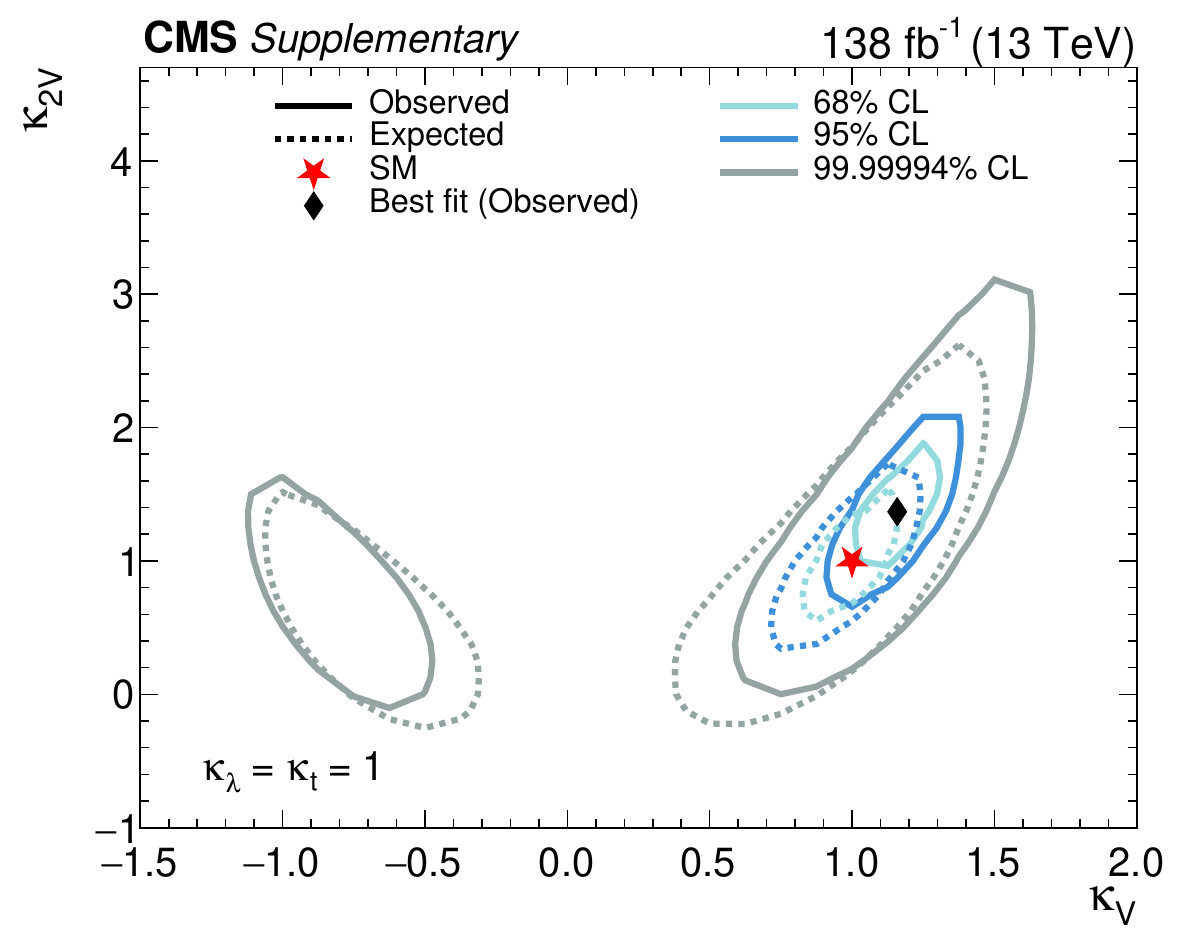}
  \caption{
    The 68\%, 95\%, and 5 standard deviation contours of $-2\Delta\ln(L)$ in the (\CV, \CVV) plane for the combination of all channels when all the other parameters are fixed to their SM values.
  }
  \label{fig:cv_cvv_extended}
\end{figure}

\begin{figure}[tbh!]
  \centering
  \includegraphics[width=0.45\textwidth]{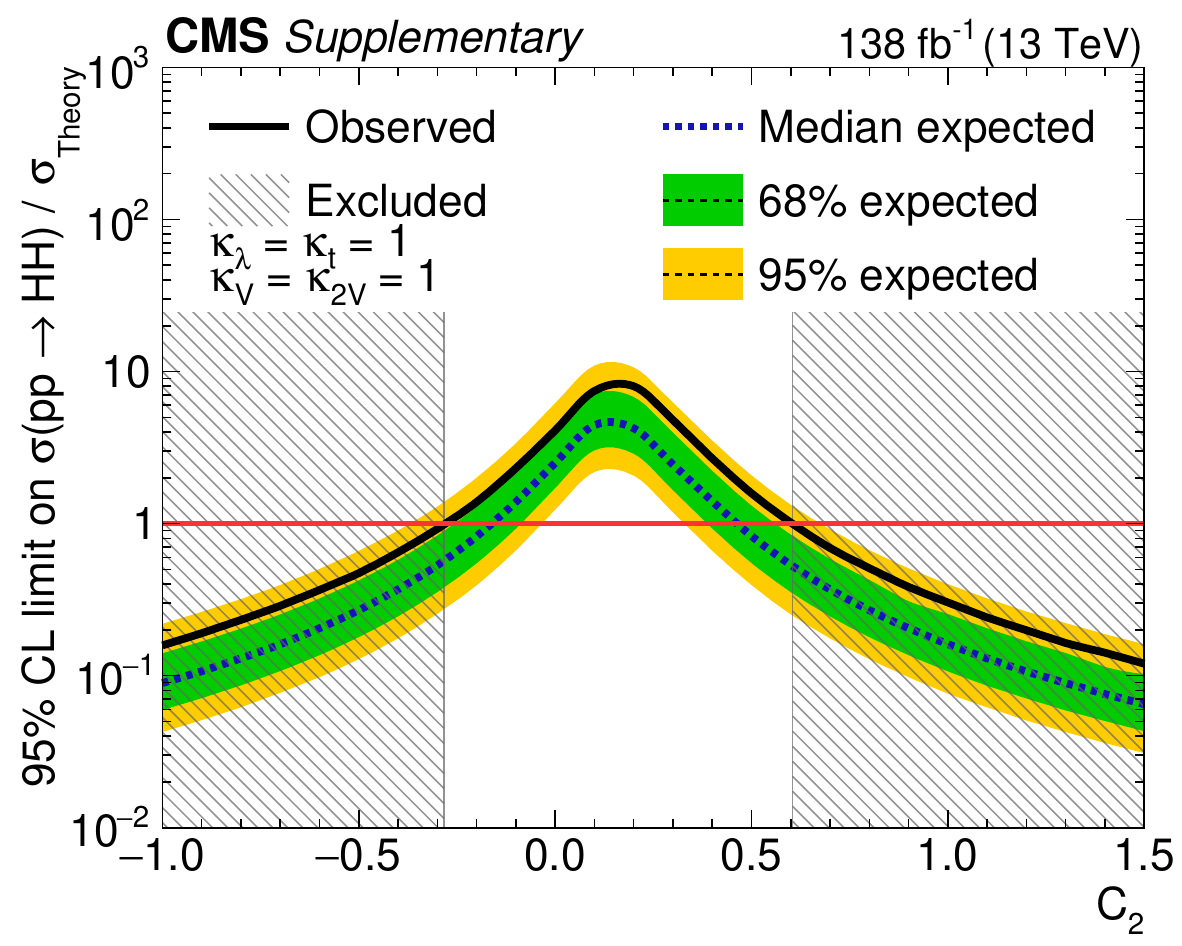}
  \caption{
    The 95\% \CL upper limits on the inclusive \HH signal strength as a function of the \ctwo coupling modifier.
    All other couplings are set to the values predicted by the SM.
    The theoretical uncertainties in the \GGF \HH signal cross sections are considered in this case.
    The inner (green) band and the outer (yellow) band indicate the 68 and 95\% \CL intervals, respectively, under the background-only hypothesis.
  }
  \label{fig:c2_limits_r_combibation}
\end{figure}

\begin{figure}[tbh!]
  \centering
  \includegraphics[width=0.45\textwidth]{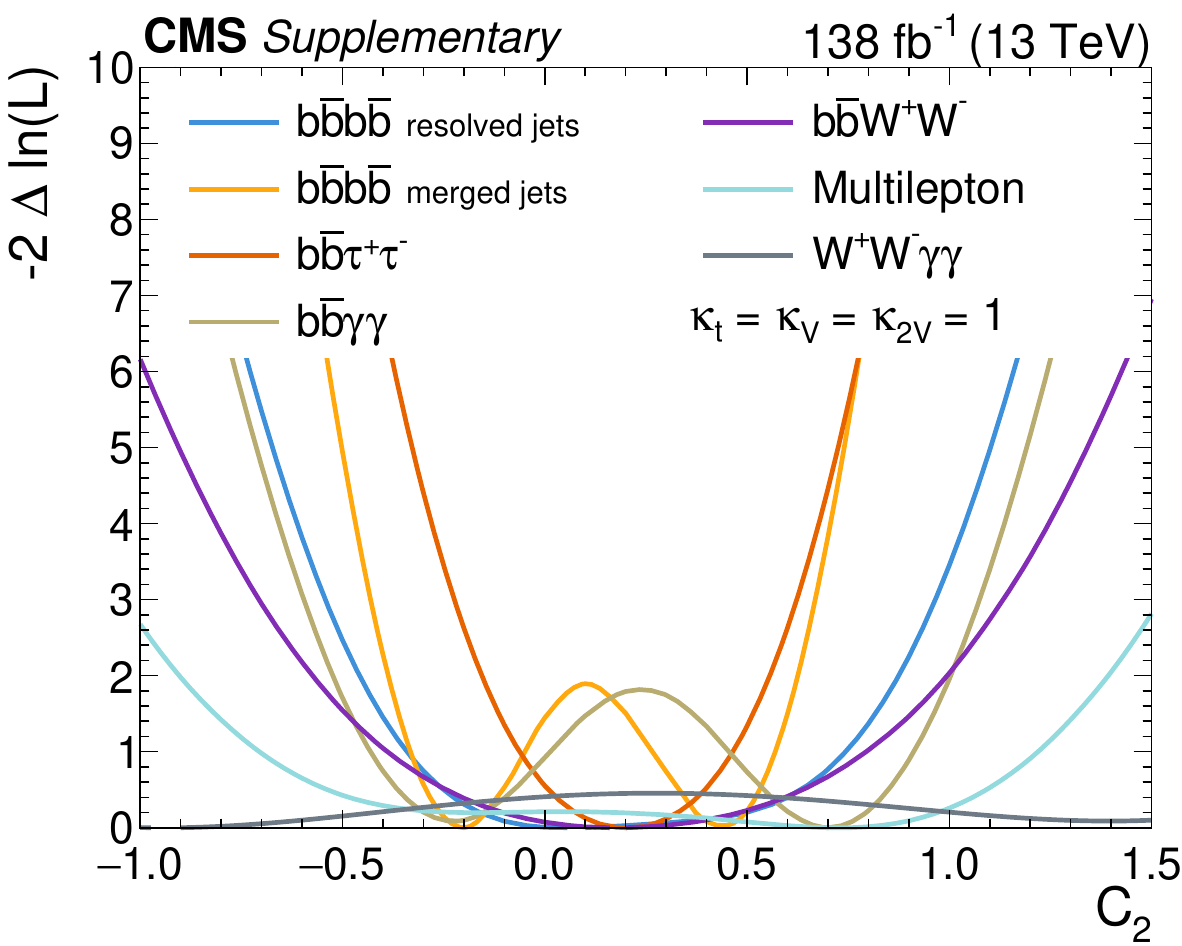}
  \caption{
    The $-2\Delta\ln(L)$ scan as a function of the \ctwo coupling modifier for all channels, when all the other parameters are fixed to their SM values.
  }
  \label{fig:c2_likeihood_channels}
\end{figure}

\begin{figure}[tbh!]
  \centering
  \includegraphics[width=0.45\textwidth]{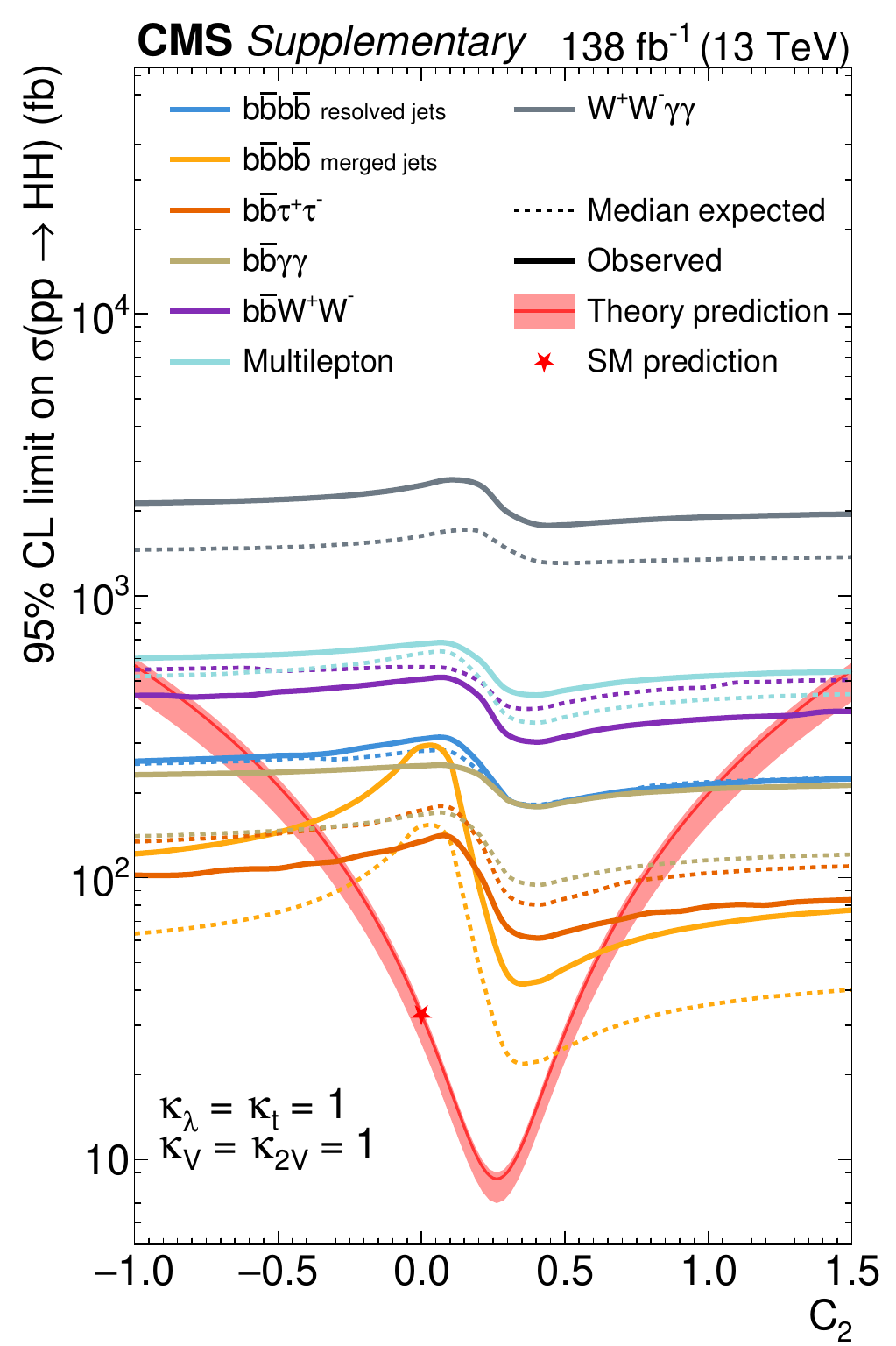}
  \caption{
    The 95\% \CL upper limits on the inclusive \HH cross section as a function of the \ctwo coupling modifier for all channels.
    All other couplings are set to the values predicted by the SM.
    The theoretical uncertainties in the \HH \GGF signal cross sections are not considered because we directly constrain the measured cross section.
  }
  \label{fig:c2_limits_xs_channels}
\end{figure}

\begin{figure}[tbh!]
  \centering
  \includegraphics[width=0.45\textwidth]{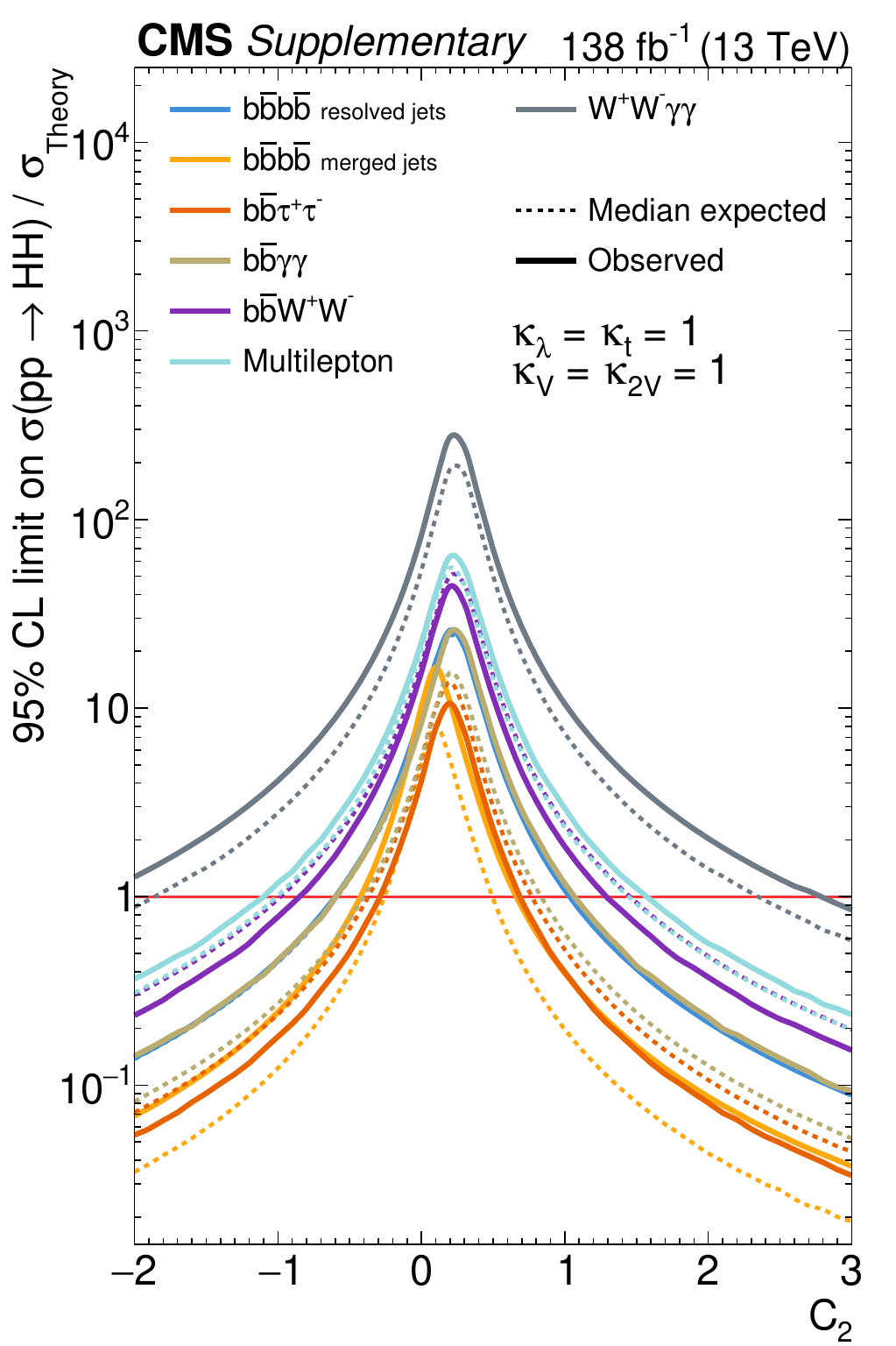}
  \caption{
    The 95\% \CL upper limits on the \HH signal strength as a function of the \ctwo coupling modifier for all channels.
    All other couplings are set to the values predicted by the SM.
    The theoretical uncertainties in the \GGF \HH signal cross sections are considered in this case.
  }
  \label{fig:c2_limits_r_channels}
\end{figure}

\begin{figure}[tbh!]
  \centering
  \includegraphics[width=0.45\textwidth]{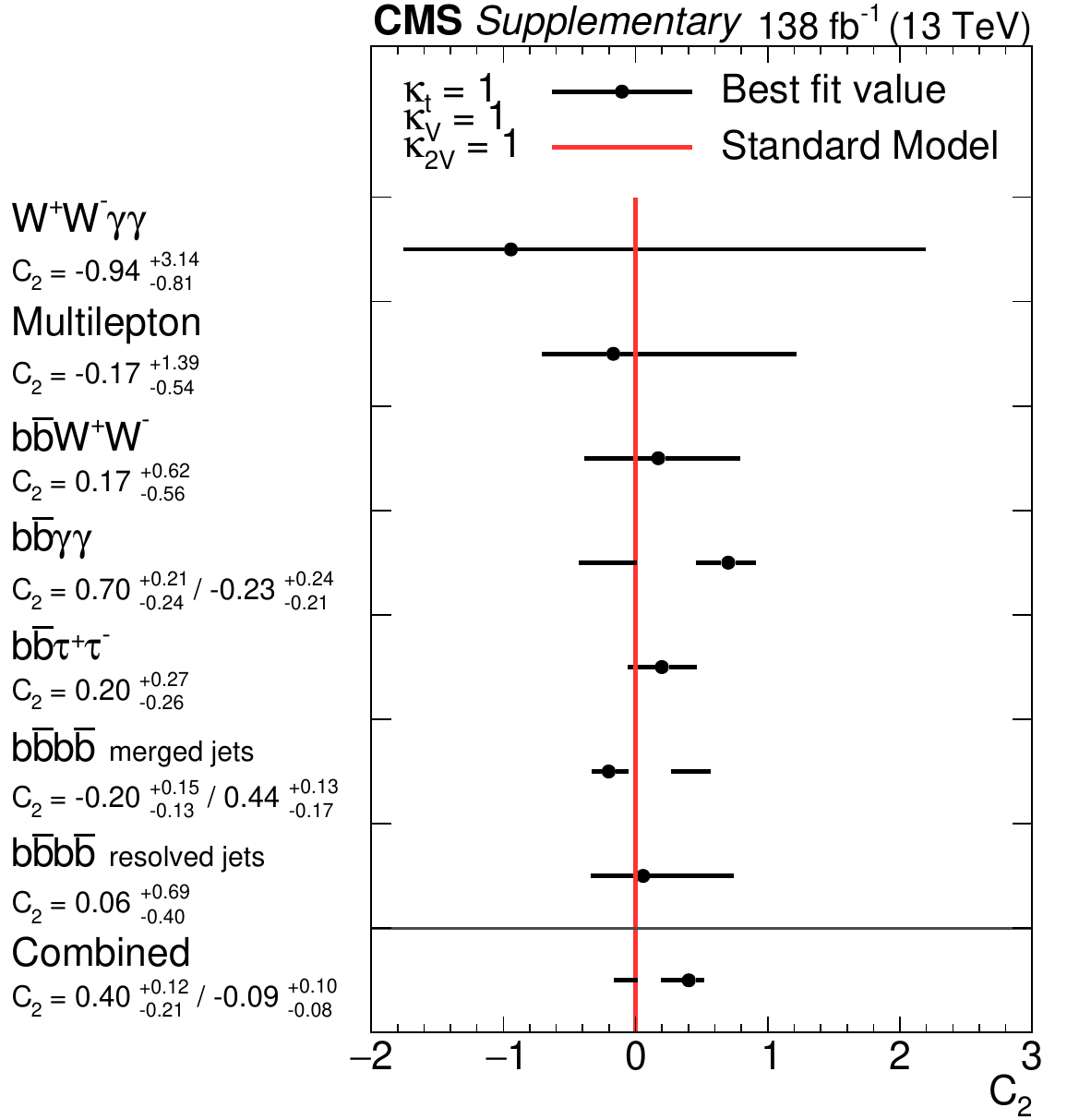}
  \caption{
    The best fit value for the \ctwo coupling modifier compared to the SM expectation for all channels and their combination, when all the other parameters are fixed to their SM values.
  }
  \label{fig:c2_bestfit_channels}
\end{figure}

\begin{figure}[tbh!]
  \centering
  \includegraphics[width=0.45\textwidth]{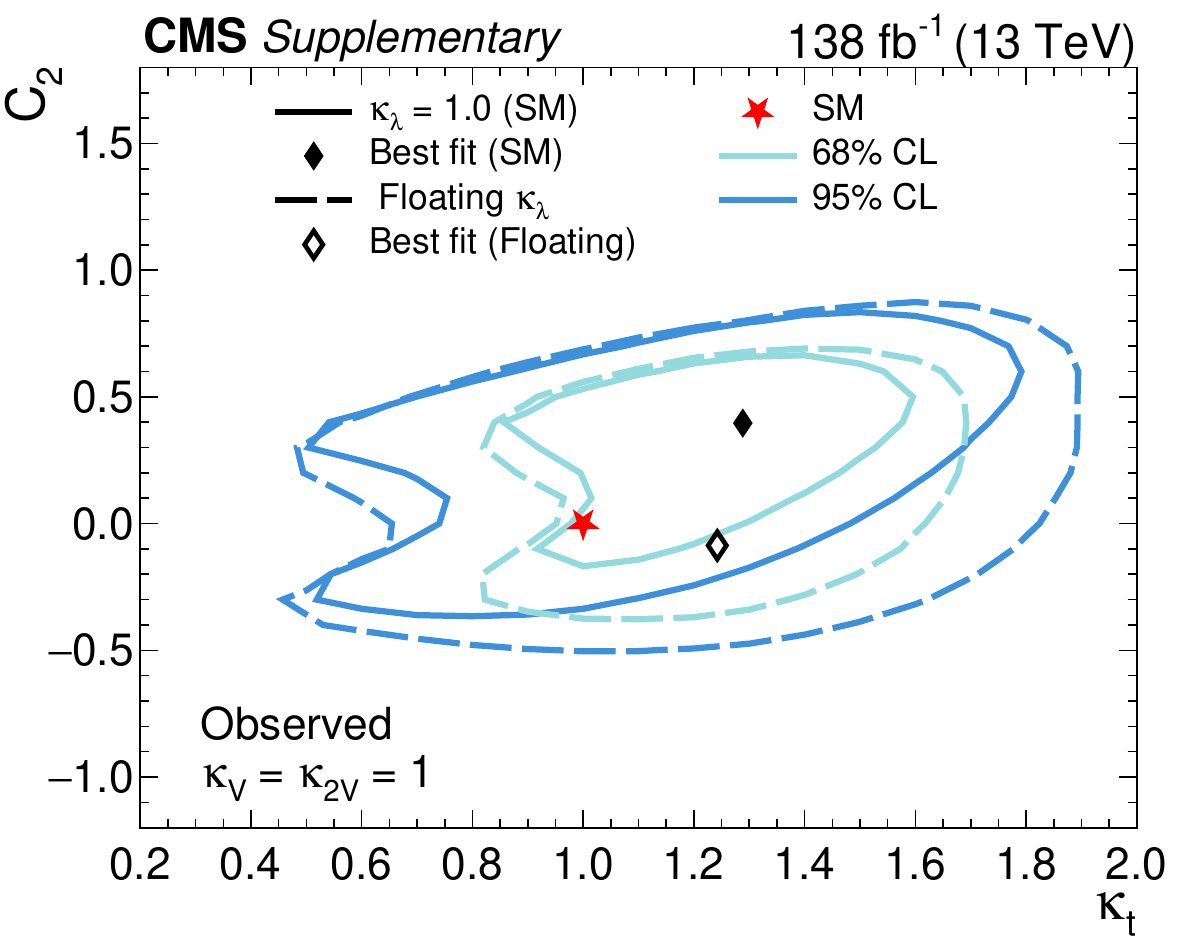}
  \caption{
    The observed 68 and 95\% \CL contours of $-2\Delta\ln(L)$ in the (\ctwo, \ktop) plane for the combination of all channels when \klambda is allowed to vary.
    The range of \klambda is set between $-15$ and 15 to avoid unphysical areas of the phase space.
    All the other parameters are fixed to their SM values.
  }
  \label{c2kt_floatkl_likelihood_2D}
\end{figure}

\begin{figure}[tbh!]
  \centering
  \includegraphics[width=0.45\textwidth]{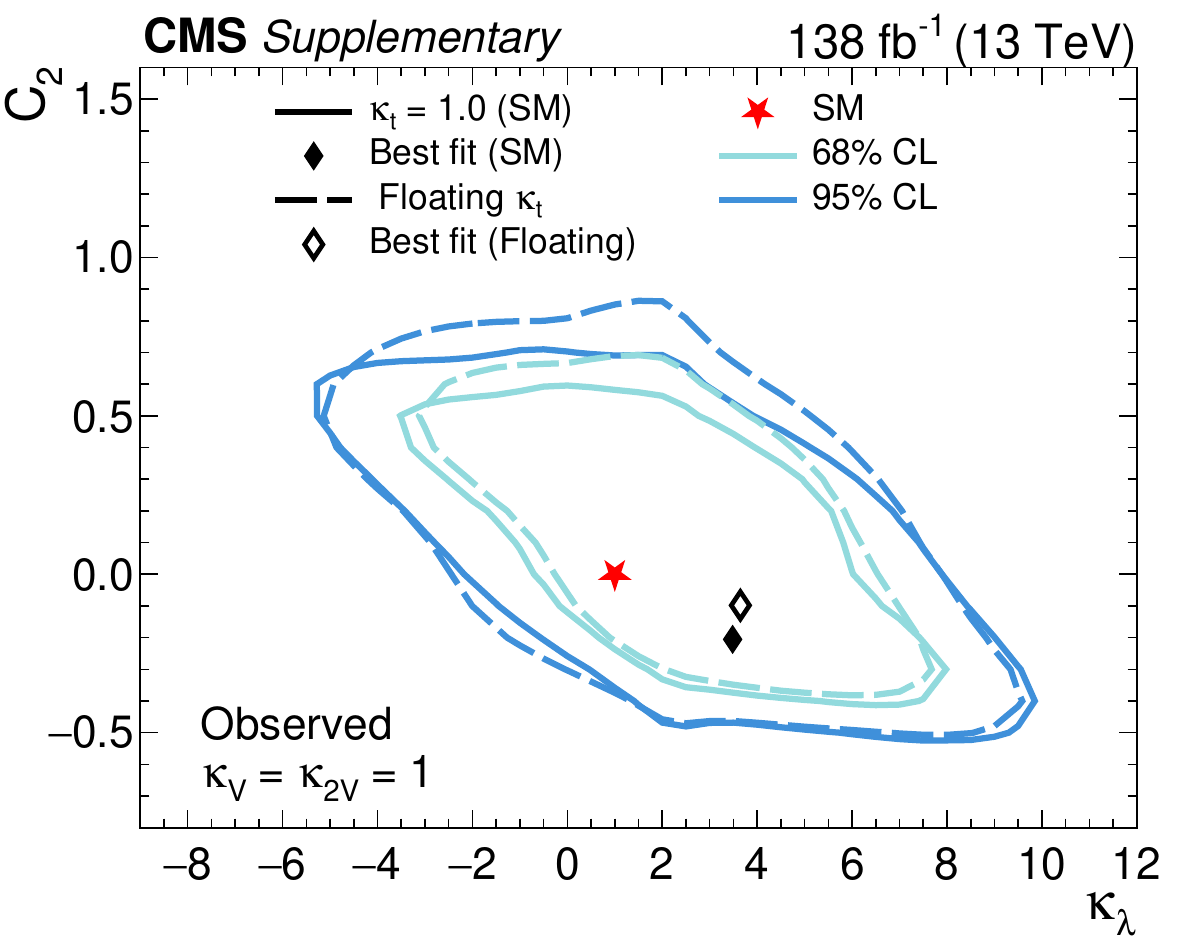}
  \caption{
    The observed 68 and 95\% \CL contours of $-2\Delta\ln(L)$ in the (\ctwo, \klambda) plane for the combination of all channels when \ktop is allowed to vary.
    The range of \ktop is set between 0.5 and 1.5 to avoid unphysical areas of the phase space.
    All the other parameters are fixed to their SM values.
  }
  \label{c2kl_floatkt_likelihood_2D}
\end{figure}

\begin{figure}[tbh!]
  \centering
  \includegraphics[width=0.45\textwidth]{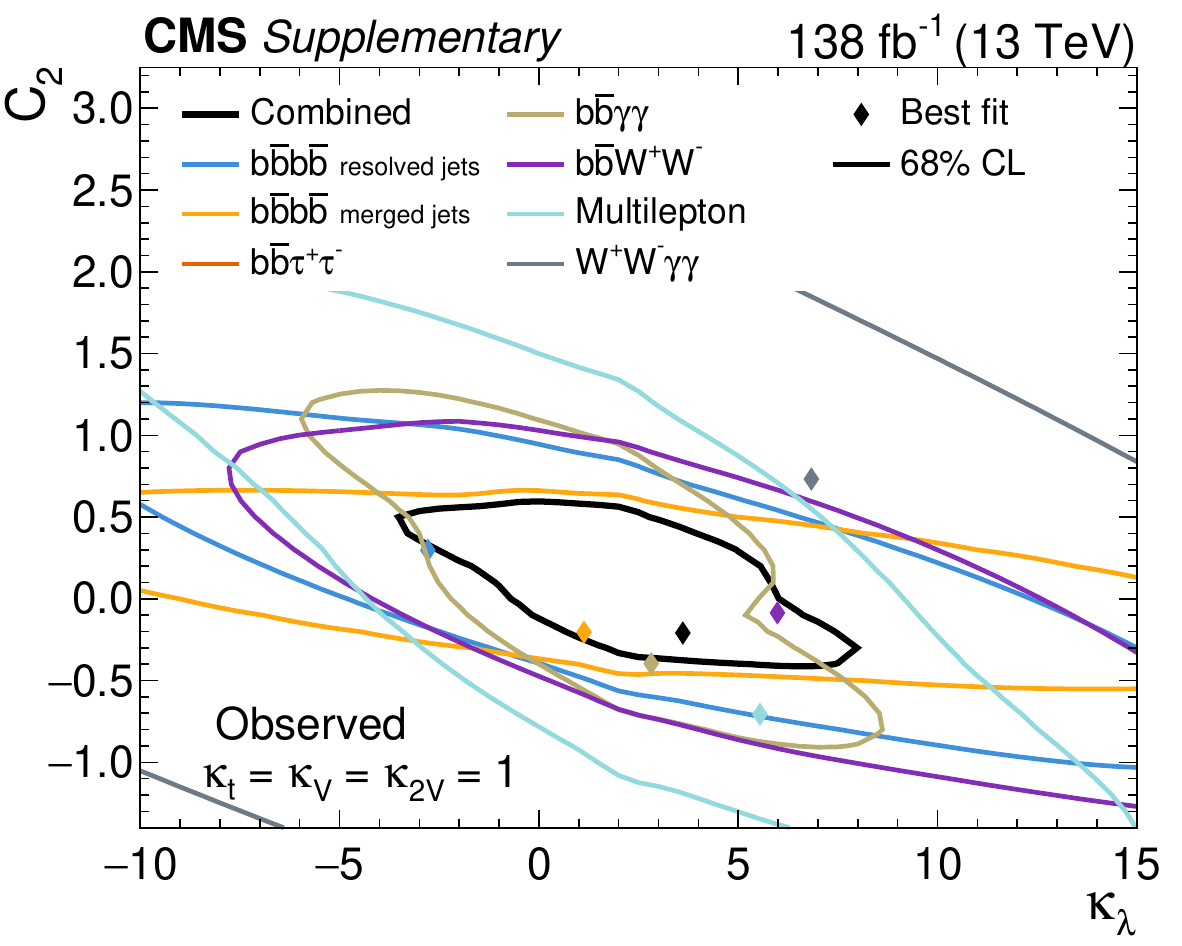}
  \caption{
    The observed 68\% \CL contours of $-2\Delta\ln(L)$ in the (\ctwo, \klambda) plane for all channels when all the other parameters are fixed to their SM value.
    The best fit value and 68\% \CL contour for the \bbtt channel are not within the range of the figure.
  }
  \label{c2kt_c2kl_likelihood_2D_channels}
\end{figure}

\clearpage

\begin{table*}[ht]
  \centering
  \topcaption{
    Upper limits on the \HH production cross section at 95\% \CL for the JHEP03(2020)91 BM1 benchmark.
    The theoretical uncertainties in the \GGF \HH signal cross section are not considered because we directly constrain the measured cross section.
  }
  \label{tab:JHEP03BM1}
  \scalebox{0.875}{
    \begin{tabular}{ l  c  c  c  c}
      \hline
      Analysis            & Observed [\text{fb}] & Expected [\text{fb}] & Expected $-$/$+$68\% [\text{fb}] & Expected $-$/$+$95\% [\text{fb}] \\
      \hline
      \bbbb resolved jets & 415.43               & 468.75               & 334.27 / 664.94                  & 250.85 / 905.55                  \\
      \bbbb merged jets   & 72.51                & 37.6                 & 25.66 / 56.78                    & 18.51 / 84.21                    \\
      \bbtt               & 146.25               & 188.96               & 129.91 / 277.85                  & 94.48 / 392.61                   \\
      \bbgg               & 362.99               & 216.8                & 141.93 / 336.92                  & 99.93 / 501.86                   \\
      \bbWW               & 787.14               & 910.16               & 656.67 / 1262.06                 & 494.19 / 1679.63                 \\
      multilepton         & 889.66               & 650.39               & 452.82 / 948.53                  & 330.28 / 1330.46                 \\
      \WWgg               & 3137.69              & 2460.94              & 1703.91 / 3589.02                & 1249.69 / 5034.18                \\
      Combination         & 62.48                & 34.67                & 23.39 / 50.97                    & 17.06 / 74.37                    \\
      \hline
    \end{tabular}
  }
\end{table*}

\begin{table*}[ht]
  \centering
  \topcaption{
    Upper limits on the \HH production cross section at 95\% \CL for the JHEP03(2020)91 BM2 benchmark.
    The theoretical uncertainties in the \HH \GGF signal cross section are not considered because we directly constrain the measured cross section.
  }
  \label{tab:JHEP03BM2}
  \scalebox{0.875}{
    \begin{tabular}{ l  c  c  c  c}
      \hline
      Analysis            & Observed [\text{fb}] & Expected [\text{fb}] & Expected $-$/$+$68\% [\text{fb}] & Expected $-$/$+$95\% [\text{fb}] \\
      \hline
      \bbbb resolved jets & 458.4                & 460.94               & 329.27 / 653.86                  & 244.87 / 896.3                   \\
      \bbbb merged jets   & 77.49                & 39.55                & 27.0 / 60.05                     & 19.47 / 88.75                    \\
      \bbtt               & 150.65               & 195.31               & 134.75 / 285.62                  & 98.42 / 402.44                   \\
      \bbgg               & 334.07               & 208.98               & 138.73 / 322.28                  & 97.96 / 476.44                   \\
      \bbWW               & 775.27               & 828.12               & 597.49 / 1151.62                 & 449.65 / 1530.63                 \\
      multilepton         & 827.61               & 630.86               & 436.8 / 920.04                   & 320.36 / 1286.61                 \\
      \WWgg               & 3009.3               & 2226.56              & 1541.63 / 3247.21                & 1130.68 / 4527.21                \\
      Combination         & 70.48                & 35.64                & 24.68 / 53.55                    & 18.1 / 77.97                     \\
      \hline
    \end{tabular}
  }
\end{table*}

\begin{table*}[ht]
  \centering
  \topcaption{
    Upper limits on the \HH production cross section at 95\% \CL for the JHEP03(2020)91 BM3 benchmark.
    The theoretical uncertainties in the \GGF \HH signal cross section are not considered because we directly constrain the measured cross section.
  }
  \label{tab:JHEP03BM3}
  \scalebox{0.875}{
    \begin{tabular}{ l  c  c  c  c}
      \hline
      Analysis            & Observed [\text{fb}] & Expected [\text{fb}] & Expected $-$/$+$68\% [\text{fb}] & Expected $-$/$+$95\% [\text{fb}] \\
      \hline
      \bbbb resolved jets & 167.16               & 189.94               & 135.69 / 268.68                  & 100.91 / 366.42                  \\
      \bbbb merged jets   & 25.99                & 13.18                & 9.48 / 20.33                     & 6.59 / 29.9                      \\
      \bbtt               & 56.04                & 74.22                & 51.03 / 109.42                   & 37.11 / 154.84                   \\
      \bbgg               & 175.39               & 90.33                & 58.46 / 142.54                   & 40.58 / 215.84                   \\
      \bbWW               & 281.9                & 351.56               & 253.65 / 487.49                  & 190.89 / 646.51                  \\
      multilepton         & 537.55               & 292.97               & 202.13 / 430.77                  & 147.63 / 608.7                   \\
      \WWgg               & 1805.28              & 1238.28              & 851.32 / 1830.58                 & 619.14 / 2593.94                 \\
      Combination         & 22.94                & 12.21                & 8.27 / 18.82                     & 6.48 / 27.11                     \\
      \hline
    \end{tabular}
  }
\end{table*}

\begin{table*}[ht]
  \centering
  \topcaption{
    Upper limits on the \HH production cross section at 95\% \CL for the JHEP03(2020)91 BM4 benchmark.
    The theoretical uncertainties in the \GGF \HH signal cross section are not considered because we directly constrain the measured cross section.
  }
  \label{tab:JHEP03BM4}
  \scalebox{0.875}{
    \begin{tabular}{ l  c  c  c  c}
      \hline
      Analysis            & Observed [\text{fb}] & Expected [\text{fb}] & Expected $-$/$+$68\% [\text{fb}] & Expected $-$/$+$95\% [\text{fb}] \\
      \hline
      \bbbb resolved jets & 185.3                & 199.22               & 142.31 / 282.6                   & 105.83 / 384.86                  \\
      \bbbb merged jets   & 74.53                & 38.57                & 26.03 / 57.33                    & 18.99 / 83.57                    \\
      \bbtt               & 78.63                & 101.56               & 70.32 / 148.93                   & 51.57 / 209.52                   \\
      \bbgg               & 211.59               & 122.07               & 80.43 / 189.7                    & 56.27 / 282.58                   \\
      \bbWW               & 370.16               & 464.84               & 334.28 / 648.28                  & 250.58 / 866.49                  \\
      multilepton         & 559.41               & 435.55               & 302.63 / 635.2                   & 222.88 / 888.28                  \\
      \WWgg               & 2394.44              & 1441.41              & 990.97 / 2119.38                 & 720.7 / 2994.8                   \\
      Combination         & 54.28                & 30.76                & 21.34 / 45.97                    & 14.9 / 64.55                     \\
      \hline
    \end{tabular}
  }
\end{table*}

\begin{table*}[ht]
  \centering
  \topcaption{
    Upper limits on the \HH production cross section at 95\% \CL for the JHEP03(2020)91 BM5 benchmark.
    The theoretical uncertainties in the \GGF \HH signal cross section are not considered because we directly constrain the measured cross section.
  }
  \label{tab:JHEP03BM5}
  \scalebox{0.875}{
    \begin{tabular}{ l  c  c  c  c}
      \hline
      Analysis            & Observed [\text{fb}] & Expected [\text{fb}] & Expected $-$/$+$68\% [\text{fb}] & Expected $-$/$+$95\% [\text{fb}] \\
      \hline
      \bbbb resolved jets & 155.28               & 170.9                & 121.87 / 242.43                  & 91.46 / 330.15                   \\
      \bbbb merged jets   & 39.62                & 20.02                & 13.66 / 30.55                    & 10.32 / 45.24                    \\
      \bbtt               & 59.63                & 78.12                & 53.9 / 114.56                    & 39.37 / 161.17                   \\
      \bbgg               & 177.03               & 94.24                & 61.69 / 147.58                   & 43.44 / 221.42                   \\
      \bbWW               & 284.75               & 360.35               & 259.99 / 501.12                  & 195.66 / 666.04                  \\
      multilepton         & 555.86               & 312.5                & 215.61 / 456.99                  & 157.47 / 643.9                   \\
      \WWgg               & 1781.23              & 1160.16              & 797.61 / 1710.46                 & 580.08 / 2427.49                 \\
      Combination         & 35.3                 & 18.07                & 11.95 / 26.71                    & 8.75 / 38.62                     \\
      \hline
    \end{tabular}
  }
\end{table*}

\begin{table*}[ht]
  \centering
  \topcaption{
    Upper limits on the \HH production cross section at 95\% \CL for the JHEP03(2020)91 BM6 benchmark.
    The theoretical uncertainties in the \HH \GGF signal cross section are not considered because we directly constrain the measured cross section.
  }
  \label{tab:JHEP03BM6}
  \scalebox{0.875}{
    \begin{tabular}{ l  c  c  c  c}
      \hline
      Analysis            & Observed [\text{fb}] & Expected [\text{fb}] & Expected $-$/$+$68\% [\text{fb}] & Expected $-$/$+$95\% [\text{fb}] \\
      \hline
      \bbbb resolved jets & 251.0                & 267.58               & 191.15 / 378.5                   & 142.15 / 516.19                  \\
      \bbbb merged jets   & 85.91                & 44.43                & 30.2 / 66.75                     & 22.22 / 98.53                    \\
      \bbtt               & 103.7                & 132.81               & 91.96 / 194.75                   & 67.44 / 273.17                   \\
      \bbgg               & 252.23               & 152.83               & 100.82 / 236.29                  & 71.64 / 350.52                   \\
      \bbWW               & 479.53               & 572.27               & 411.53 / 798.09                  & 308.49 / 1066.73                 \\
      multilepton         & 659.53               & 517.58               & 361.6 / 754.83                   & 264.85 / 1055.58                 \\
      \WWgg               & 2583.26              & 1734.38              & 1196.62 / 2557.06                & 873.96 / 3607.74                 \\
      Combination         & 66.25                & 37.6                 & 25.55 / 54.68                    & 18.8 / 78.45                     \\
      \hline
    \end{tabular}
  }
\end{table*}

\begin{table*}[ht]
  \centering
  \topcaption{
    Upper limits on the \HH production cross section at 95\% \CL for the JHEP03(2020)91 BM7 benchmark.
    The theoretical uncertainties in the \HH \GGF signal cross section are not considered because we directly constrain the measured cross section.
  }
  \label{tab:JHEP03BM7}
  \scalebox{0.875}{
    \begin{tabular}{ l  c  c  c  c}
      \hline
      Analysis            & Observed [\text{fb}] & Expected [\text{fb}] & Expected $-$/$+$68\% [\text{fb}] & Expected $-$/$+$95\% [\text{fb}] \\
      \hline
      \bbbb resolved jets & 128.91               & 143.07               & 102.2 / 202.37                   & 76.0 / 275.99                    \\
      \bbbb merged jets   & 34.6                 & 18.07                & 11.95 / 26.42                    & 8.75 / 38.45                     \\
      \bbtt               & 50.3                 & 65.92                & 45.32 / 96.66                    & 32.96 / 135.99                   \\
      \bbgg               & 158.07               & 82.52                & 53.61 / 129.56                   & 37.39 / 194.98                   \\
      \bbWW               & 261.75               & 333.98               & 240.17 / 465.78                  & 180.04 / 618.27                  \\
      multilepton         & 525.55               & 280.27               & 194.06 / 410.98                  & 142.33 / 578.19                  \\
      \WWgg               & 1640.12              & 1093.75              & 751.95 / 1621.28                 & 546.88 / 2307.14                 \\
      Combination         & 28.59                & 15.14                & 11.15 / 23.34                    & 8.04 / 32.91                     \\
      \hline
    \end{tabular}
  }
\end{table*}

\begin{table*}[ht]
  \centering
  \topcaption{
    Upper limits on the \HH production cross section at 95\% \CL for the JHEP04(2016)126 BM1 benchmark.
    The theoretical uncertainties in the \HH \GGF signal cross section are not considered because we directly constrain the measured cross section.
  }
  \label{tab:JHEP04BM1}
  \scalebox{0.875}{
    \begin{tabular}{ l  c  c  c  c}
      \hline
      Analysis            & Observed [\text{fb}] & Expected [\text{fb}] & Expected $-$/$+$68\% [\text{fb}] & Expected $-$/$+$95\% [\text{fb}] \\
      \hline
      \bbbb resolved jets & 144.69               & 162.11               & 115.6 / 229.31                   & 86.75 / 312.73                   \\
      \bbbb merged jets   & 48.55                & 24.9                 & 16.99 / 37.21                    & 12.84 / 54.36                    \\
      \bbtt               & 59.6                 & 77.64                & 53.56 / 114.46                   & 39.12 / 161.5                    \\
      \bbgg               & 188.92               & 99.12                & 64.39 / 156.41                   & 44.91 / 235.71                   \\
      \bbWW               & 357.64               & 408.2                & 295.97 / 569.29                  & 221.64 / 760.91                  \\
      multilepton         & 468.68               & 353.52               & 246.13 / 514.16                  & 179.52 / 720.09                  \\
      \WWgg               & 1894.82              & 1339.84              & 921.14 / 1980.72                 & 669.92 / 2806.69                 \\
      \ttgg               & 683.75               & 626.95               & 410.44 / 984.32                  & 288.99 / 1481.42                 \\
      Combination         & 38.85                & 21.0                 & 14.96 / 31.37                    & 10.83 / 44.31                    \\
      \hline
    \end{tabular}
  }
\end{table*}

\begin{table*}[ht]
  \centering
  \topcaption{
    Upper limits on the \HH production cross section at 95\% \CL for the JHEP04(2016)126 BM2 benchmark.
    The theoretical uncertainties in the \HH \GGF signal cross section are not considered because we directly constrain the measured cross section.
  }
  \label{tab:JHEP04BM2}
  \scalebox{0.875}{
    \begin{tabular}{ l  c  c  c  c}
      \hline
      Analysis            & Observed [\text{fb}] & Expected [\text{fb}] & Expected $-$/$+$68\% [\text{fb}] & Expected $-$/$+$95\% [\text{fb}] \\
      \hline
      \bbbb resolved jets & 129.78               & 146.97               & 104.81 / 208.49                  & 78.65 / 285.79                   \\
      \bbbb merged jets   & 11.48                & 6.35                 & 4.12 / 9.18                      & 2.78 / 13.29                     \\
      \bbtt               & 32.86                & 43.46                & 29.66 / 63.9                     & 21.39 / 91.36                    \\
      \bbgg               & 101.35               & 52.25                & 33.68 / 83.48                    & 23.27 / 127.66                   \\
      \bbWW               & 171.98               & 198.24               & 143.27 / 274.89                  & 106.86 / 364.56                  \\
      multilepton         & 204.62               & 158.69               & 108.71 / 235.23                  & 78.73 / 335.71                   \\
      \WWgg               & 1240.76              & 929.69               & 636.89 / 1381.79                 & 461.21 / 1974.57                 \\
      \ttgg               & 461.27               & 431.64               & 282.58 / 677.68                  & 198.96 / 1019.92                 \\
      Combination         & 10.45                & 5.37                 & 3.9 / 8.45                       & 3.02 / 11.9                      \\
      \hline
    \end{tabular}
  }
\end{table*}

\begin{table*}[ht]
  \centering
  \topcaption{
    Upper limits on the \HH production cross section at 95\% \CL for the JHEP04(2016)126 BM3 benchmark.
    The theoretical uncertainties in the \HH \GGF signal cross section are not considered because we directly constrain the measured cross section.
  }
  \label{tab:JHEP04BM3}
  \scalebox{0.875}{
    \begin{tabular}{ l  c  c  c  c}
      \hline
      Analysis            & Observed [\text{fb}] & Expected [\text{fb}] & Expected $-$/$+$68\% [\text{fb}] & Expected $-$/$+$95\% [\text{fb}] \\
      \hline
      \bbbb resolved jets & 195.41               & 204.1                & 146.29 / 289.53                  & 109.23 / 396.88                  \\
      \bbbb merged jets   & 85.84                & 43.46                & 29.88 / 65.28                    & 21.73 / 95.32                    \\
      \bbtt               & 82.28                & 106.45               & 73.7 / 156.09                    & 54.05 / 218.94                   \\
      \bbgg               & 216.47               & 126.46               & 83.64 / 196.53                   & 58.79 / 292.02                   \\
      \bbWW               & 350.21               & 480.47               & 347.24 / 670.07                  & 259.0 / 895.62                   \\
      multilepton         & 562.38               & 449.22               & 312.13 / 655.14                  & 229.87 / 913.38                  \\
      \WWgg               & 2053.45              & 1507.81              & 1040.3 / 2223.03                 & 759.8 / 3136.46                  \\
      \ttgg               & 738.4                & 662.11               & 435.12 / 1034.24                 & 307.78 / 1562.05                 \\
      Combination         & 60.49                & 34.67                & 24.0 / 50.42                     & 17.6 / 71.48                     \\
      \hline
    \end{tabular}
  }
\end{table*}

\begin{table*}[ht]
  \centering
  \topcaption{
    Upper limits on the \HH production cross section at 95\% \CL for the JHEP04(2016)126 BM4 benchmark.
    The theoretical uncertainties in the \HH \GGF signal cross section are not considered because we directly constrain the measured cross section.
  }
  \label{tab:JHEP04BM4}
  \scalebox{0.875}{
    \begin{tabular}{ l  c  c  c  c}
      \hline
      Analysis            & Observed [\text{fb}] & Expected [\text{fb}] & Expected $-$/$+$68\% [\text{fb}] & Expected $-$/$+$95\% [\text{fb}] \\
      \hline
      \bbbb resolved jets & 265.75               & 278.32               & 198.82 / 394.81                  & 147.86 / 541.2                   \\
      \bbbb merged jets   & 177.25               & 91.31                & 63.0 / 136.8                     & 46.01 / 198.98                   \\
      \bbtt               & 122.23               & 153.81               & 106.49 / 224.93                  & 78.11 / 315.02                   \\
      \bbgg               & 267.77               & 172.36               & 114.42 / 265.11                  & 80.8 / 390.58                    \\
      \bbWW               & 585.6                & 619.14               & 445.23 / 863.46                  & 333.76 / 1154.11                 \\
      multilepton         & 677.65               & 599.61               & 418.91 / 869.69                  & 306.83 / 1212.4                  \\
      \WWgg               & 2663.19              & 1890.62              & 1304.42 / 2772.35                & 952.7 / 3900.26                  \\
      \ttgg               & 948.48               & 839.84               & 555.43 / 1311.86                 & 390.4 / 1961.96                  \\
      Combination         & 108.41               & 62.01                & 43.41 / 89.2                     & 31.49 / 124.52                   \\
      \hline
    \end{tabular}
  }
\end{table*}

\begin{table*}[ht]
  \centering
  \topcaption{
    Upper limits on the \HH production cross section at 95\% \CL for the JHEP04(2016)126 BM5 benchmark.
    The theoretical uncertainties in the \HH \GGF signal cross section are not considered because we directly constrain the measured cross section.
  }
  \label{tab:JHEP04BM5}
  \scalebox{0.875}{
    \begin{tabular}{ l  c  c  c  c}
      \hline
      Analysis            & Observed [\text{fb}] & Expected [\text{fb}] & Expected $-$/$+$68\% [\text{fb}] & Expected $-$/$+$95\% [\text{fb}] \\
      \hline
      \bbbb resolved jets & 143.19               & 161.62               & 115.25 / 229.27                  & 86.49 / 312.23                   \\
      \bbbb merged jets   & 24.86                & 13.18                & 8.37 / 19.49                     & 6.18 / 28.18                     \\
      \bbtt               & 49.4                 & 65.92                & 45.32 / 96.66                    & 32.96 / 136.8                    \\
      \bbgg               & 156.84               & 80.57                & 52.14 / 126.81                   & 36.19 / 192.36                   \\
      \bbWW               & 252.93               & 318.36               & 229.7 / 441.45                   & 172.86 / 587.51                  \\
      multilepton         & 439.15               & 262.7                & 181.24 / 386.25                  & 132.37 / 545.8                   \\
      \WWgg               & 1541.96              & 1109.38              & 762.7 / 1644.44                  & 554.69 / 2340.09                 \\
      \ttgg               & 586.92               & 533.2                & 349.07 / 832.88                  & 245.77 / 1257.94                 \\
      Combination         & 22.15                & 11.23                & 8.27 / 17.32                     & 5.97 / 25.47                     \\
      \hline
    \end{tabular}
  }
\end{table*}

\begin{table*}[ht]
  \centering
  \topcaption{
    Upper limits on the \HH production cross section at 95\% \CL for the JHEP04(2016)126 BM6 benchmark.
    The theoretical uncertainties in the \HH \GGF signal cross section are not considered because we directly constrain the measured cross section.
  }
  \label{tab:JHEP04BM6}
  \scalebox{0.875}{
    \begin{tabular}{ l  c  c  c  c}
      \hline
      Analysis            & Observed [\text{fb}] & Expected [\text{fb}] & Expected $-$/$+$68\% [\text{fb}] & Expected $-$/$+$95\% [\text{fb}] \\
      \hline
      \bbbb resolved jets & 253.42               & 274.41               & 195.69 / 389.27                  & 146.85 / 530.13                  \\
      \bbbb merged jets   & 100.46               & 51.76                & 35.18 / 77.75                    & 25.88 / 112.9                    \\
      \bbtt               & 112.05               & 145.02               & 100.05 / 212.65                  & 73.08 / 299.17                   \\
      \bbgg               & 299.11               & 176.76               & 116.46 / 275.4                   & 81.47 / 410.54                   \\
      \bbWW               & 607.27               & 675.78               & 485.97 / 942.46                  & 364.29 / 1259.69                 \\
      multilepton         & 740.39               & 583.98               & 405.77 / 849.35                  & 298.84 / 1185.92                 \\
      \WWgg               & 2841.96              & 2085.94              & 1439.17 / 3058.75                & 1051.12 / 4303.18                \\
      \ttgg               & 1118.94              & 1015.62              & 669.17 / 1590.49                 & 468.14 / 2397.94                 \\
      Combination         & 76.13                & 42.48                & 29.74 / 62.46                    & 21.57 / 88.78                    \\
      \hline
    \end{tabular}
  }
\end{table*}

\begin{table*}[ht]
  \centering
  \topcaption{
    Upper limits on the \HH production cross section at 95\% \CL for the JHEP04(2016)126 BM7 benchmark.
    The theoretical uncertainties in the \HH \GGF signal cross section are not considered because we directly constrain the measured cross section.
  }
  \label{tab:JHEP04BM7}
  \scalebox{0.875}{
    \begin{tabular}{ l  c  c  c  c}
      \hline
      Analysis            & Observed [\text{fb}] & Expected [\text{fb}] & Expected $-$/$+$68\% [\text{fb}] & Expected $-$/$+$95\% [\text{fb}] \\
      \hline
      \bbbb resolved jets & 1154.73              & 1156.25              & 821.74 / 1644.79                 & 614.26 / 2266.09                 \\
      \bbbb merged jets   & 675.53               & 344.73               & 237.0 / 516.49                   & 172.36 / 755.37                  \\
      \bbtt               & 537.5                & 619.14               & 431.06 / 902.95                  & 314.41 / 1266.54                 \\
      \bbgg               & 528.29               & 449.22               & 306.64 / 673.04                  & 221.1 / 968.1                    \\
      \bbWW               & 2447.17              & 2234.38              & 1606.78 / 3116.1                 & 1204.47 / 4164.99                \\
      multilepton         & 1087.9               & 1160.16              & 816.12 / 1664.22                 & 602.74 / 2297.57                 \\
      \WWgg               & 4653.42              & 3546.88              & 2503.64 / 5102.05                & 1856.57 / 7033.55                \\
      \ttgg               & 2853.68              & 2437.5               & 1618.06 / 3778.31                & 1142.58 / 5637.78                \\
      Combination         & 326.77               & 213.87               & 149.93 / 308.49                  & 110.28 / 430.01                  \\
      \hline
    \end{tabular}
  }
\end{table*}

\begin{table*}[ht]
  \centering
  \topcaption{
    Upper limits on the \HH production cross section at 95\% \CL for the JHEP04(2016)126 BM8 benchmark.
    The theoretical uncertainties in the \HH \GGF signal cross section are not considered because we directly constrain the measured cross section.
  }
  \label{tab:JHEP04BM8}
  \scalebox{0.875}{
    \begin{tabular}{ l  c  c  c  c}
      \hline
      Analysis            & Observed [\text{fb}] & Expected [\text{fb}] & Expected $-$/$+$68\% [\text{fb}] & Expected $-$/$+$95\% [\text{fb}] \\
      \hline
      \bbbb resolved jets & 156.42               & 174.8                & 124.65 / 247.97                  & 93.55 / 337.7                    \\
      \bbbb merged jets   & 42.29                & 21.97                & 14.89 / 32.83                    & 10.64 / 48.23                    \\
      \bbtt               & 61.37                & 80.08                & 55.56 / 117.74                   & 40.35 / 165.4                    \\
      \bbgg               & 181.29               & 97.17                & 63.61 / 152.17                   & 44.79 / 228.3                    \\
      \bbWW               & 317.11               & 386.72               & 279.02 / 539.32                  & 209.98 / 715.89                  \\
      multilepton         & 495.07               & 334.96               & 232.74 / 488.51                  & 171.41 / 685.21                  \\
      \WWgg               & 1893.77              & 1230.47              & 845.95 / 1814.13                 & 615.23 / 2574.61                 \\
      \ttgg               & 620.95               & 564.45               & 369.53 / 881.69                  & 260.18 / 1331.66                 \\
      Combination         & 35.98                & 19.04                & 12.99 / 28.46                    & 9.82 / 40.88                     \\
      \hline
    \end{tabular}
  }
\end{table*}

\begin{table*}[ht]
  \centering
  \topcaption{
    Upper limits on the \HH production cross section at 95\% \CL for the JHEP04(2016)126 BM8a benchmark.
    The theoretical uncertainties in the \HH \GGF signal cross section are not considered because we directly constrain the measured cross section.
  }
  \label{tab:JHEP04BM8a}
  \scalebox{0.875}{
    \begin{tabular}{ l  c  c  c  c}
      \hline
      Analysis            & Observed [\text{fb}] & Expected [\text{fb}] & Expected $-$/$+$68\% [\text{fb}] & Expected $-$/$+$95\% [\text{fb}] \\
      \hline
      \bbbb resolved jets & 159.9                & 179.2                & 127.79 / 254.2                   & 95.9 / 346.19                    \\
      \bbbb merged jets   & 55.29                & 28.81                & 19.99 / 43.05                    & 13.95 / 62.19                    \\
      \bbtt               & 62.93                & 81.54                & 56.46 / 120.22                   & 41.41 / 169.62                   \\
      \bbgg               & 191.66               & 103.52               & 67.51 / 162.11                   & 47.31 / 244.41                   \\
      \bbWW               & 342.36               & 412.11               & 296.35 / 571.45                  & 222.15 / 760.52                  \\
      multilepton         & 608.09               & 352.54               & 245.45 / 515.55                  & 179.02 / 722.05                  \\
      \WWgg               & 1854.52              & 1351.56              & 929.2 / 1992.66                  & 675.78 / 2827.98                 \\
      \ttgg               & 677.86               & 619.14               & 407.94 / 969.59                  & 285.39 / 1461.82                 \\
      Combination         & 44.25                & 23.93                & 16.32 / 34.99                    & 12.34 / 49.45                    \\
      \hline
    \end{tabular}
  }
\end{table*}

\begin{table*}[ht]
  \centering
  \topcaption{
    Upper limits on the \HH production cross section at 95\% \CL for the JHEP04(2016)126 BM9 benchmark.
    The theoretical uncertainties in the \HH \GGF signal cross section are not considered because we directly constrain the measured cross section.
  }
  \label{tab:JHEP04BM9}
  \scalebox{0.875}{
    \begin{tabular}{ l  c  c  c  c}
      \hline
      Analysis            & Observed [\text{fb}] & Expected [\text{fb}] & Expected $-$/$+$68\% [\text{fb}] & Expected $-$/$+$95\% [\text{fb}] \\
      \hline
      \bbbb resolved jets & 140.98               & 155.76               & 111.27 / 220.95                  & 82.75 / 300.91                   \\
      \bbbb merged jets   & 36.6                 & 19.04                & 12.91 / 28.46                    & 9.22 / 41.8                      \\
      \bbtt               & 53.85                & 70.8                 & 48.68 / 103.82                   & 35.4 / 146.06                    \\
      \bbgg               & 167.12               & 88.38                & 57.42 / 138.4                    & 40.05 / 208.67                   \\
      \bbWW               & 284.46               & 357.42               & 257.88 / 497.04                  & 194.07 / 660.63                  \\
      multilepton         & 540.83               & 297.85               & 206.23 / 436.76                  & 151.25 / 614.45                  \\
      \WWgg               & 1790.22              & 1167.97              & 802.98 / 1731.29                 & 583.98 / 2449.47                 \\
      \ttgg               & 586.01               & 535.16               & 352.6 / 838.06                   & 246.67 / 1263.53                 \\
      Combination         & 31.4                 & 16.11                & 11.72 / 24.59                    & 8.31 / 34.89                     \\
      \hline
    \end{tabular}
  }
\end{table*}

\begin{table*}[ht]
  \centering
  \topcaption{
    Upper limits on the \HH production cross section at 95\% \CL for the JHEP04(2016)126 BM10 benchmark.
    The theoretical uncertainties in the \HH \GGF signal cross section are not considered because we directly constrain the measured cross section.
  }
  \label{tab:JHEP04BM10}
  \scalebox{0.875}{
    \begin{tabular}{ l  c  c  c  c}
      \hline
      Analysis            & Observed [\text{fb}] & Expected [\text{fb}] & Expected $-$/$+$68\% [\text{fb}] & Expected $-$/$+$95\% [\text{fb}] \\
      \hline
      \bbbb resolved jets & 543.52               & 603.52               & 430.37 / 856.11                  & 322.98 / 1173.55                 \\
      \bbbb merged jets   & 145.61               & 75.68                & 51.74 / 112.79                   & 37.84 / 164.59                   \\
      \bbtt               & 207.82               & 263.67               & 181.27 / 388.74                  & 131.84 / 550.09                  \\
      \bbgg               & 458.7                & 297.85               & 196.98 / 460.5                   & 138.45 / 681.37                  \\
      \bbWW               & 1286.37              & 1441.41              & 1039.97 / 1998.72                & 782.64 / 2660.01                 \\
      multilepton         & 989.32               & 855.47               & 596.49 / 1247.61                 & 441.1 / 1739.4                   \\
      \WWgg               & 3944.23              & 3171.88              & 2223.64 / 4587.92                & 1635.5 / 6346.12                 \\
      \ttgg               & 2251.15              & 2015.62              & 1324.6 / 3148.47                 & 936.95 / 4708.71                 \\
      Combination         & 112.78               & 64.94                & 45.46 / 96.0                     & 32.98 / 136.83                   \\
      \hline
    \end{tabular}
  }
\end{table*}

\begin{table*}[ht]
  \centering
  \topcaption{
    Upper limits on the \HH production cross section at 95\% \CL for the JHEP04(2016)126 BM11 benchmark.
    The theoretical uncertainties in the \HH \GGF signal cross section are not considered because we directly constrain the measured cross section.
  }
  \label{tab:JHEP04BM11}
  \scalebox{0.875}{
    \begin{tabular}{ l  c  c  c  c}
      \hline
      Analysis            & Observed [\text{fb}] & Expected [\text{fb}] & Expected $-$/$+$68\% [\text{fb}] & Expected $-$/$+$95\% [\text{fb}] \\
      \hline
      \bbbb resolved jets & 336.07               & 367.19               & 261.84 / 520.87                  & 196.5 / 709.35                   \\
      \bbbb merged jets   & 68.5                 & 35.64                & 24.05 / 53.26                    & 17.54 / 78.67                    \\
      \bbtt               & 120.66               & 156.74               & 108.14 / 230.46                  & 78.98 / 325.65                   \\
      \bbgg               & 315.53               & 184.08               & 120.51 / 286.07                  & 84.85 / 427.2                    \\
      \bbWW               & 672.8                & 744.14               & 536.89 / 1031.86                 & 404.05 / 1373.26                 \\
      multilepton         & 793.05               & 572.27               & 398.43 / 834.59                  & 290.6 / 1170.65                  \\
      \WWgg               & 2748.0               & 2148.44              & 1487.54 / 3150.4                 & 1091.0 / 4405.69                 \\
      \ttgg               & 1204.53              & 1082.03              & 712.92 / 1690.17                 & 498.75 / 2527.74                 \\
      Combination         & 58.28                & 31.74                & 21.82 / 47.17                    & 15.87 / 68.37                    \\
      \hline
    \end{tabular}
  }
\end{table*}

\begin{table*}[ht]
  \centering
  \topcaption{
    Upper limits on the \HH production cross section at 95\% \CL for the JHEP04(2016)126 BM12 benchmark.
    The theoretical uncertainties in the \HH \GGF signal cross section are not considered because we directly constrain the measured cross section.
  }
  \label{tab:JHEP04BM12}
  \scalebox{0.875}{
    \begin{tabular}{ l  c  c  c  c}
      \hline
      Analysis            & Observed [\text{fb}] & Expected [\text{fb}] & Expected $-$/$+$68\% [\text{fb}] & Expected $-$/$+$95\% [\text{fb}] \\
      \hline
      \bbbb resolved jets & 571.6                & 544.92               & 389.27 / 775.16                  & 289.49 / 1061.08                 \\
      \bbbb merged jets   & 560.66               & 284.18               & 195.37 / 424.64                  & 142.09 / 618.64                  \\
      \bbtt               & 257.3                & 310.55               & 216.21 / 452.9                   & 157.7 / 635.27                   \\
      \bbgg               & 380.51               & 288.09               & 193.84 / 438.52                  & 137.29 / 638.47                  \\
      \bbWW               & 1216.94              & 1082.03              & 778.11 / 1509.02                 & 583.28 / 2030.88                 \\
      multilepton         & 896.73               & 898.44               & 629.85 / 1295.95                 & 463.26 / 1800.82                 \\
      \WWgg               & 3890.59              & 2742.19              & 1905.34 / 3977.33                & 1403.23 / 5527.62                \\
      \ttgg               & 1559.83              & 1359.38              & 896.74 / 2112.55                 & 637.21 / 3170.48                 \\
      Combination         & 214.08               & 140.62               & 98.59 / 202.84                   & 72.51 / 280.99                   \\
      \hline
    \end{tabular}
  }
\end{table*}

\clearpage

\begin{figure*}[tbh!]
  \centering
  \includegraphics[width=0.8\textwidth]{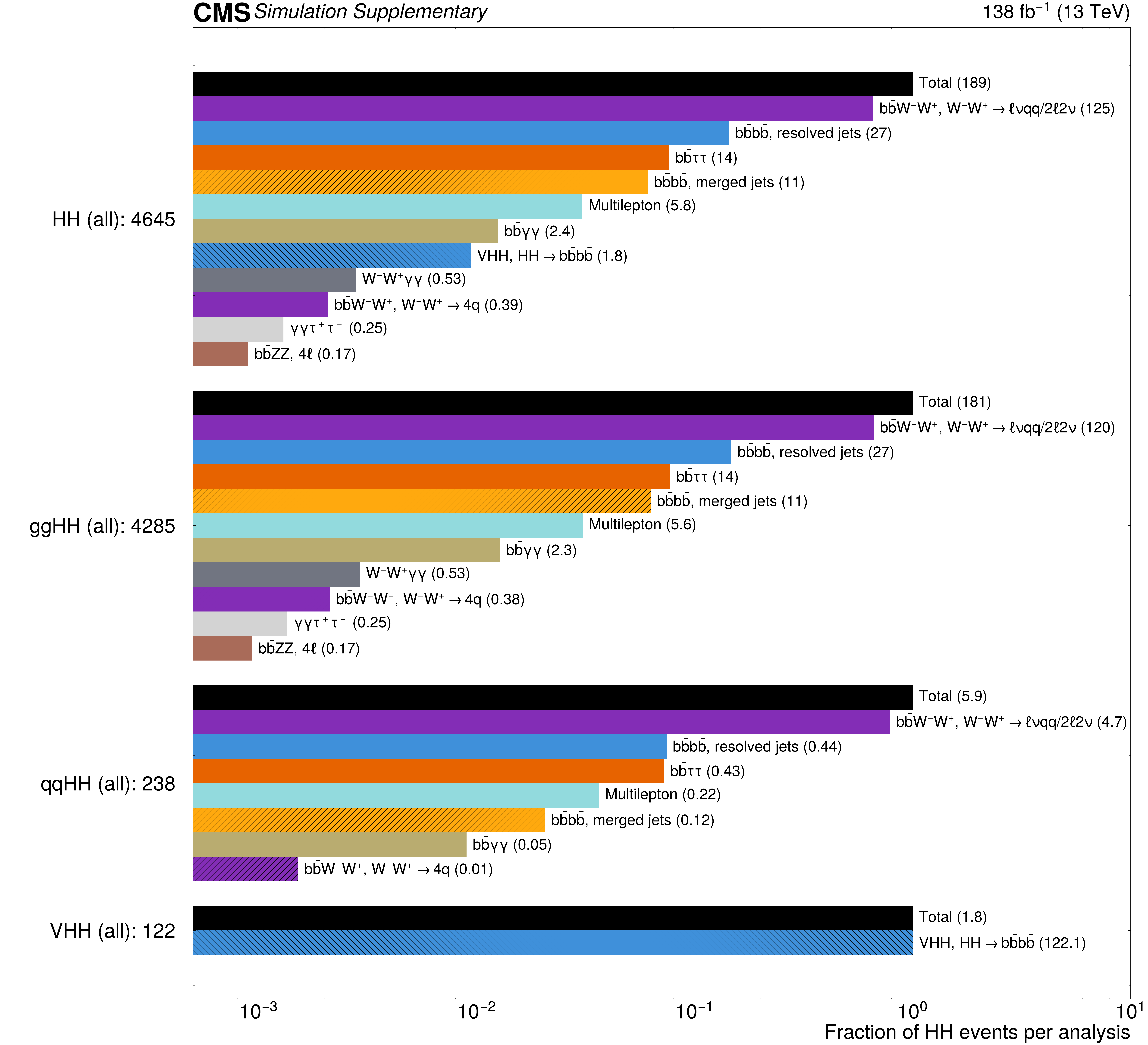}
  \caption{
    A bar chart showing the fraction of all \HH events that are selected in each analysis.
    The numbers shown, correspond to the total of selected events per analysis.
    The results are calculated inclusively and per production mode.
  }
  \label{barchart1}
\end{figure*}

\begin{figure*}[tbh!]
  \centering
  \includegraphics[width=0.7\textwidth]{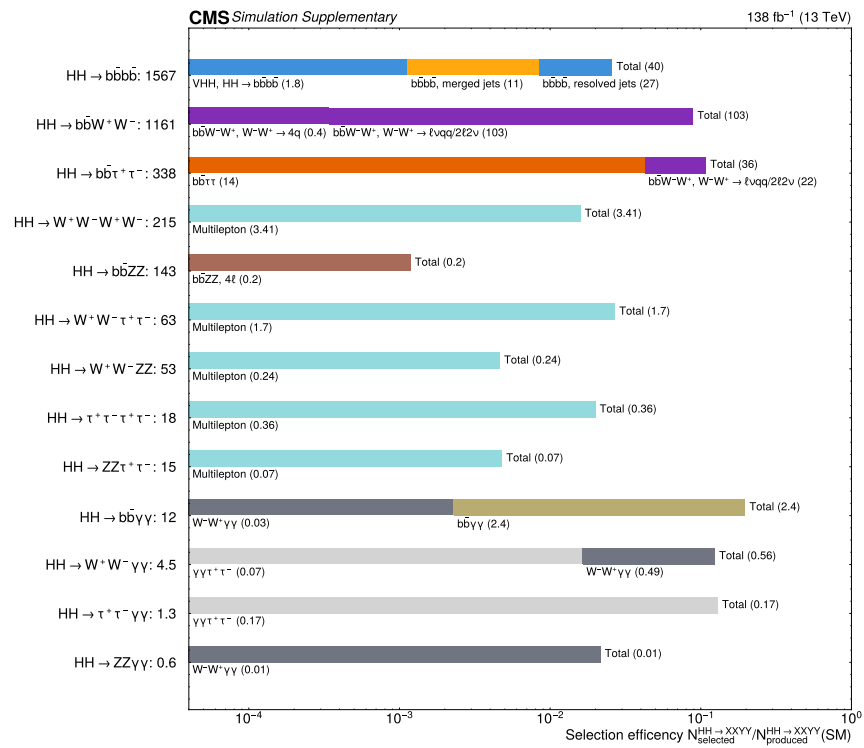}
  \caption{
    A bar chart showing the selection efficiency for the \HH signal per channel.
    The number of selected events per channel is also shown.
  }
  \label{barchart2}
\end{figure*}

}
\cleardoublepage \section{The CMS Collaboration \label{app:collab}}\begin{sloppypar}\hyphenpenalty=5000\widowpenalty=500\clubpenalty=5000\cmsinstitute{Yerevan Physics Institute, Yerevan, Armenia}
{\tolerance=6000
A.~Hayrapetyan, V.~Makarenko\cmsorcid{0000-0002-8406-8605}, A.~Tumasyan\cmsAuthorMark{1}\cmsorcid{0009-0000-0684-6742}
\par}
\cmsinstitute{Institut f\"{u}r Hochenergiephysik, Vienna, Austria}
{\tolerance=6000
W.~Adam\cmsorcid{0000-0001-9099-4341}, J.W.~Andrejkovic, L.~Benato\cmsorcid{0000-0001-5135-7489}, T.~Bergauer\cmsorcid{0000-0002-5786-0293}, M.~Dragicevic\cmsorcid{0000-0003-1967-6783}, C.~Giordano, P.S.~Hussain\cmsorcid{0000-0002-4825-5278}, M.~Jeitler\cmsAuthorMark{2}\cmsorcid{0000-0002-5141-9560}, N.~Krammer\cmsorcid{0000-0002-0548-0985}, A.~Li\cmsorcid{0000-0002-4547-116X}, D.~Liko\cmsorcid{0000-0002-3380-473X}, M.~Matthewman, I.~Mikulec\cmsorcid{0000-0003-0385-2746}, J.~Schieck\cmsAuthorMark{2}\cmsorcid{0000-0002-1058-8093}, R.~Sch\"{o}fbeck\cmsAuthorMark{2}\cmsorcid{0000-0002-2332-8784}, D.~Schwarz\cmsorcid{0000-0002-3821-7331}, M.~Shooshtari, M.~Sonawane\cmsorcid{0000-0003-0510-7010}, W.~Waltenberger\cmsorcid{0000-0002-6215-7228}, C.-E.~Wulz\cmsAuthorMark{2}\cmsorcid{0000-0001-9226-5812}
% ...
\par}
\vskip\cmsinstskip
\dag:~Deceased\\
$^{1}$Also at Yerevan State University, Yerevan, Armenia\\
$^{2}$Also at TU Wien, Vienna, Austria\\
% ...
\end{sloppypar}
\end{document}